\documentclass[fleqn,usenatbib]{mnras}

\usepackage{newtxtext,newtxmath}

\usepackage[T1]{fontenc}
\usepackage{graphicx}
\usepackage{float}
\usepackage{caption}
\usepackage{subcaption}
\usepackage{breqn}
\usepackage{mathtools}
\usepackage{amsmath,amsfonts,bm}
\usepackage{textcomp}

\DeclareRobustCommand{\VAN}[3]{#2}
\let\VANthebibliography\thebibliography
\def\thebibliography{\DeclareRobustCommand{\VAN}[3]{##3}\VANthebibliography}

\usepackage{graphicx}	
\usepackage{amsmath}	
\usepackage{xcolor}

\defcitealias{deVries2023}{Paper 1}

\title[Elliptical instability and convection]{Tidal dissipation due to the elliptical instability and turbulent viscosity in convection zones in rotating giant planets and stars}

\author[N. B. de Vries et al.]{
Nils B. de Vries,$^{1}$\thanks{E-mail: mmnbdv@leeds.ac.uk (NBV)}
Adrian J. Barker,$^{1}$
Rainer Hollerbach$^{1,2}$
\\
$^{1}$ School of Mathematics, University of Leeds, Leeds LS2 9JT, UK\\
$^{2}$ Isaac Newton Institute for Mathematical Sciences, 20 Clarkson Road, Cambridge CB3 0EH
}

\date{Accepted XXX. Received YYY; in original form ZZZ}

\pubyear{2023}

\begin{document}
\label{firstpage}
\pagerange{\pageref{firstpage}--\pageref{lastpage}}
\maketitle

\begin{abstract}
Tidal dissipation in star-planet systems can occur through various mechanisms, among which is the elliptical instability. This acts on elliptically deformed equilibrium tidal flows in rotating fluid planets and stars, and excites inertial waves in convective regions if the dimensionless tidal amplitude ($\epsilon$) is sufficiently large. We study its interaction with turbulent convection, and attempt to constrain the contributions of both elliptical instability and convection to tidal dissipation. For this, we perform an extensive suite of Cartesian hydrodynamical simulations of rotating Rayleigh-B\'{e}nard convection in a small patch of a planet.  We find that tidal dissipation resulting from the elliptical instability, when it operates, is consistent with $\epsilon^3$, as in prior simulations without convection. Convective motions also act as an effective viscosity on large-scale tidal flows, resulting in continuous tidal dissipation (scaling as $\epsilon^2$). We derive scaling laws for the effective viscosity using (rotating) mixing-length theory, and find that they predict the turbulent quantities found in our simulations very well. In addition, we examine the reduction of the effective viscosity for fast tides, which we observe to scale with tidal frequency ($\omega$) as $\omega^{-2}$. We evaluate our scaling laws using interior models of Hot Jupiters computed with MESA. We conclude that rotation reduces convective length scales, velocities and effective viscosities (though not in the fast tides regime). We estimate that elliptical instability is efficient for the shortest-period Hot Jupiters, and that effective viscosity of turbulent convection is negligible in giant planets compared with inertial waves.
\end{abstract}

\begin{keywords}
Hydrodynamics -- planet-star interactions -- instabilities -- convection -- planets and satellites: gaseous planets
\end{keywords}



\section{Introduction}
Tidal deformations and the corresponding dissipation of tidal flows lead to transfers of angular momentum and energy from one body to its companion. This can result in many long-term effects in exoplanetary and close binary systems, such as tidal circularisation of orbits \citep[e.g.][]{binarycirc}, spin-orbit synchronisation \citep[e.g.][]{tidalcircularizationDobbs, Binarysync} and tidal heating \citep[potentially leading to radius inflation, e.g.][]{Bodenheimer_2001}. Perhaps the most extreme outcome is orbital decay and inspiral of a short-period exoplanet, which has potentially been observed for WASP-12b \citep[e.g.][]{M2016,Orbitaldecaysummary,wasp12bdecay}. Indeed, considerable study has gone into understanding the effects of tides in stars and planets, a review of which can be found in \citet{Ogilviereview}. Tidal effects are thought to be especially strong in Hot Jupiters and other short-period exoplanets due to their close proximities to their stars. 

The tidal response in a star or planet is usually split up into an equilibrium or non-wave-like tide, and a dynamical or wave-like tide \citep[e.g.][]{Zahn1977tidesplit,OgilvieIW}. The equilibrium tide is the quasi-hydrostatic fluid bulge rotating around the body \citep[e.g.][]{Zahn1977tidesplit}, while the dynamical tide consists of waves generated by resonant tidal forcing (such as inertial waves in convection zones or internal gravity -- or gravito-inertial -- waves in radiation zones). The equilibrium tide is thought to be dissipated through its interaction with turbulence, usually of a convective nature \citep[][]{Zahn1966,GoldreichNicholson1977,Zahn1989Turbvisceqtide,Goodmaneffvisc,Penev2007,Penev2009a,Penev2009b,Ogilvieeffvisc,Vidal_Barker_2019,Vidal_Barker_2020,Craig2019effvisc,Craig2020effvisc}, or by instabilities of the equilibrium tide itself (which could involve the excitation of waves \citep[e.g.][]{Cebron2010,librationellip,CebronellipHJ,Barker2013,BBO2016,Barker2016}. In this paper we primarily focus on the equilibrium tide and study tidal dissipation due to both the elliptical instability of this flow in convective regions of stars and planets \citep[e.g.][]{Waleffeellipinstab,Ellipticalinstability}, as well as the interaction of the equilibrium flow with the turbulent convection itself. 

The net effect of the equilibrium tide is to deform the body into an ellipsoidal shape (more correctly: prolate spheroidal in the absence of a rotational bulge) that approximately follows the companion. Recently, such a tidal deformation was observed directly for the first time in the Hot Jupiter WASP-103b using the transit method \citep[][]{tidaldeformation}. The elliptical deformation of body 1 due to a second body is represented by the ellipticity, or (dimensionless) tidal amplitude parameter:
\begin{equation}
    \epsilon=\left(\frac{m_2}{m_1}\right)\left(\frac{R_1}{a}\right)^3,
\end{equation} 
where $m_1$ and $m_2$ are the masses of bodies 1 and 2, i.e. the planet and host star, respectively, $R_1$ is the radius of body 1, and $a$ is the orbital separation (semi-major axis). This is essentially a measure of the maximum dimensionless radial displacement in the equilibrium tide. The largest estimated elliptical deformation is $\epsilon\approx 0.06$ for WASP-19b \citep[with its $0.78$ day orbit, e.g.][]{wasp19bdeformation}, and it can be similarly large with values $\epsilon\gtrsim 0.01$ for other Hot Jupiters with short orbital periods (or in the very closest binary stars).

This elliptical deformation of the streamlines allows the elliptical instability to operate \citep[][]{Waleffeellipinstab,Ellipticalinstability}. This elliptical deformation, no matter how small, can potentially excite pairs of inertial waves inside the planet. These waves couple with the deformation \citep[][]{Waleffeellipinstab}, leading to exponential growth of their amplitudes. This mechanism is in essence a triadic (three-wave) resonance interaction. To excite these inertial waves in planets, energy must be extracted from the tidal flow. Thus, rotational or orbital energy is transferred into these waves and when these waves dissipate this energy is then converted into heat. In this way, the instability results in tidal dissipation. 

\begin{figure}
    \centering
    \includegraphics[width=\linewidth,height=70mm]{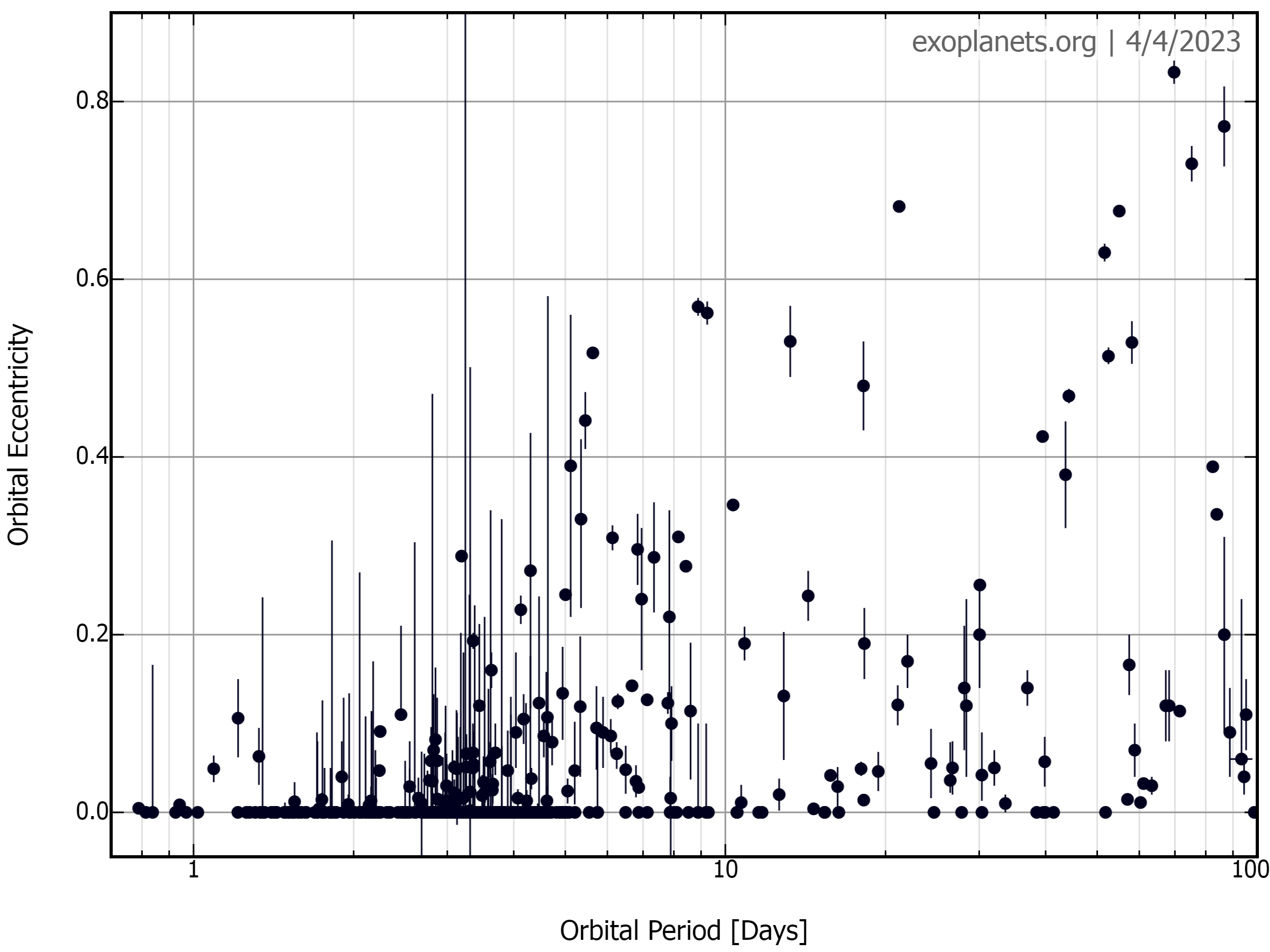}
\caption{Eccentricity distribution of exoplanets with $P_{\mathrm{orb}}<100$ days and masses $M>0.3M_J$. Those with $P_{\mathrm{orb}}<10$ days are referred to as ``Hot Jupiters". Exoplanets with periods $P_{\mathrm{orb}}<3$ days have eccentricities $e<0.2$, but most of these planets have $e\approx 0$, whereas those with $P_{\mathrm{orb}}>10$ days exhibit a wide range of eccentricities. Figure produced from \url{exoplanets.org} \citep{exoplanets.org}.}
    \label{fig:eccentricitydistribution}
\end{figure}

However, if the waves are viscously damped -- by either the (tiny) molecular viscosity of the fluid or by a turbulent viscosity -- before they can grow, the instability cannot operate. Larger deformations $\epsilon$ result in faster growth of the waves and means that they can overcome larger viscosities. An easily deformable, close-in planet is therefore favoured for occurrence of this instability, which suggests why we are considering it as a potential tidal mechanism for Hot Jupiters. Specifically, it is thought that the elliptical instability could be one of the processes responsible for circularisation of planets with very short orbital periods up to 3 days and tidal locking, i.e. tidal spin-orbit synchronisation, for planets with orbits up to 15 days \citep[][]{Barker2013,Barker2016}. We show the eccentricity distribution of these planets as a function of their orbital period from observations in Fig.~\ref{fig:eccentricitydistribution}. Nearly all Hot Jupiters with periods $P_{\mathrm{orb}}<3$ days have eccentricities $e\approx 0$, and those with $P_{\mathrm{orb}}<10$ days have a strong preference for circular orbits or small $e$ values, whereas those with $P_{\mathrm{orb}}>10$ days have a wide range of eccentricities. This distribution is thought to result from tidal dissipation inside these planets, but based on prior theoretical results it does not appear to be explained by the elliptical instability in isolation. We thus appear to require a more efficient mechanism of tidal dissipation in Hot Jupiters.

To parameterise the rate of tidal dissipation we often use the (modified) tidal quality factor $Q'$, first defined when considering tidal evolution in the solar system \citep[][]{Goldreich1963firstQ}. $Q'$ is a measure of the total energy stored in the tide ($E_0$) divided by the energy dissipated in one tidal period, i.e.,
\begin{equation}
    Q'=\frac{3}{2k_2}\frac{2\pi E_0}{\int |\dot{E}|dt}.
\end{equation}
Here, $\dot{E}$ is the rate at which energy is dissipated and $k_2$ is the Love number, which is related to the density distribution (being smaller for more centrally-condensed bodies, with $k_2=3/2$ for a homogeneous fluid body). A higher value of $Q'$ corresponds to lower tidal dissipation and vice versa. Thus lower values of $Q'$ correspond to shorter tidal evolutionary timescales. However, the actual tidal dissipation timescales depend on both the process in question and the periods and masses of the planet and companion. The factor $Q'$ is not a constant parameter, and will depend on tidal frequency and amplitude as well as the internal structure and rotation of the body. However, it is thought to take values of approximately $10^1-10^2$ for rocky planets \citep[][]{GOLDREICH}, approximately $10^4-10^5$ for Jupiter \citep[][]{Lainey2009JupiterQ} and Saturn \citep[][]{Lainey2012SaturnQ,Lainey2017SaturnQ_Cassini}, and approximately $10^6$ or smaller for Hot Jupiters \citep[e.g.][]{Ogilviereview}.

The effect of the elliptical instability on tidal dissipation has been studied previously in simulations using a local Cartesian box model located within the convection zone of a planet or star, both with \citep[][]{Barker2014} and without \citep[][]{Barker2013} weak magnetic fields. The former study found that the elliptical instability leads to bursty behaviour, where the inertial waves generated by the instability interact with geostrophic columnar vortical flows produced by their nonlinear interactions. Similar behaviour features in global hydrodynamical simulations of the elliptical instability \citep[][]{Barker2016}, where zonal flows take the place of columnar vortices in the resulting dynamics. Such dynamics might be referred to as ``predator-prey" dynamics, where columnar vortices or zonal flows can be thought of as the predators and the inertial waves as the prey. In this analogy the columnar vortices feed off the inertial waves, and as the energy in these vortices increases inertial waves become suppressed. Once the energy in the inertial waves decreases, the vortices also consequently decay until inertial waves can grow again, and the cycle starts anew. Upon taking magnetic fields into account in the local model, the behaviour changed from bursts to sustained energy input into the flow, as magnetic fields break up or prevent formation of strong vortices \citep[][]{Barker2014}. Similar sustained behaviour is observed if vortices are damped by an artificial frictional force mimicking Ekman friction due to rigid (no-slip) boundaries \citep[e.g.][]{LeReun2017}. 

These prior studies set out to analyse the elliptical instability in the convective regions of planetary (or stellar) interiors, but did not incorporate convection explicitly (except perhaps by motivating a choice of viscosity). The interaction of the elliptical instability with convection has been studied within linear theory \citep[e.g.][]{Ellipticalinstability,LeBars2006_linellipcylinder}, experimentally in cylindrical containers \citep[e.g.][]{Lavorelexperimentalellip} and using idealised laminar global simulations in a triaxial ellipsoid \citep[e.g.][]{Cebron2010}. However, these studies mainly focused on heat transport instead of tidal dissipation, which is our focus in this work.

Due to the introduction of convection another mechanism of tidal dissipation arises in the system in addition to the elliptical instability. If convection is sufficiently turbulent, it is expected that it will damp the tidal flow, which can be parameterised as an effective viscosity $\nu_{\mathrm{eff}}\gg \nu$ (where $\nu$ is the tiny molecular viscosity). The efficiency of this effective viscosity as a tidal dissipation mechanism has long been a subject of debate, particularly in the fast tides regime when the tidal frequency $\omega$ exceeds the dominant convective frequency $\omega_c$. In this case, the effective viscosity is expected to be reduced, but its scaling behaviour with $\omega$ is debated. Based on arguments stemming from mixing-length theory (MLT), \citet[][]{Zahn1966,Zahn1989Turbvisceqtide} argued that it is expected that the effective viscosity is proportional to the distance travelled by an eddy, i.e. the characteristic convective length scale. However, if the convective timescale exceeds the tidal timescale, the convective eddies can only interact with the tidal flow on the length scales an eddy can travel in a tidal period. Following this argument, the length scale, and thus the effective viscosity, is reduced according to $\nu_{\mathrm{eff}}\propto \omega_c/\omega$. \citet[][]{GoldreichNicholson1977} on the other hand argued that only convective eddies with a frequency similar to the tidal frequency, i.e. $\omega_c\sim\omega$, could contribute. These so-called `resonant' eddies would then require both a smaller velocity and smaller length scale to achieve this `resonant' frequency. Following a Kolmogorov scaling argument, this results in an effective viscosity scaling as $\nu_{\mathrm{eff}}\propto (\omega_c/\omega)^{-2}$.

Many works have been devoted to finding the correct scaling using numerical and asymptotic methods. The initial works of \citet[][]{Penev2007,Penev2009a,Penev2009b} found evidence for the $\omega^{-1}$ scaling, but did not probe very far into the fast tides regime (i.e.~they considered $\omega/\omega_c=\mathcal{O}(1)$). Subsequent works \citep[][]{Ogilvieeffvisc, Vidal_Barker_2019,Vidal_Barker_2020, Craig2019effvisc,Craig2020effvisc} found strong evidence to favour the $\omega^{-2}$ scaling for fast tides ($\omega \gtrsim 10\omega_c$), although a weaker ``intermediate scaling" closer to $\omega^{-1}$ (with exponent between $-1$ and $-1/2$) has been observed for $\omega\sim \omega_c$ \citep[][]{Vidal_Barker_2019,Craig2020effvisc,Vidal_Barker_2020}. In this paper we build upon \citet[][]{Craig2019effvisc,Craig2020effvisc}, which used local box simulations to examine the effective viscosity of convective turbulence acting on the tidal flow. Here we also take into account the influence of rapid rotation on the convection, which is expected to be important in giant planets and young rapidly-rotating stars. We also use an elliptical background flow that corresponds more closely with the equilibrium tide, compared with the oscillating shear flow used in e.g.~\citet{Craig2019effvisc,Craig2020effvisc}, which is stable to elliptical instability.

\begin{figure}
    \centering
    \includegraphics[width=0.8\linewidth]{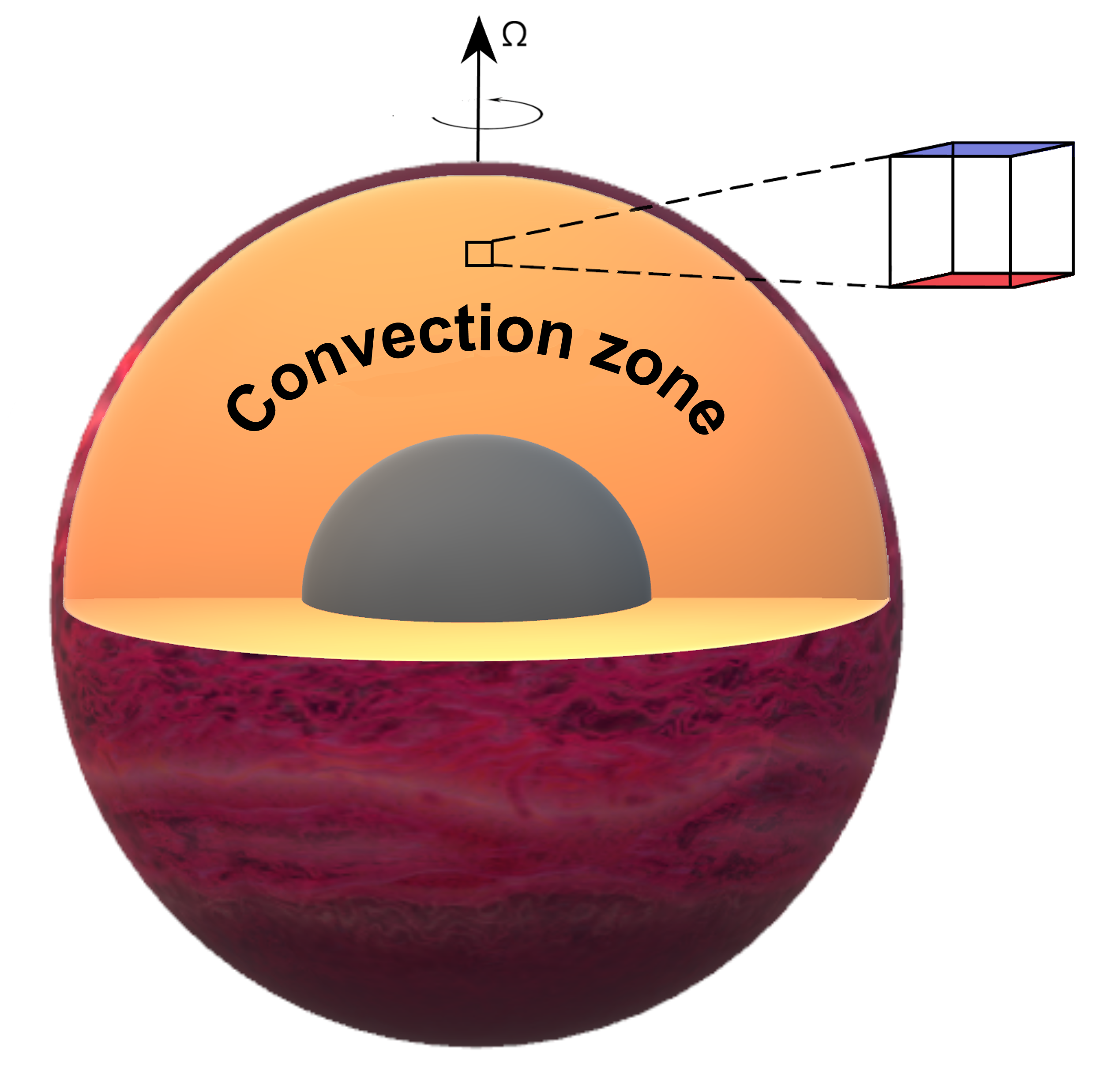}
    \caption{Location of the local box in the convection zone of a Hot Jupiter. We indicate the rotation axis and the local temperature gradient, which is represented by the red (hot) and blue (cold) sides of the box.}
    \label{fig:HJlocalbox}
\end{figure}

In \citet[][]{deVries2023}, hereafter \citetalias[][]{deVries2023}, the non-linear interactions of the elliptical instability and convection were studied. We found evidence for both energy injection by the elliptical instability, as well as from the effective viscosity arising from the interaction of turbulent convection with the equilibrium tide. On the other hand, the generation of convective Large Scale Vortices (LSVs), which on a planet may instead correspond with zonal flows at mid to low latitudes \cite{CurrieconvRMLT}, was found to inhibit the elliptical instability for the Ekman numbers (ratio of viscous to Coriolis forces) we considered.

In \citetalias[][]{deVries2023} we focused on exploring the fluid dynamical interactions of the elliptical instability and convection. Here we build upon \citetalias[][]{deVries2023} by endeavouring to quantify the tidal dissipation that arises from the elliptical instability as well as the effective viscosity of the convection acting on the equilibrium tide. To this end we will derive temperature-based scaling laws using mixing-length theory and rotating mixing-length theory for key convective quantities such as the vertical convective velocity, dominant length scale and frequency, and verify that they agree with our simulation results. \citet[][]{Craig2020effvisc} obtained empirically three regimes for the effective viscosity (as a function of the ratio of tidal to convective frequencies) in non-rotating simulations based on the aforementioned convective quantities. Here we apply rotating mixing-length theory to their scaling laws to derive corresponding expressions for the effective viscosity in the rapidly rotating regime (relevant for giant planets). We compare these predictions with simulations to validate using these prescriptions for rotating convection. If these agree, we might be able to use these expressions to compute the effective viscosity using realistic values of the Rayleigh number, Ekman number, viscosity and tidal deformation for giant planets and stars.
To this end we continue to explore the local box model \citep[][]{Barker2013,Barker2014,LeReun2017} -- representing a small patch of the polar regions of a planet or star (see Fig.~\ref{fig:HJlocalbox}) -- from \citetalias[][]{deVries2023}. We extend the range of parameters they surveyed by running additional simulations varying the Ekman number, Rayleigh number and ellipticity. Finally, we will apply our scaling laws to make predictions for $Q'$ -- based on interior models of Hot Jupiters obtained using the MESA code -- due to the elliptical instability and turbulent effective viscosity and compare these to the linearly-excited inertial waves. 

In Section \ref{sec:modelsetup} we will describe the model used and discuss the scaling law predictions obtained using RMLT. In Section \ref{sec:scalinglaws} we derive scaling laws from our numerical simulations and compare them with our theoretical predictions. In Section \ref{sec:astroapplication} we outline the astrophysical implications of our results, by generating interior profiles of a Jupiter-like and a Hot Jupiter planet using the MESA code, which we use to evaluate the dissipation of the equilibrium tide and that due to inertial waves. We finally present a discussion and our conclusions in Section \ref{sec:discussion}.

\section{Model Setup}
\label{sec:modelsetup}

\subsection{The elliptical instability}
\label{sec:ellip}

We build upon the results of \citetalias[][]{deVries2023}, using the same setup, so we only give a brief overview of our model here. (See \citetalias[][]{deVries2023} for a more detailed description.) In the frame rotating with the tidal bulge, the equilibrium tide is an elliptical flow inside the planet. We define the rotation rate $\gamma$ of this flow as the difference of the planetary spin $\Omega$ and the orbital rotation rate $n$, i.e.~$\gamma\equiv\Omega-n$. We work in the frame rotating with the planet at the rate $\Omega$, modelling a small patch of an equilibrium tidal flow, which we treat as a background flow $\textbf{U}_0$. 
Following \cite{Barker2013}, the equilibrium tide can be written in this frame as:
\begin{equation}
    \textbf{U}_0=\mathrm{A}\textbf{x}=-\gamma \epsilon\begin{pmatrix}
\sin(2\gamma t) & \cos(2\gamma t) & 0\\
\cos(2\gamma t) & -\sin(2\gamma t) & 0\\
0 & 0 & 0
\end{pmatrix}\textbf{x},
\label{eq:backgroundflow}
\end{equation}
where $\textbf{x}$ represents the position vector from the centre of the planet in the frame rotating with the planet. This represents the exact equilibrium tide of a uniformly rotating incompressible fluid body perturbed by an orbiting companion \citep{Chandrasekhar_ellipsoidal,BBO2016}, and also approximates the main features of the equilibrium tide in more realistic models \citep[e.g.][]{OgilvieIW,Barker2020}.

The elliptical instability operates when two inertial waves have frequencies that approximately add up to the tidal frequency $2\gamma$ \citep{Ellipticalinstability}. In the short wavelength limit, this occurs for two waves with frequencies $\omega=\pm\gamma$. These waves must also satisfy the inertial wave dispersion relation:
\begin{equation}
    \omega=\pm2\Omega\cos(\theta),
    \label{eq:dispersionrelation}
\end{equation}
where $\theta$ is the angle between the wavevector and rotation axis, which therefore allows us to determine that the elliptical instability can only operate in the interval $n=[-\Omega,3\Omega]$. Outside this interval no inertial waves exist that satisfy both the dispersion relation and $\omega=\pm\gamma$. Finally, it is known that the elliptical instability grows exponentially (in linear theory) at a rate proportional to $\epsilon\gamma$ \citep{Ellipticalinstability}. For clarity of presentation $\gamma=\Omega$ is chosen in this work, unless otherwise mentioned, resulting in $n=0$, i.e.~strictly representing the unphysical case where there is no rotation of the bulge. The body in question is not rotating around its companion which causes the tidal effects. However, it turns out that for simulations the only linear effect of choosing a different value of $\Omega$, and therefore a non-zero value of $n$, would be to modify the fastest growing mode, and also its growth rate \citep[e.g.][]{Ellipticalinstability,Barker2013,deVries2023}.

\subsection{Governing equations and setup of the simulations}
\label{sec:governeq}
\label{sec:simsetup}

We use Rotating Rayleigh-B\'enard Convection (RRBC) as our model to study the convective instability, as it is the simplest model of rotating convection \citep[][]{Chandrsekharbook} which allows us to study its interaction with the elliptical instability. In addition, we use the Boussinesq approximation, which is appropriate for studying small-scale convective (and wavelike) flows. Using the Boussinesq approximation is valid if the vertical size of our simulated domain $d$ is much smaller than a pressure or density scale height and the flows in the simulation are much slower than the sound speed \citep[][]{bousapprox}. However, by choosing this approximation we neglect variations in properties such as the density and temperature. Furthermore, since we require small vertical scales, we cannot model the largest-scale convective flows using this approximation.

The box in our current setup represents a polar region, which we have illustrated in Fig.~\ref{fig:HJlocalbox}. This location arises from our choice of rotation axis, which points in the $z$-direction, and temperature profile, which solely depends on $z$. By making this choice the local rotation and gravity vectors are either aligned or anti-aligned (depending on the sign of $\Omega$) and thus we are located at the poles. The aforementioned temperature profile of the conduction state, i.e.~the temperature gradient introduced by the hot and cold plates at the bottom and top of our box, respectively, and about which we perturb, is given by:
 \begin{equation}
     \alpha g (T-T_0)=\frac{z N^2}{d},
 \end{equation}
 where $g$ is the local gravitational acceleration (assumed constant), $\alpha$ is the (constant) thermal expansion coefficient and $N^2$ is the (constant) squared Brunt-V\"ais\"al\"a (or buoyancy) frequency, which is (negative) positive for (un)stable stratification. We choose $T_0=0$ without loss of generality. As a result, the temperature at the bottom is $T(z=0)=0$, while the temperature at the top is $T(z=d)=N^2/(\alpha g)$, such that the temperature difference is $\Delta T =-N^2/\alpha g$. Note that the introduction of buoyancy modifies the (gravito-)inertial wave dispersion relation to:
\begin{equation}
    \omega^2=4\Omega^2\cos^2(\theta)+N^2\sin^2(\theta).
    \label{eq:dispersionrelation+buoy}
\end{equation}

To non-dimensionalise the governing equations we scale lengths by the vertical domain size $d$ (representing the distance between the plates), times by the thermal timescale $d^2/\kappa$, and we consequently scale velocities with $\kappa/d$. Finally, we use $T=\Delta T\theta$ to scale the temperature (i.e. the temperature is scaled by the temperature difference between the plates). Using these non-dimensionalisations and the Boussinesq approximation, the governing equations, in the frame rotating at the rate $\Omega$ about $z$, for the dimensionless perturbations $\textbf{u}$ and $\theta$ to the background flow $\textbf{U}_0$ and temperature profile $T(z)$ are:
\begin{align}
    &\frac{D\textbf{u}}{Dt}+\textbf{u}\cdot\nabla \textbf{U}_0+\frac{ \text{Pr}}{ \text{Ek}}\hat{\textbf{z}}\times \textbf{u} = -\nabla p +  \text{PrRa}\theta \hat{\textbf{z}} + \text{Pr}\nabla^2\textbf{u},
    \label{eq:maingovern} \\
    &\nabla \cdot \textbf{u}=0,
    \label{eq:govern2} \\
    &\frac{D\theta}{Dt}-u_z=\nabla^2\theta,
    \label{eq:govern3}
\end{align}
where
\begin{equation}
    \frac{D}{Dt}\equiv\frac{\partial}{\partial t}+\textbf{U}_0\cdot \nabla+ \textbf{u}\cdot \nabla,
\end{equation}
with $\textbf{u}=(u_x,u_y,u_z)$ and $p$ being the perturbation to the pressure.
The non-dimensional parameters describing the convection are the Rayleigh, Ekman and Prandtl numbers:
\begin{equation}
 \mathrm{Ra} = \frac{\alpha g (-N^2) d^4}{\nu\kappa}, \qquad \mathrm{Ek} = \frac{\nu}{2\Omega d^2}, \qquad \mathrm{Pr}=\frac{\nu}{\kappa},
\end{equation}
where $\nu$ and $\kappa$ are the constant kinematic viscosity and thermal diffusivity. Due to the equilibrium tidal background flow there are two additional dimensionless numbers in the system: $\epsilon$ and $\gamma$ (and there would also be $n$ if we allowed rotation of the bulge). Finally, we can relate the Rayleigh number and dimensional squared buoyancy frequency: $N^2 = -\text{Ra\,Pr}\,\kappa^2/(\alpha g d^4)$. Upon setting $\text{Pr}=1$ we find in dimensionless (thermal time) units: $N^2=-\text{Ra}$. 

Our simulations are executed in a small Cartesian box of dimensionless size $[L_x,L_y,1]$ with $L_x=L_y=L$. As in \citetalias[][]{deVries2023}, to fully resolve bursts of the elliptical instability in tandem with the convective LSV we set $L=4$ in most simulations. However, the simulations that measure properties unrelated to the elliptical instability are executed in a smaller box with $L=2$. This box size ensures the LSV is still present, and the results are therefore similar to those with $L=4$. From the appendix of \citetalias[][]{deVries2023}{}{} we infer that the effective viscosity (without elliptical instability) is unaffected by this variation of the box size. The boundary conditions are periodic in the horizontal directions, and stress-free and impermeable in the vertical direction. We have chosen these boundary conditions because they are probably more relevant in the deep interior of a planet, far removed from any boundaries, than no-slip boundary conditions. The vertical boundary conditions are therefore: $u_z(z=0)=u_z(z=1)=0$, $\partial_z u_x(z=0)=\partial_z u_x(z=1)=\partial_z u_y(z=0)=\partial_z u_y(z=1)=0$. By choosing impermeable vertical boundaries the convection in our box represents a single convection cell in the vertical. Finally, vertical boundary conditions for the temperature perturbation are chosen to be perfectly conducting, with $\theta(z=0)=\theta(z=1)=0$.

The simulations are performed using the Snoopy code \citep{snoopycode}, which implements a Fourier pseudo-spectral method using FFTW3 in a local Cartesian box. We use a sine-cosine decomposition in $z$ and shearing waves (i.e.\ time-dependent Fourier modes) in $x$ and $y$ to account for the linear spatial dependence of the background flow. A 3rd-order Runge-Kutta scheme is used for the time-stepping, together with a CFL safety factor to ensure the timesteps are small enough to capture non-linear effects, usually set to 1.5. The anti-aliasing in the code uses the standard 2/3 rule \citep{boyd2001chebyshev}. A variety of different Rayleigh numbers were analysed using the simulations. The values of the Rayleigh number are typically reported using the supercriticality $R\equiv\text{Ra}/\text{Ra}_c$ for clarity, where $\text{Ra}_c$ is the onset Rayleigh number (determined numerically). The range of the studied supercriticalities at $\text{Ek}=5\cdot10^{-5.5}$ is from $2$ to $20$. The studied values of $\epsilon$ range from $0.01$ to $0.20$, and the Ekman number ranges from $5\cdot10^{-4.5}$ to $5\cdot10^{-6}$.

\subsection{Energetic analysis of simulations}
\label{sec:diagnostics}
 
Following \citetalias[][]{deVries2023} we derive the kinetic energy equation by taking the dot product of $\textbf{u}$ with Eq.~\ref{eq:maingovern} and subsequently volume-averaging all quantities, where the latter is defined as: $\langle X\rangle=\frac{1}{L^2d}\int_V X\text{ }dV$. We obtain: 
 \begin{equation}
     \frac{d}{dt} K= I + \langle \text{PrRa}\theta u_z\rangle - D_{\nu},
     \label{eq:kin_energy}
 \end{equation}
where we have defined the total kinetic energy $K$, the energy transfer rate from the background tidal flow $I$ and the mean viscous dissipation rate $D_\nu$ according to:
\begin{equation}
     K\equiv\frac{1}{2}\langle|\textbf{u}|^2\rangle, \quad I\equiv-\langle\textbf{u}\text{A}\textbf{u}\rangle,  \quad D_{\nu}\equiv-\text{Pr} \langle \textbf{u} \cdot \nabla^2 \textbf{u} \rangle.
 \end{equation}

 To obtain an equation for the thermal (potential) energy when $\textrm{Ra}>0$, we multiply Eq. \ref{eq:govern3} by $\textrm{PrRa}\theta$ and average over the box to obtain:
\begin{equation}
    \frac{d}{dt}P=\langle\text{PrRa}\theta u_z\rangle-D_{\kappa},
    \label{eq:therm_energy}
\end{equation}
where we have defined the mean thermal energy $P$ and the mean thermal dissipation rate $D_\kappa$ as:
\begin{equation}
    P\equiv\text{PrRa}\frac{1}{2}\langle\theta^2\rangle, \quad D_\kappa\equiv-\text{PrRa}\langle\theta\nabla^2\theta\rangle. 
\end{equation}     

The total energy is $E=K+P$, which thus obeys:
\begin{equation}
    \frac{d}{dt}E=I+2\langle\text{PrRa}\theta u_z\rangle-D_\nu-D_\kappa=I+2\langle\text{PrRa}\theta u_z\rangle-D,
    \label{eq:Energ_equation}
\end{equation}
where $D=D_\nu + D_\kappa$ is the total dissipation rate. In a steady state, i.e. no change in time of the total energy, it is expected that the (time-averaged value of the) energy injected together with the buoyancy work balances the total dissipation. Since there are two energy injection terms, the total dissipation cannot be used directly to infer tidal dissipation rates. However, the energy injected by the tide must be dissipated if a steady state is to be maintained. Therefore, to interpret the tidal energy dissipation rate we examine the tidal energy injection rate $I$. (When $\textrm{Ra}<0$, the thermal energy is $-P$ and a minus sign is introduced into both terms on the RHS of Eq.~\ref{eq:therm_energy}. The buoyancy work terms then cancel between Eq.~\ref{eq:kin_energy} and Eq.~\ref{eq:therm_energy}, leaving only $I$ and $D$ in Eq.~\ref{eq:Energ_equation} such that in steady state $I\approx D$.)

Since we know both the elliptical instability \citep{Barker2013} and convection \citep[e.g.][]{CelineLSV,FavierLSV} in isolation can produce geostrophic flows such as vortices, we introduce further diagnostics to analyse these flows and their role in any possible bursty dynamical behaviour. To do this, we decompose the total energy injection from the background flow according to
\begin{equation}
    I = I_{2D}+I_{3D},
\end{equation}
where the barotropic energy injection is defined as $I_{2D}=-\langle\textbf{u}_{2D}\text{A}\textbf{u}_{2D}\rangle$
and the baroclinic energy injection is defined as $I_{3D}=-\langle\textbf{u}_{3D}\text{A}\textbf{u}_{3D}\rangle$.
$I_{2D}$ (and $\textbf{u}_{2D}$) are defined to include all (geostrophic) modes where the wavevector has only non-vanishing $x$ and $y$ components, with $k_z=0$, and $I_{3D}$ (and $\textbf{u}_{3D}$) includes all the modes with $k_z\neq0$. Because pure inertial waves with $k_z=0$ have $\omega=0$, and this work is concerned with convectively unstable simulations, i.e. no gravity waves exist which could have non-zero frequencies even when $k_z=0$, this decomposition can be crudely thought of as a decomposition into waves/convective eddies ($I_{3D}$) and geostrophic vortices ($I_{2D}$). We have found that at small ellipticities the time-averaged energy input into the vortical motions $I_{2D}$ is approximately zero \citep[or small, see also][]{Barker2013}, but that the input into the waves $I_{3D}$ is on average non-zero (which it must be when the elliptical instability operates) and clearly demonstrates any bursty behaviour observed. Based on this observation, only results derived from $I_{3D}$ will be plotted in this paper.

Arguments to describe scaling laws for the dissipation due to the elliptical instability were first proposed in \citet[][]{Barker2013} by (crudely) picturing the instability saturation as involving the most unstable single mode whose amplitude saturates when its growth rate ($\sigma$) balances its nonlinear cascade rate. Thus, if the most important mode of the elliptical instability satisfies $\sigma\sim k u$, where $k$ is its wave number magnitude and $u$ is its velocity amplitude, then we find $u\sim\epsilon\gamma/k$. The total dissipation rate $D$ therefore scales as $D\sim u^2\sigma\sim \epsilon^3\gamma^3/k^2$. Thus, in such a statistically-steady state the dissipation and energy injection rate are expected to scale as
\begin{equation}
    D=I\propto \epsilon^3.
    \label{eq:epsiloncubed}
\end{equation}
If this scaling law holds, the dissipation falls off rapidly as the orbital period of the planet increases, since $\epsilon \propto P_{\mathrm{orb}}^{-2}$, resulting in $Q'\propto P_{\mathrm{orb}}^4$. The result of crudely applying this is that circularisation of Hot Jupiters would only be predicted out to about three-day orbital periods. In \citetalias[][]{deVries2023} we observed that, when the elliptical instability operates, the energy injection is consistent with either scaling as $\epsilon^{3}$ or possibly as the steeper $\epsilon^6$. We will explore this issue further here using simulations, and also determine the astrophysical implications of these results.

We can also interpret the energy transfer rates $I$ and $I_{3D}$ in terms of an effective viscosity like in \citetalias[][]{deVries2023}, obtaining $\nu_{\mathrm{eff}}$ and $\nu_{\mathrm{eff,}3D}$ respectively. This interpretation is most commonly used to measure the interaction between turbulent convection and the equilibrium tide, but also applies for the elliptical instability. To calculate the effective viscosity, we assume that the tidal flow is viscously dissipated by some spatially and temporally constant kinematic viscosity $\nu_{\mathrm{eff}}$, which will depend in principle upon $\mathrm{Ra}, \mathrm{Ek}, \mathrm{Pr}, \gamma$ and $\epsilon$ (and also $n$, if that was varied), as well as $L$. This viscous dissipation rate should then equal the rate of work done on the convective flow by the tidal flow. Following \citet{Goodmaneffvisc,Ogilvieeffvisc,Craig2019effvisc}, we note that the rate of work done on the convective flow is:
\begin{equation}
    I=-\frac{1}{V}\int_V \textbf{u}\cdot(\textbf{u}\cdot\nabla)\textbf{U}_0 dV.
    \label{eq:I_effvisc}
\end{equation}
To obtain the rate of energy dissipation we define the strain rate tensor for the tidal flow as $e_{ij}^0\equiv\frac{1}{2}(\partial_iU_{0,j}+\partial_jU_{0,i})$, resulting in:
\begin{equation}
    \frac{2\nu_{\mathrm{eff}}}{V}\int_V e_{ij}^0e_{ij}^0 dV=4\nu_{\mathrm{eff}}\gamma^2\epsilon^2.
    \label{eq:strain_effvisc}
\end{equation}
The effective viscosity is then \textit{defined by}
\begin{equation}
    \nu_{\mathrm{eff}}=I/(4\gamma^2\epsilon^2).
\end{equation}

In \citetalias[][]{deVries2023} we found that when the elliptical instability does not operate the convection can still interact with the tidal flow to provide $I\propto\epsilon^2$ such that $\nu_{\mathrm{eff}}$ is independent of $\epsilon$. Our interpretation of this regime as ``convective turbulent viscosity damping the tidal flow" can be understood from crudely applying classical eddy viscosity arguments to the Reynolds stress component $\langle u_iu_j\rangle $ that appears in Eq.~\ref{eq:I_effvisc}. In this approach, the velocity correlation would be proportional to the tidal velocity shear, i.e., $\langle u_iu_j\rangle\propto \nabla\textbf{U}_0$ \citep[see for example Eq.~19 in][]{Terquem2021}{}{} and $|\nabla\textbf{U}_0|\sim\epsilon$, thus leading to $I\propto\epsilon^2$.

In our model we do not consider the evolution of the tidal flow $\textbf{U}_0$. Instead we treat it as a fixed (but time-dependent) background flow. The energy in this background flow is considered to be much larger than the energy in the perturbations. As such any energy transferred from this flow to the perturbations (or vice versa) is negligible compared to the energy in the background flow. Therefore, the background flow itself is not modified in our simulations. As a consequence, our results apply to a snapshot in the evolution of our system in time. This is a reasonable approximation, considering that  timescales of tidal evolution are typically much longer than convective or rotational timescales.

\subsection{Scalings of the effective viscosity using mixing-length theory}
\label{sec:RMLTscalings}

We concluded in \citetalias[][]{deVries2023} that turbulent convection acts to damp the equilibrium tidal flow like an effective viscosity (independently of $\epsilon$). In \citet[][]{Craig2020effvisc}, who studied the effective viscosity in a non-rotating local box model of convection, three different regimes with associated scaling laws for the effective viscosity were observed. The scalings they obtained depend on the convective velocity $u_c$, the convective length scale $l_c$ and the ratio of the tidal frequency $\omega=2\gamma$ to the convective frequency $\omega_c$, and are given by: 
\begin{equation}
  \nu_{\mathrm{eff}} =
    \begin{cases}
      5u_{c}l_{c} & \frac{| \omega|}{\omega_c}\lesssim 10^{-2},\\
      \frac{1}{2}u_{c}l_{c}\bigg(\frac{\omega_c}{\omega}\bigg)^{\frac{1}{2}} & \frac{| \omega|}{\omega_c}\in[10^{-2},5],\\
      \frac{25}{\sqrt{20}}u_{c}l_{c}\bigg(\frac{\omega_c}{\omega}\bigg)^2 & \frac{| \omega|}{\omega_c}\gtrsim 5.
    \end{cases} 
    \label{eq:effviscCraig}
\end{equation}
We have reported the (upper bound) numerical coefficients from \citet[][]{Craig2020effvisc}{}{} here, but wish to clarify that rotation and our different background flow might modify these. Note that the choice of scaling laws for the convective quantities $u_c$, $l_c$ and $\omega_c$ will depend on rotation (and perhaps magnetic fields etc.). Therefore, before we can apply the above scalings, we must derive appropriate scaling laws for these quantities depending on which regime the flow is in and verify that these regimes apply in our numerical simulations. In non-rotating simulations, it is reasonable to set $l_c=d$, pretending that $d$ is the Boussinesq equivalent of a pressure scale height (or mixing length i.e.~multiple of a pressure scale height). However, it is not clear whether this is appropriate for rapid rotation, where we might imagine using a shorter horizontal length scale for $l_c$ would be more appropriate instead, which would reduce the turbulent viscosity. Which of these is appropriate may depend on the intended application, i.e.\ the effective viscosity is not a property of the fluid, but a way to model the interaction between a particular fluid flow and convective flow. From now on we choose $l_c$ to represent a horizontal convective length scale, which is therefore modified by rotation, and we will show that this is a suitable choice to match our simulation results. 
 
We can apply mixing-length theory \citep[MLT,][]{MLTBohmVitense} to predict the scaling laws of convective properties such as convective velocities, length scales, turnover times and effective viscosities. MLT has been applied to non-rotating cases previously \citep[e.g.][]{Zahn1966,Craig2019effvisc,Craig2020effvisc}, but our cases are sufficiently rapidly rotating that we must account for modifications of convective properties by rotation. To do so, we use rotating mixing-length theory \citep[RMLT;][]{StevensonRMLT} to predict scaling laws for rotating convection \citep[following e.g.][]{Barker2014RMLT,Mathis2016RMLT,CurrieconvRMLT}. Within RMLT, the vertical convective velocity, which is expected to be roughly equal to the horizontal velocity on the relevant scales, is given by:
\begin{equation}
    u_{c}\sim d^{1/5}F^{2/5}\Omega^{-1/5},
\end{equation}
where $F$ is the vertical heat flux (more specifically a buoyancy flux with units of $L^2T^{-3}$). We may write this in terms of the standard dimensionless numbers by converting the Rayleigh number to a flux-based Rayleigh number $\text{Ra}_F$, which are related by
\begin{equation}
    \text{Ra}\sim \text{Ra}_F^{2/5} \text{Pr}^{1/5} \text{Ek}^{-4/5}\sim F^{2/5}d^{8/5}\kappa^{-1}\nu^{-1/5}\text{Ek}^{-4/5},
\end{equation}
since $N^2\sim F^{2/5}\Omega^{4/5} d^{-4/5}$ and by definition $\text{Ra}_F=\text{Nu}\text{Ra}$, where $\mathrm{Nu}=F/(-\kappa N^2)$ is a Nusselt number (ratio of total heat flux to conductive flux). Converting to the Rayleigh number (based on a fixed temperature drop or $N^2$) from the flux-based Rayleigh number (based on a fixed heat flux $F$) entails a switch from flux-based scalings to temperature-based (and by extension $N^2$-based) scalings. This switch is necessary as the simulations are executed using a constant temperature difference, i.e. they are temperature-based rather than flux-based. After this switch, RMLT predicts for the convective velocity:
\begin{equation}
    u_{c}\sim \text{RaEk}\frac{\kappa}{d}.
    \label{eq:uc_scaling}
\end{equation}
Furthermore, the dominant horizontal length scale of convection is predicted to scale as
\begin{equation}
    l_c\sim\Omega^{-3/5} F^{1/5} d^{3/5}\sim \frac{\text{Ra}^{1/2}\text{Ek}}{\text{Pr}^{1/2}}d.
    \label{eq:lc_scaling}
\end{equation}
Finally, the convective turnover frequency (based on the horizontal length scale) according to RMLT is
\begin{equation}
    \omega_c\sim\frac{u_{c}}{l_c}\sim \text{Ra}^{1/2}\text{Pr}^{1/2}\frac{\kappa}{d^2}.
\end{equation}
These are the RMLT scalings written in terms of Rayleigh, Ekman and Prandtl numbers. These scalings agree with those found in \citet[][]{Celine2019,Aurnou2020_scalings}, and with many others, indicating that the results found from the Coriolis-Inertia-Archimedean (CIA) balance are in agreement with the predictions of RMLT following \citet[][]{StevensonRMLT}. The three effective viscosity scaling laws in Eq.~\ref{eq:effviscCraig} can be written using these predictions from RMLT as:
\begin{equation}
  \nu_{\mathrm{eff}} \propto
    \begin{cases}
      \text{Ra}^{3/2}\text{Ek}^2\text{Pr}^{-1/2}\kappa& \textrm{low frequency},\\
\text{Ra}^{7/4}\text{Ek}^{2}\text{Pr}^{-1/4}\kappa^{3/2}d^{-1}\omega^{-1/2} & \textrm{intermediate freq.},\\
    \text{Ra}^{5/2}\text{Ek}^{2}\text{Pr}^{1/2}\kappa^{3}d^{-4}\omega^{-2}& \textrm{high frequency}.
    \end{cases} 
    \label{eq:effviscscalings}
\end{equation}
The first of these regimes occurs when the tidal frequency is low, while the rotation rate is high (so that we use RMLT rather than MLT). Naively, this situation seems counter-intuitive because the tidal frequency is related to the rotation rate, but it can occur if the body is close to spin-orbit synchronisation. We have not supplied ranges of $\omega/\omega_c$ for which these apply as we will determine these based on our simulations. Instead we elect to refer to these regimes as the low, intermediate and high frequency regimes, where the frequency in question is the tidal frequency (compared with the convective frequency). Note that these regimes have not been previously verified with simulations of rotating convection interacting with tidal flows (unlike in the non-rotating case).

We can use the scalings in Eqs.~\ref{eq:uc_scaling}, \ref{eq:lc_scaling} and \ref{eq:effviscscalings} to analyse our results as a function of both Rayleigh and Ekman numbers, in regimes attainable by simulations. To analyse our simulation results in terms of the Ekman number we used two approaches: fixing the Rayleigh number and fixing the supercriticality $R={\text{Ra}}/{\text{Ra}_c}$. The second approach modifies the power of the Ekman number scaling as the critical Rayleigh number scales as $\text{Ra}_c\approx3({\pi^2}/{2})^{2/3}\text{Ek}^{-4/3}$ for rapid rotation, which results in $u_c\sim R \textrm{Ek}^{-1/3}$ and $l_c\sim R^{1/2}\text{Ek}^{1/3}$, omitting all parameters which are set to one. This leads to the following changes to $\nu_{\mathrm{eff}}$ scalings: 
\begin{equation}
  \nu_{\mathrm{eff}} \propto
    \begin{cases}
      R^{3/2}\text{Ek}^0 & \textrm{low frequency},\\
     R^{7/4}\text{Ek}^{-1/3}\omega^{-1/2} & \textrm{intermediate freq.},\\
      R^{5/2}\text{Ek}^{-4/3}\omega^{-2}& \textrm{high frequency}.
    \end{cases} 
    \label{eq:effviscscalings_constR}
\end{equation}

For completeness, since some of our simulations enter the regime where rotation is no longer rapid, we include here the scalings of the relevant quantities using non-rotating MLT in terms of Rayleigh and Prandtl numbers:
\begin{equation}
    u_c\sim\textrm{Ra}^{1/2}\textrm{Pr}^{1/2}\frac{\kappa}{d},
\end{equation}
and the relevant length scale in this regime is likely to be comparable with the vertical length scale $d$, i.e. $l_c=d$. It follows that:
\begin{equation}
\omega_c\sim\textrm{Ra}^{1/2}\textrm{Pr}^{1/2}\frac{\kappa}{d^2},
\end{equation}
which is the same scaling obtained previously using RMLT. The three regimes we expect for the effective viscosity using MLT are then:
\begin{equation}
  \nu_{\mathrm{eff}} \propto
    \begin{cases}
      \textrm{Ra}^{1/2}\textrm{Pr}^{1/2}\kappa & \textrm{low frequency},\\
      \textrm{Ra}^{3/4}\textrm{Pr}^{3/4}\kappa^{3/2}d^{-1}\omega^{-1/2} & \textrm{intermediate freq.},\\
      \textrm{Ra}^{3/2}\textrm{Pr}^{3/2}\kappa^{3}d^{-4}\omega^{-2} & \textrm{high frequency}.
    \end{cases} 
    \label{eq:effviscscalings_MLT}
\end{equation}
\begin{figure*}
    \centering\begin{minipage}[b]{0.45\textwidth}
         \centering
         \includegraphics[width=\textwidth]{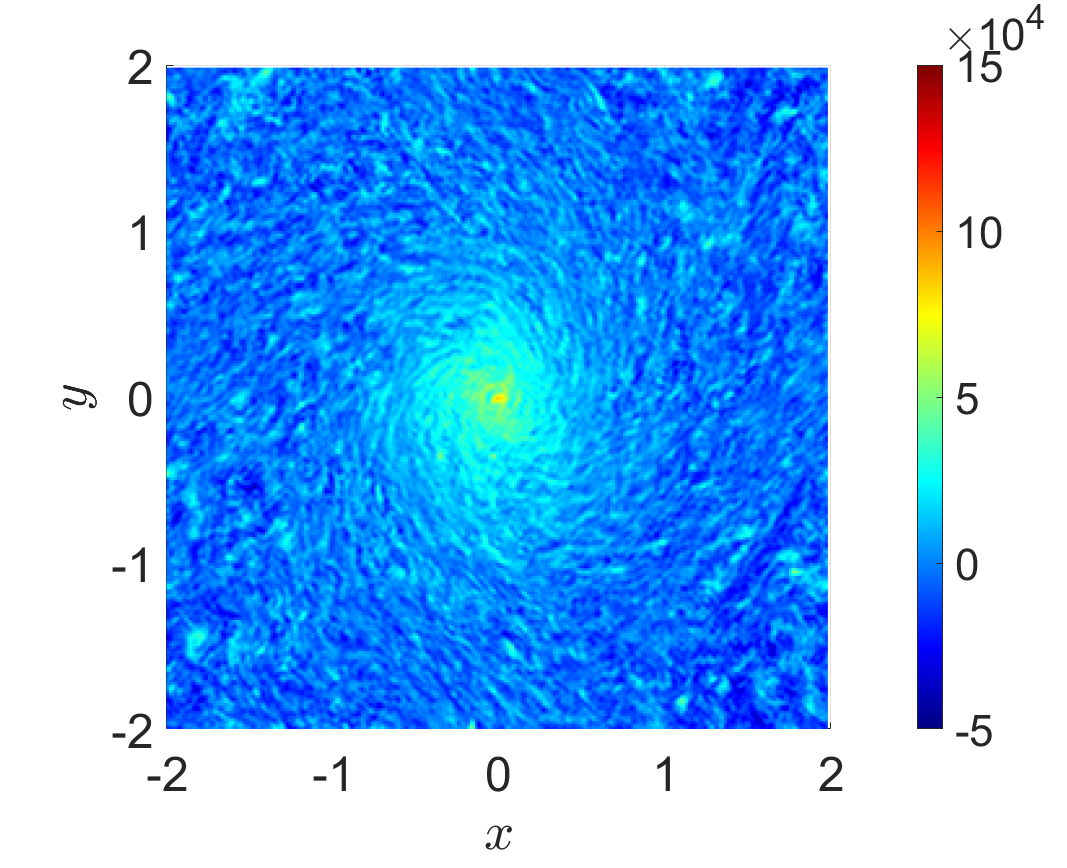}
         \label{fig:vorticityellip}
     \end{minipage}
     \hfill
     \begin{minipage}[b]{0.45\textwidth}
         \centering
         \includegraphics[width=\textwidth]{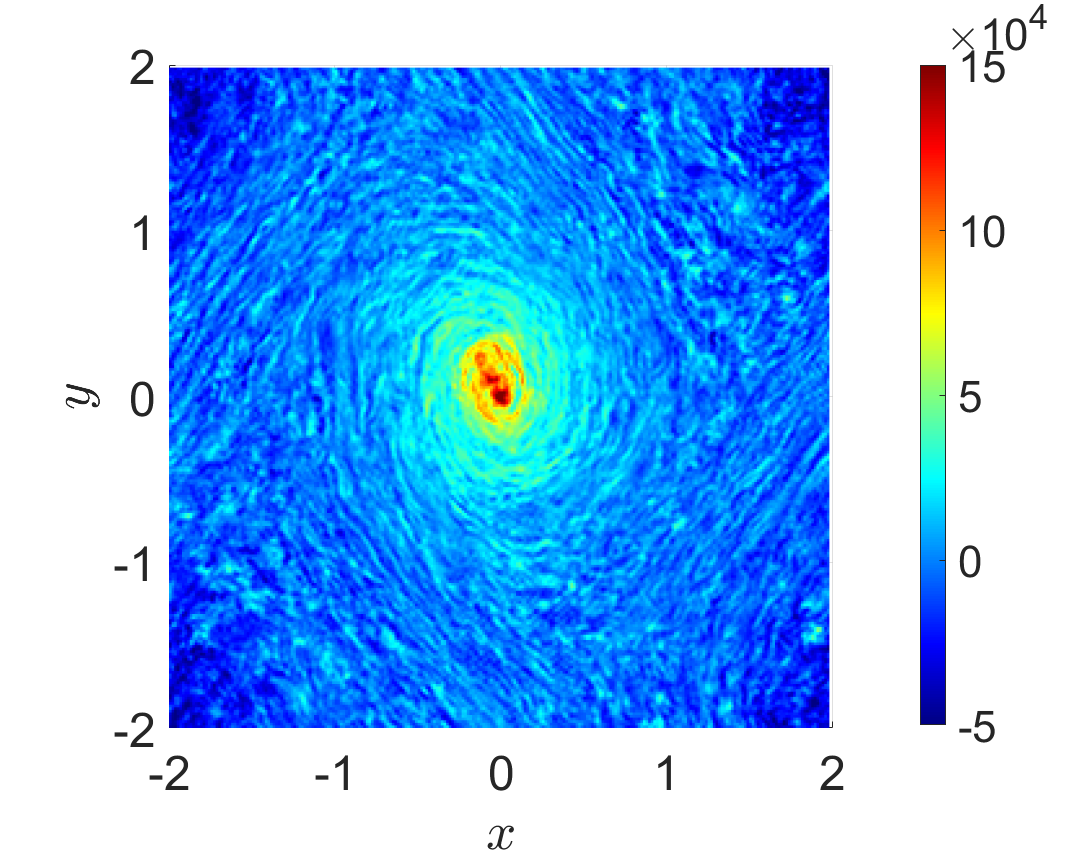}
         \label{fig:vorticityellip+conv}
     \end{minipage}
     \caption{The vertical vorticity perturbation averaged over $z$ ($\langle\omega_z\rangle_z$) of the flow. The cyclonic vortex is centred for clarity in both images. Left: Convection on top of the equilibrium tide in the regime without the elliptical instability $\textrm{Ra}=6\textrm{Ra}_c$, $\epsilon=0.04$, $\text{Ek}=5\cdot10^{-5.5}$ at $t=0.12$. Right: Convection on top of the equilibrium tide in the regime with a strong elliptical instability with $\text{Ra}=6\text{Ra}_c$, $\epsilon=0.1$, $\text{Ek}=5\cdot10^{-5.5}$ at $t=0.12$.}
     \label{fig:vorticity_flowpictures}
\end{figure*}
The high frequency regime within non-rotating MLT is unlikely to occur in our simulations as that regime only applies when the tidal frequency is high, yet the rotation rate is low. It is however likely to be important in reality, for example inside spun-down Hot Jupiter host stars, due to for example magnetic braking \citep[e.g.][]{Benbakoura2019}. If a Hot Jupiter host star is spun down, and is thus slowly rotating, but there is a large orbital frequency due to the short-period Hot Jupiter companion, the tidal frequency is also high (and in the fast tides regime), indicating that this regime is relevant there \citep[e.g.][]{Craig2020effvisc,Barker2020}.

From this multitude of scalings a new question arises: for a given system, which scalings (if any!) are the correct ones? This question in reality consists of two separate questions. The first part of the question is related to whether MLT or RMLT (or neither) predictions should be used, and the second part relates to which tidal frequency regime is applicable. One of our key aims in this paper is to test these scalings and to determine the appropriate ones for astrophysical extrapolation.

We can quantify the transition from MLT to RMLT using the convective Rossby number: 
\begin{equation}
    \textrm{Ro}_c\equiv\left(\frac{u_c}{2\Omega l_c}\right)=\left(\frac{\omega_c}{2\Omega}\right),
\end{equation}
which is based on the spin of the planet, and the convective velocities and frequencies. Fortunately, using these temperature-based definitions, regardless of whether the regime in question is MLT or RMLT, the expression for the Rossby number in terms of the diffusion-free scalings is the same because $\omega_c$ has the same form in both regimes. This useful result was also found previously \citep[e.g.][]{Aurnou2020_scalings}, and leads to the expression for the convective Rossby number:
\begin{equation}
    \textrm{Ro}_c\sim\textrm{Ra}^{1/2}\textrm{Pr}^{-1/2}\textrm{Ek}.
\end{equation}
On the other hand, the transitions between the different frequency regimes for $\nu_{\mathrm{eff}}$ depend on the ratio $\omega/\omega_c$, which we can write as:
\begin{equation}
\frac{\omega}{\omega_c}=\frac{\omega}{u_c/l_c}=\frac{1}{2}\frac{2\omega l_c}{u_c}\equiv\frac{1}{2}\textrm{Ro}_\omega^{-1}.
\end{equation}
We have defined this quantity as a ``tidal convective Rossby number", Ro$_\omega$. The two Rossby numbers are related via the factor $\Omega/\omega$. In this work, the two Rossby numbers differ by a factor of $1/2$, because $\Omega=\gamma=\frac{1}{2}\omega$ is set for the simulations with a given Ek. The regime transitions are thus expected to occur at roughly the same value of the rotation rate. Using the tidal frequency transitions obtained in \cite{Craig2020effvisc}, where the transition from intermediate to high frequency regimes occurs around $\frac{\omega}{\omega_c}\approx5$, this may be expected to occur here at $\textrm{Ro}_\omega\approx0.1$. The transition from MLT to RMLT on the other hand is likely to start at $\textrm{Ro}_c\approx0.1$ \citep[e.g. Fig.~4 of][]{Barker2014RMLT}.

\subsection{Illustrative simulations}

To illustrate the flow observed in our simulations, we plot snapshots of the vertically-averaged vertical vorticity perturbation (to the elliptical flow) $\langle\omega_z\rangle_z$ at $\textrm{Ek}=5\cdot10^{-5.5}$, $t=0.12$ in Fig.~\ref{fig:vorticity_flowpictures}. In the figure on the left we plot the simulation with $\textrm{Ra}=6\textrm{Ra}_c$, $\epsilon=0.04$. In this simulation the equilibrium tide is present (as a background flow, but is not shown explicitly), but the ellipticity is sufficiently small such that the convective LSV inhibits the elliptical instability \citepalias[][]{deVries2023}. The observed behaviour is a cyclonic convective LSV embedded in an anticyclonic background. However, the cyclone appears very noisy due to the presence of many small-scale convective eddies. In the figure on the right we plot the simulation with $\textrm{Ra}=6\textrm{Ra}_c$, $\epsilon=0.1$. This is in the regime with a strong elliptical instability, albeit with a slightly larger $\epsilon$ than realistically expected for a Hot Jupiter. For illustration we have chosen a snapshot during a burst of the elliptical instability. The cyclonic vortex is stronger than the one in the left panel. Furthermore, the surrounding background is more strongly anticyclonic as a result. 

Our subsequent analysis of the contributions of the elliptical instability to the energy injection rate (and hence tidal dissipation rate) is based on flows more like the one on the right of Fig.~\ref{fig:vorticity_flowpictures}, while the analysis of the effective viscosity of convection originates primarily from quantities measured from flows like the one shown on the left.

\section{Scaling laws for the elliptical instability and rotating convection}
\label{sec:scalinglaws}

Our simulations necessarily use dimensionless parameters that are far from the astrophysical ones, except perhaps for $\epsilon$ for the hottest Jupiters. Hence, we now turn to obtain scaling laws for the energy injection due to the elliptical instability to compare with the heuristic arguments in \S~\ref{sec:diagnostics}, as well as scaling laws for the convective velocity and effective viscosity by testing the prescriptions obtained in \S~\ref{sec:RMLTscalings}. For the latter, we choose parameters in the strongly rotationally-constrained regime, with fast tides, and thus we expect to observe the high frequency RMLT scaling for the effective viscosity in our simulations. We will also justify this regime as being the most relevant in giant planets later in \S~\ref{sec:astroapplication}.

\subsection{Energy injection due to elliptical instability}

\begin{figure*}
    \centering
    \includegraphics[width=0.8\linewidth]{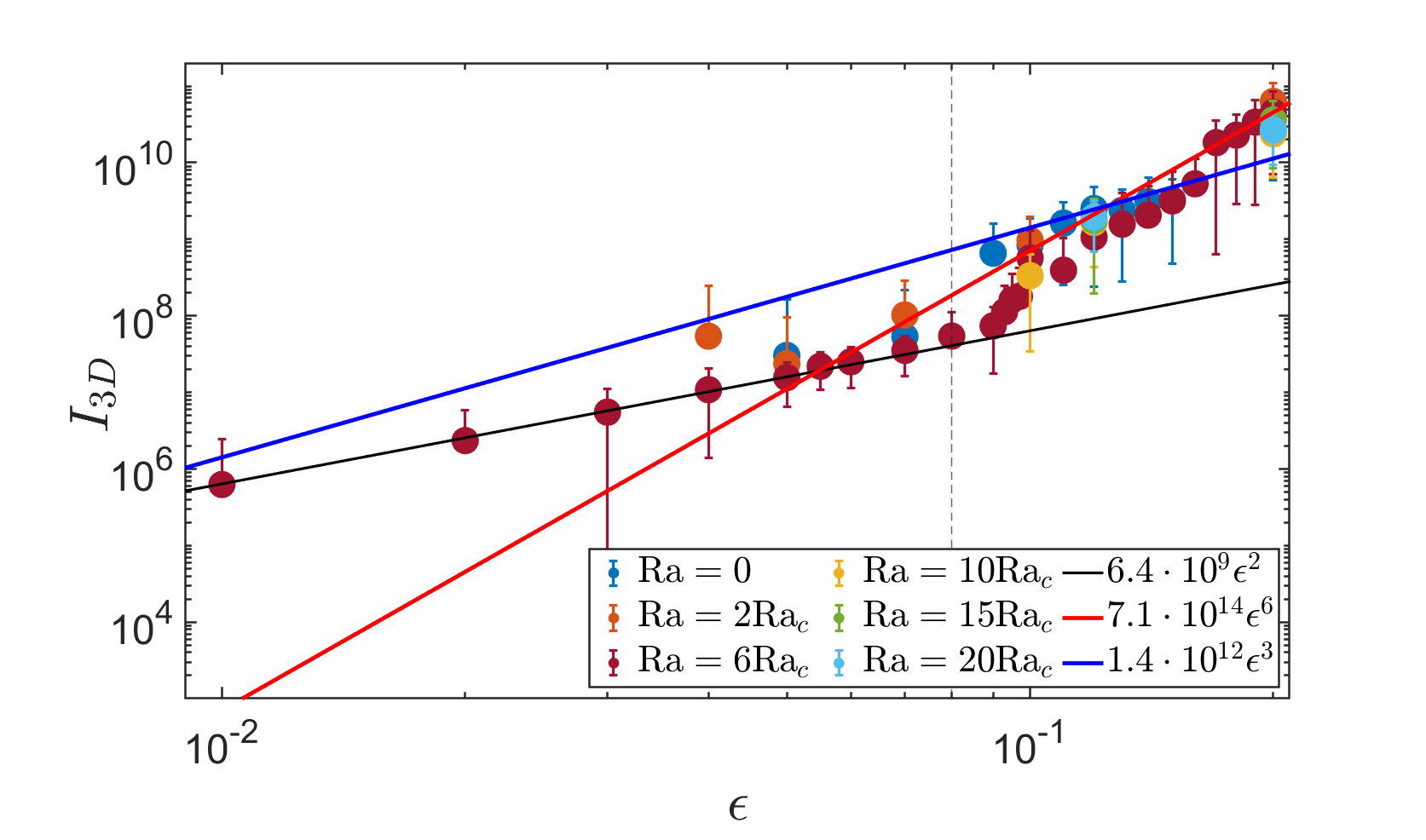}
    \caption{Energy injection rate (into 3D modes) $I_{3D}$ as a function of $\epsilon$ for various Rayleigh numbers. The vertical dashed line at $\epsilon=0.08$ marks the transition between sustained behaviour on the left, and bursts in addition to sustained behaviour on the right. Three lines are fitted to the data at $\text{Ra}=6\text{Ra}_c$. The sustained behaviour is consistent with an $\epsilon^2$ scaling for $\epsilon\lesssim 0.08$, represented by the black line. Bursts of the elliptical instability contribute on top of this sustained energy injection, resulting in a much larger energy injection for larger $\epsilon$. The sustained+bursts energy injection is fitted using an $\epsilon^3$ fit in blue, and an $\epsilon^6$ fit in red.}
    \label{fig:I3Dfitepsilonsqr}
\end{figure*}

When the flow is sufficiently turbulent, the energy injection rate ($I_{3D}$) due to the elliptical instability on its own scales consistently with $\epsilon^3$ \citep[][]{Barker2013,Barker2014}. However, the energy injection we observe in our simulations doesn't result from the elliptical instability alone. We plot the energy injection $I_{3D}$ as a function of $\epsilon$ for various values of Ra at fixed $\textrm{Ek}=5\cdot10^{-5.5}$ in Fig.~\ref{fig:I3Dfitepsilonsqr}, which we divide into two regimes by a vertical dashed line. This vertical dashed line is located at $\epsilon=0.08$. As we found in \citetalias[][]{deVries2023}, the points to the left of this line represent simulations without visible bursts of elliptical instability for $\textrm{Ra}\gtrsim2\textrm{Ra}_c$, for which it appears to have been largely suppressed. The $\textrm{Ra}=6\textrm{Ra}_c$ data points in burgundy are fitted using an $\epsilon^2$ scaling. The data agrees very well with this scaling for $\epsilon$ below the transition, indicating that the energy injection here corresponds to an effective viscosity that is independent of $\epsilon$. This presumably results from the action of convective turbulence in damping the tidal flow rather than from the elliptical instability, as we will justify further in \S~\ref{RMLTfits}.

The points on the right of the vertical dashed line feature bursts of instability in which the kinetic energy and energy transfer rates repeatedly grow to large values, indicating that the elliptical instability operates in this regime. The operation of the elliptical instability appears to be in addition to the effective viscosity resulting from convective turbulence, but the energy injection rate due to the elliptical instability is much larger, as we illustrate by the strong departure of these points from the black $\epsilon^2$ line. We fit these with our (naive) theoretically predicted $\epsilon^3$ fit (solid-blue line), and a previously observed $\epsilon^6$ fit \citep[solid-red line][]{Barker2013}. Both fits are consistent with the data on the right hand side (over such a narrow range of $\epsilon$), and are inconsistent with data on the left. Furthermore, the data and fits are consistent at all values of Ra, indicating that this scaling is independent of the Rayleigh number. The energy injection rate of the elliptical instability would remain greater than that of the effective viscosity due to convection for $\epsilon\gtrsim0.01$ if we extrapolate the former with an $\epsilon^3$ scaling. Following \citet[][]{Barker2014}, we use the naive theoretical prediction to obtain a proportionality constant $\chi$ from our fit to the data shown in Fig.~\ref{fig:I3Dfitepsilonsqr} such that $I_{3D}\equiv \chi \epsilon^3\gamma^3$. We find $\chi\approx0.044$ for the plotted blue line, with $\chi\approx0.18$ as an upper estimate when fitting to the top right clump of data points. If instead we calculate based on $I_{3D}\propto\chi_2 \epsilon^6\gamma^3$, we obtain $\chi\equiv\chi_2\epsilon^3\approx22.45\cdot\epsilon^3$. To illustrate the efficiency of this $\epsilon^6$ scaling we insert the highest-inferred ellipticity of a Hot Jupiter, $\epsilon=0.06$, and find $\chi=4.8\cdot10^{-3}$. Hence, the elliptical instability is considerably weaker if this steeper scaling applies. The $\epsilon^3$ scaling can thus be viewed as an ``upper bound" on the energy transfer rates resulting from the elliptical instability for small $\epsilon$.

\subsection{Comparison of RMLT predictions to the simulations}
\label{RMLTfits}

\begin{figure*}
    \centering\begin{subfigure}[b]{0.45\textwidth}
         \centering
         \includegraphics[width=\textwidth,
    trim=1.0cm 0cm 1.2cm 0cm,clip=true]{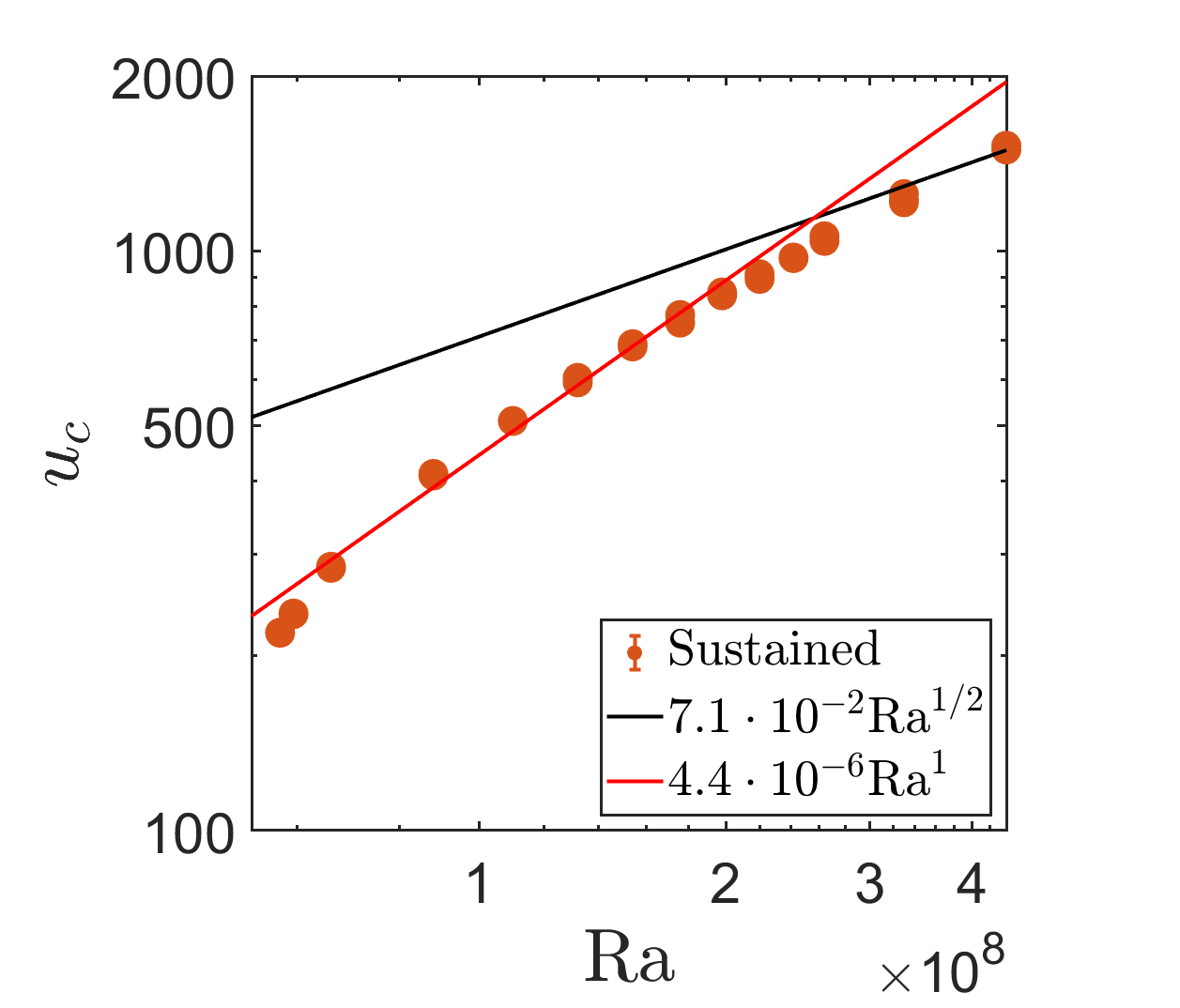}
         \label{fig:uconvscalings_Ra}
     \end{subfigure}
     \hfill
     \begin{subfigure}[b]{0.45\textwidth}
         \centering
         \includegraphics[width=\textwidth,
    trim=1.0cm 0cm 1.2cm 0cm,clip=true]{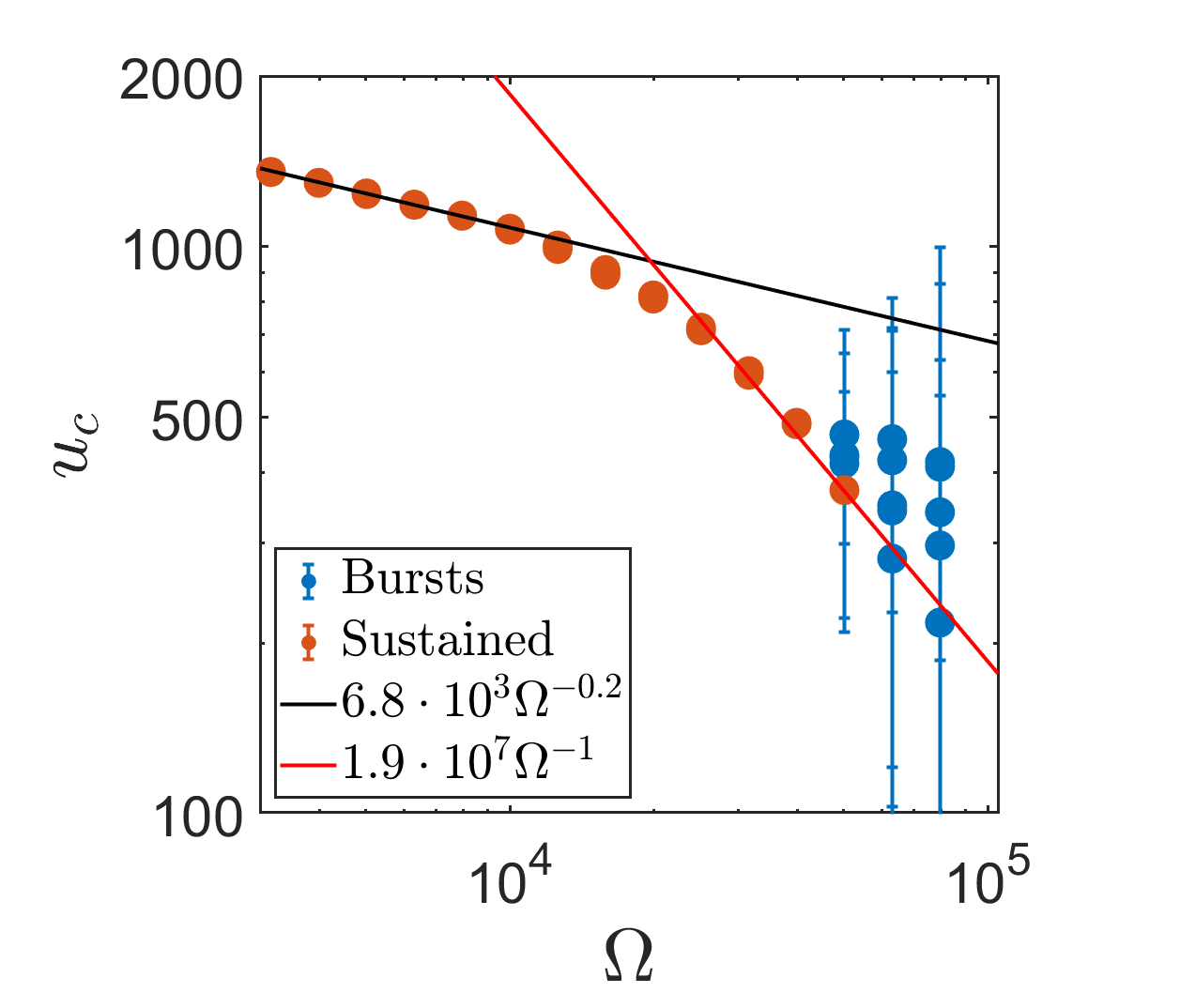}
         \label{fig:uconvscalings_Omega}
     \end{subfigure}
     \caption{Left: Scaling of the vertical convective velocity compared with the predictions of MLT and RMLT at fixed $\textrm{Ek}=5\cdot10^{-5.5}$. Only those simulations with sufficiently small ellipticities are used such that no bursts of the elliptical instability are present, as indicated by the orange data points. At these ellipticities the vertical velocity is negligibly impacted by the ellipticity. We observe the RMLT scaling with Ra in red, and a hint for the non-rotating MLT scaling with $\textrm{Ra}^{1/2}$ in black. Right: Scaling of the vertical convective velocity at fixed $\textrm{Ra}=1.3\cdot10^8$ and $\epsilon\in[0.02,0.05]$. The blue data points correspond to simulations with bursts of the elliptical instability. We retrieve the RMLT scaling in red at large $\Omega$, and find that the scaling tends to the MLT prediction to be independent of rotation rate as $\Omega$ becomes small, here illustrated by the black-solid line, which follows $\Omega^{-0.2}$.}
     \label{fig:uconv_scalings}
\end{figure*}

In this section we explore further the regime for $\epsilon\lesssim 0.08$ that we have identified, and we will demonstrate that it results from convective turbulence damping the background tidal flow. First, we fit the convective velocities as a function of Rayleigh number in the left panel of Fig.~\ref{fig:uconv_scalings} to verify our predictions based on RMLT. The data is obtained from simulations at fixed $\text{Ek}=5\cdot10^{-5.5}$, and with such values of $\epsilon$ that only sustained energy injection is present without visible bursts of elliptical instability (which tend to produce larger vertical velocities when they occur). These values of $\epsilon$ that contain no visible bursts of the elliptical instability vary with Rayleigh number as stronger convective driving results in stronger suppression of the elliptical instability; for example at $\mathrm{Ra}=4\mathrm{Ra}_c\approx0.9\cdot10^8$ values up to $\epsilon=0.04$ are used, while at $\mathrm{Ra}=10\mathrm{Ra}_c\approx2.2\cdot10^8$ we use up to $\epsilon=0.075$, and at $\mathrm{Ra}=20\mathrm{Ra}_c\approx4.4\cdot10^8$ we use up to $\epsilon=0.1$. The same values of $\epsilon$ are used for all subsequent figures as a function of $\mathrm{Ra}$. In this and subsequent figures, orange circles represent simulations without bursts of the elliptical instability and blue circles represent those in which there are visible bursts. We plot the best fit RMLT scaling in solid-red and for stronger convection (i.e.\ relatively weaker rotation), we fit the non-rotating MLT scaling in solid-black. The RMLT scaling is in very good  agreement with our data for $\mathrm{Ra}\lesssim 3\cdot 10^8$, indicating that RMLT is the appropriate description of rotating convection in our simulations.

We separately fit the convective velocities as a function of the rotation rate $\Omega$ in the right panel of Fig.~\ref{fig:uconv_scalings} at constant $\textrm{Ra}=1.3\cdot10^8$ at $\epsilon\in[0.02,0.05]$. These values of $\epsilon$ are used in all subsequent figures at fixed Ra. We have elected to plot these results as a function of $\Omega$ instead of Ekman number because $\Omega$ has a more direct relation to the tidal frequency $\omega$ than the Ekman number, particularly in real bodies where $\nu,d\neq1$. In these simulations we have set $\nu=d=1$, however, so $\Omega=(1/2)\textrm{Ek}^{-1}$. The simulations at high rotation rate do feature bursts of the elliptical instability, because the associated high tidal frequency strengthens the elliptical instability whilst weakening the convective driving because the Rayleigh number is fixed. The data points at strong rotation, $\Omega\geq10^{4.4}$, fit the RMLT prediction of $\Omega^{-1}$ well. The data points at weaker rotation rates become more weakly dependent on $\Omega$ as they begin to approach the non-rotating MLT prediction. The black-solid line fitted to the left-most data points scales only weakly as $\Omega^{-0.2}$. It is expected that at even smaller rotation rates, or larger Rayleigh numbers, this scaling would become fully independent of rotation. This figure indicates that the transition from MLT to RMLT is indeed gradual, instead of abrupt. From both figures we find -- according to RMLT -- that the convective velocity is well-described by
\begin{equation}
\label{eq:ucresultRMLT}
     u_c=0.28\textrm{RaEk}\frac{\kappa}{d},
\end{equation}
for rapid rotation, and for weaker rotation it follows the non-rotating MLT scaling
\begin{equation}
     u_c=7.1\cdot10^{-2}\textrm{Ra}^{1/2}\textrm{Pr}^{1/2}\frac{\kappa}{d}.
\end{equation}
Note that both scalings are in fact diffusion-free but have been written using the standard dimensionless numbers from our fits.

\begin{figure*}
    \centering\begin{subfigure}[b]{0.45\textwidth}
         \centering
         \includegraphics[width=\textwidth,
    trim=0.0cm 0cm 2cm 0cm,clip=true]{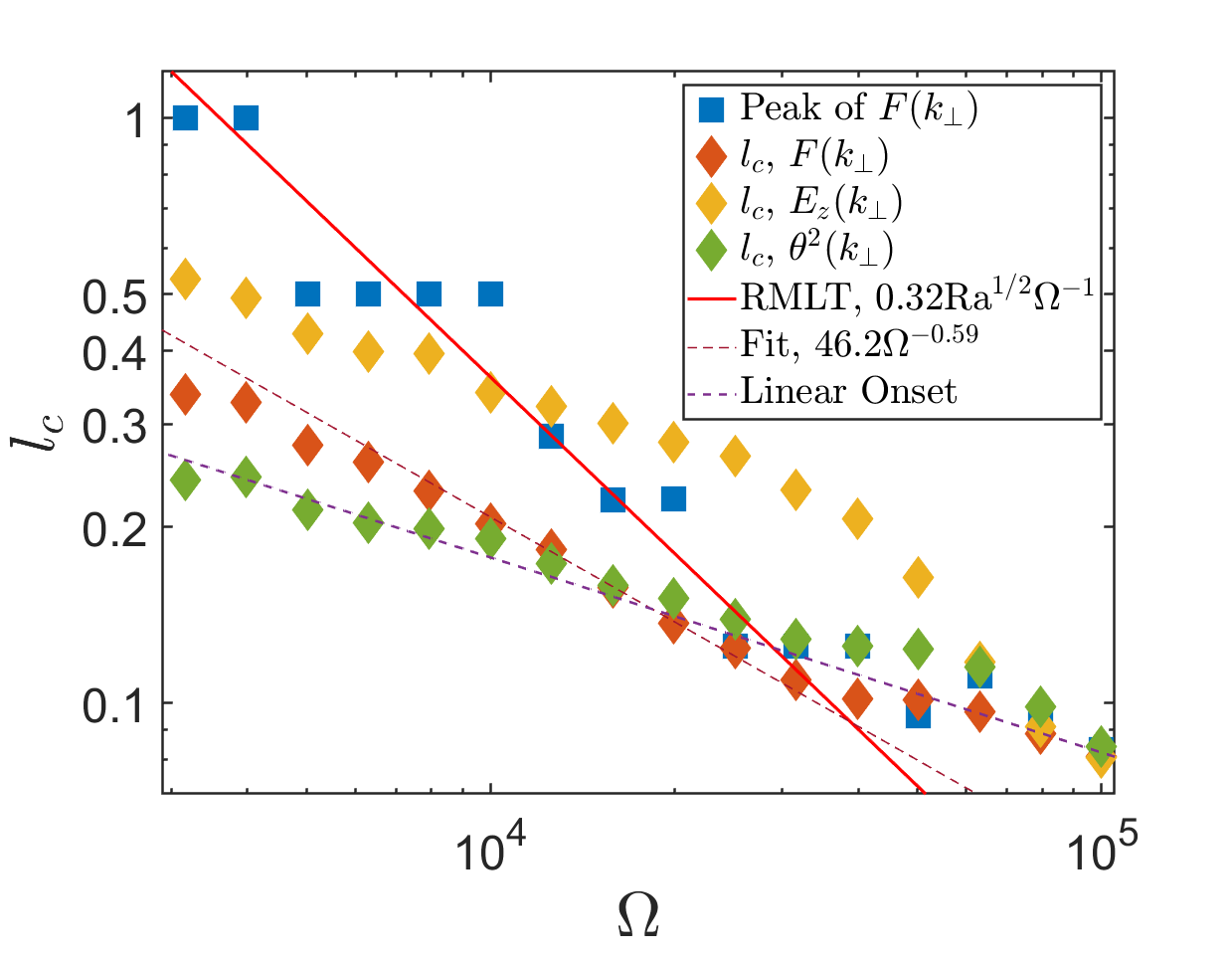}
    \caption{Fixed Ra}
         \label{fig:lconvscalings_Omega_Ra}
     \end{subfigure}
     \hfill
     \begin{subfigure}[b]{0.45\textwidth}
         \centering
         \includegraphics[width=\textwidth,
    trim=0.0cm 0cm 2cm 0cm,clip=true]{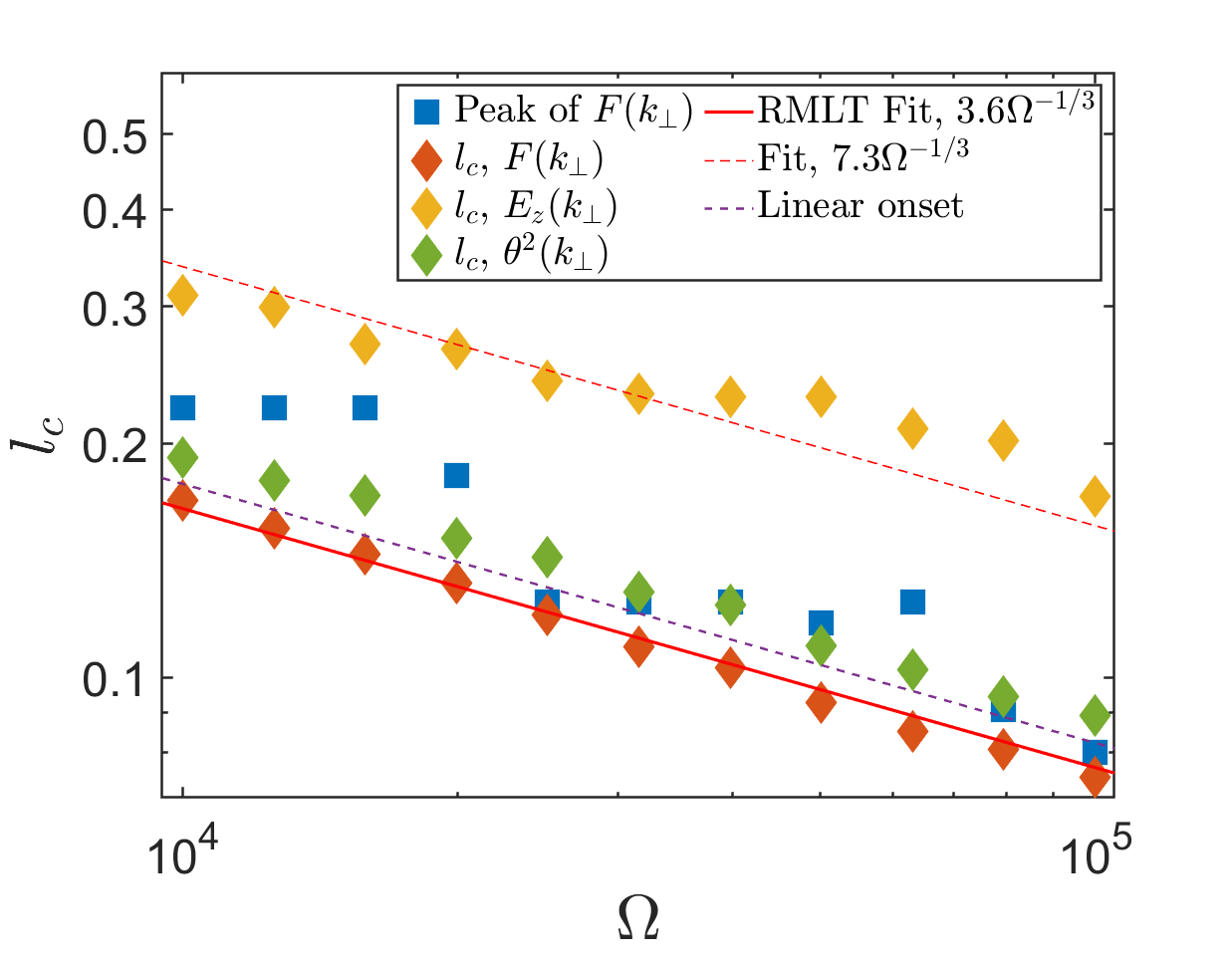}
     \caption{Fixed R (supercriticality)}
         \label{fig:lconvscalings_Omega_R}
     \end{subfigure}
     \caption{Horizontal convective length scale as a function of rotation rate, calculated using the integration methods in Eqs.~\ref{eq:lc_Currie} and \ref{eq:lc_Parodi}, which agree well. Data points are calculated based on the heat flux (orange diamonds), vertical kinetic energy (yellow diamonds) and the squared temperature perturbations (green diamonds). The peaks of the heat flux spectrum are included in blue squares. The solid-red fit is the RMLT prediction of $\textrm{Ra}^{1/2}\Omega^{-1}$ (Eq.\ref{eq:lc_sims}), and the dashed-purple line is the linear onset length scale. Left: results at fixed $\textrm{Ra}=1.3\cdot10^8$. The peaks of the heat flux agree well with the RMLT prediction. The steepest fit to the heat flux length scale is plotted in dashed-burgundy, which probably differs from the RMLT prediction in solid-red because of the modest supercriticalities involved. The linear onset scaling is plotted in dashed-purple, which only agrees with our data for the three right-most points with the smallest supercriticalities. Right: Same but at fixed supercriticality $R=6$. The heat flux data agree well with the RMLT prediction in solid-red. The linear onset scaling is plotted in dashed-purple, and differs from our simulation results.}
     \label{fig:lconv_scalings}
\end{figure*}

Next we obtain the horizontal length scale from simulations at fixed Rayleigh number $\textrm{Ra}=1.3\cdot10^8$ and at fixed supercriticality $R=6$ as a function of $\Omega$. We use two different methods to calculate a dominant $l_c$, illustrated here using the heat flux spectrum $F(k_\perp)= \mathrm{Re}(\hat{u}_z\hat{T}^*)$ as a function of horizontal wave number $k_\perp=\sqrt{k_x^2+k_y^2}$, where hats indicate a 2D $(k_x,k_y)$ Fourier transform and we have averaged over the inner vertical 1/3 of the box, i.e. between $z=1/3$ and $z=2/3$, and subsequently summed up the contribution from all modes within an integer bin of $k_\perp$. The first prescription was used by \citet[][]{Barker2014RMLT,CurrieconvRMLT}, and is obtained by:
\begin{equation}
    l_c=2\pi\left(\frac{\int k_\perp F(k_\perp)dk_\perp}{\int F(k_\perp)dk_\perp}\right)^{-1},
    \label{eq:lc_Currie}
\end{equation}
and the second was used by \citet[][]{Parodi2004convlengthscale}:
\begin{equation}
    l_c=2\pi\frac{\int (k_\perp)^{-1} F(k_\perp)dk_\perp}{\int F(k_\perp)dk_\perp}.
    \label{eq:lc_Parodi}
\end{equation}
In our simulations both methods agree very well when based on the same quantity. However, vastly different results are obtained if the energy spectrum \citep[as used by][]{Parodi2004convlengthscale}{}{} is used instead of $F(k_\perp)$ \citep[as used by][]{Barker2014RMLT,CurrieconvRMLT}{}{}, as we show in both panels of Fig.~\ref{fig:lconv_scalings}. The length scales calculated using the heat flux according to Eq.~\ref{eq:lc_Currie} are plotted in orange diamonds and the length scales according to Eq.~\ref{eq:lc_Parodi} but for the vertical kinetic energy spectrum $E_z(k_\perp)=\frac{1}{2} |\hat{u}_z|^2$ instead of $F(k_\perp)$ are plotted in yellow diamonds. We have opted to calculate length scales based on the ``vertical kinetic energy" spectrum $E_z(k_\perp)$ in the latter instead of the total kinetic energy spectrum because the total kinetic energy spectrum is strongly dominated by the large scale horizontal motions of the LSV. This forces the power to be concentrated on the largest scales, while these horizontal motions are unlikely to contribute substantially to heat transport or provide the dominant contribution to the effective viscosity. For completeness, the length scale obtained from the temperature fluctuation spectrum, i.e. $|\hat{\theta}(k_\perp)|^2$, is also plotted in green diamonds. Furthermore, we have added the length scale corresponding to the highest peak of the heat flux spectrum as a proxy for the dominant length scale in blue squares. The length scales corresponding to the peaks in the vertical kinetic energy and temperature perturbation spectra are omitted, because they tend to be located at the box scale, likely due to influence of the LSV, and then rapidly decrease and eventually align with the linear onset scale for $\Omega\gtrsim10^{4.6}$. Finally, fits to the data are included, with the RMLT prediction fit in solid-red and the linear onset length scale in dashed-purple. 

The left panel displays $l_c$ as a function of $\Omega$ at fixed $\textrm{Ra}=1.3\cdot10^8$, on the same range as the right panel of Fig.~\ref{fig:uconv_scalings}. We find that the blue squares, i.e.\ the peaks of the heat flux spectrum, follow a fit proportional to $\Omega^{-1}$ in solid-red. Note that the blue squares do not agree with this fit when $\Omega\gtrsim10^{4.5}$, which is probably because the simulations are not turbulent enough then to follow RMLT and instead lie more closely to the linear onset length scale. In terms of the length scales as obtained from the integrals there are substantial differences between those calculated based on different quantities. All three quantities match together close to linear onset for the three right-most data points, which have supercriticalities of 2.4, 1.8 and 1.3 from left to right, but they diverge for $\Omega\lesssim10^{4.8}$, coinciding with the generation of the LSV as the supercriticality of the system increases. The length scale corresponding with squared temperature perturbations in green diamonds stays close to the linear onset scale, i.e. it scales as roughly $\Omega^{-1/3}$. The length scale based on the kinetic energy is much larger than the other two, but also follows a scaling roughly similar to $\Omega^{-1/3}$ (fit not shown) in the interval $\Omega=[10^{3.5}, 10^{4.6}]$. These two scalings do not match our predictions according to RMLT and also do not display a transition to become independent of rotation when $\Omega\lesssim10^{4.4}$. The length scale calculated using the heat flux on the other hand is steeper than the other two in the range $\Omega=[10^{4.3}, 10^{4.6}]$. RMLT is expected to apply in this range because the flow is turbulent and strongly rotationally constrained. The slope fitted within this range in dashed-burgundy scales as $\Omega^{-0.6}$, which should be compared with the temperature-based RMLT scaling as $\Omega^{-1}$. This disagreement is likely to arise from the narrow range of Ra considered and because these simulations are not turbulent enough to match the RMLT scaling fully. However, it is much steeper than the result obtained from the other two quantities and tapers off at small $\Omega$ as expected.

In the right panel we demonstrate that with fixed supercriticality $R$, our results are consistent with the RMLT prediction of $l_c\propto\Omega^{-1/3}$ regardless of which quantity or method is used to compute the length scale. The solid-red line, with the same parameters as the solid-red line in the left panel matches the heat flux data well. The length scale obtained from the temperature fluctuations is slightly larger, and the length scale obtained from the vertical kinetic energy is much larger. Interestingly, the peaks in blue squares do not follow the solid-red RMLT prediction as closely as they do in the left panel. We attribute this difference to fluctuations in the spectrum causing the peak to shift around, particularly as the spectrum near the peak of the heat flux is quite broad (see Fig.~\ref{fig:lc_spectracheck} in the appendix), so the length scale based on integrals may be better suited here. Furthermore, while these data superficially seem to follow the linear onset scaling, each of these follows a distinct scaling with a different prefactor than the onset scaling. Note that when the supercriticality is fixed (equivalent to plotting results as a function of $\textrm{RaEk}^{4/3}$) instead of the Rayleigh number, the predictions of RMLT have the same dependence on $\Omega$ as the linear onset scaling, but this does not imply that the length scale is controlled by viscosity.

Based on these results, we use the length scale obtained from the integral heat flux method in the rest of this work, i.e. the dark orange diamonds, and use the solid-red RMLT fit whenever it is expected to apply. From the solid-red fit of both panels of Fig.~\ref{fig:lconv_scalings}, if we reintroduce $\textrm{Ra}$ using the definition  $\textrm{Ra}_c\approx8.7\textrm{Ek}^{-4/3}$, we obtain:
\begin{equation}
    l_c=0.63\textrm{Ra}^{1/2}\textrm{Ek}\textrm{Pr}^{-1/2}d.
    \label{eq:lc_sims}
\end{equation}
Using this scaling together with Eq.~\ref{eq:ucresultRMLT} we obtain a scaling law for the convective frequency
\begin{equation}
    \omega_c\approx0.44\textrm{Ra}^{1/2}\textrm{Pr}^{1/2}\frac{\kappa}{d^2}.
    \label{eq:omega_c_sims}
\end{equation}

\begin{figure}
    \centering
    \includegraphics[width=\linewidth,
    trim=1.7cm 0cm 2cm 0cm,clip=true]{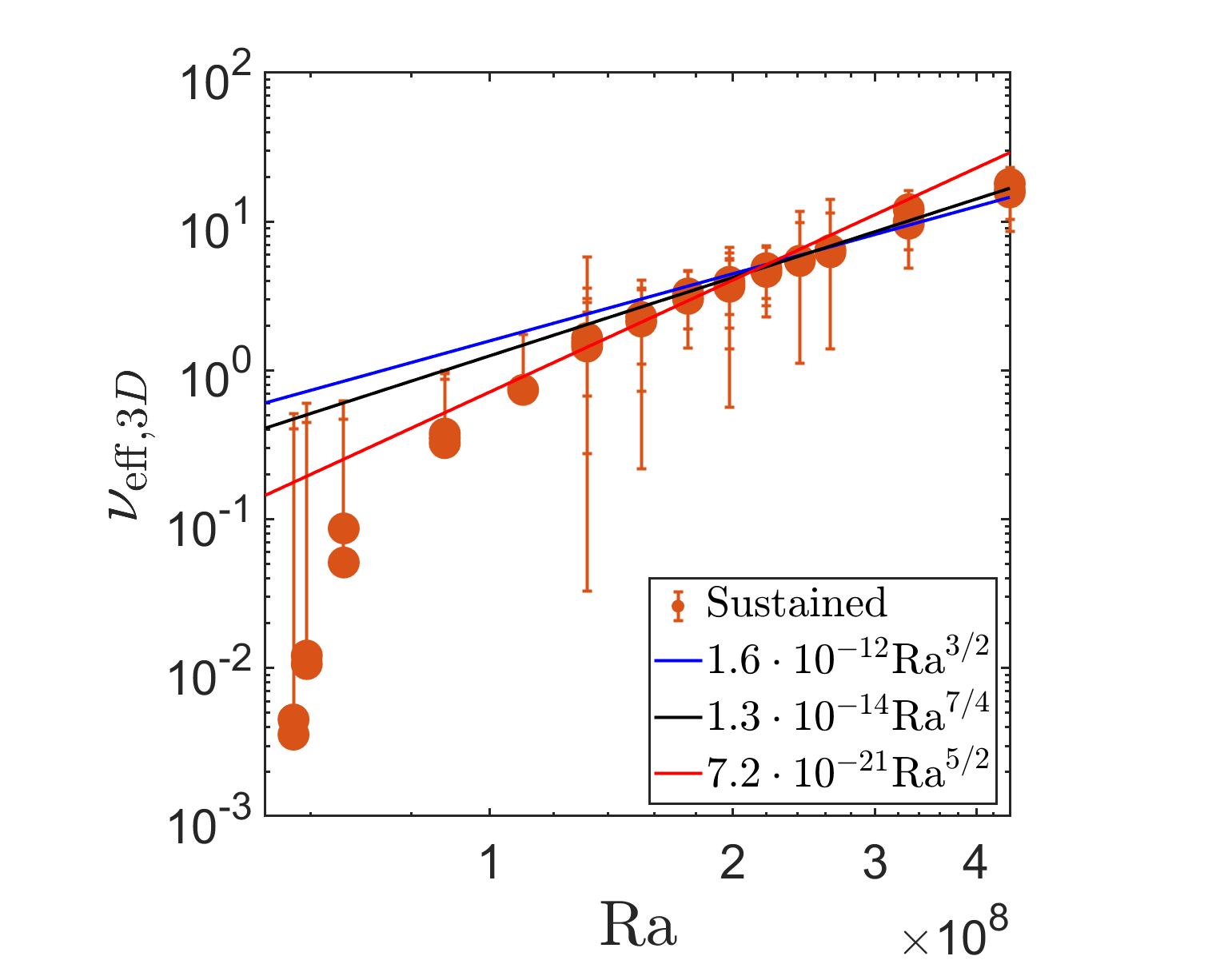}
    \caption{Effective viscosity as a function of $\text{Ra}$ at $\textrm{Ek}=5\cdot10^{-5.5}$. Only simulations featuring sustained energy injection are plotted. In addition, all three scaling law regimes predicted using RMLT are plotted.}
    \label{fig:RMLTRascaling}
\end{figure}

We examine the scaling of the effective viscosity with convection strength (Ra) in Fig.~\ref{fig:RMLTRascaling} for simulations with $\text{Ek}=5\cdot10^{-5.5}$. Only results from simulations with sustained energy injection are plotted in this figure. There is a minimum value of $\text{Ra}\approx 2.5\text{Ra}_c$ for which using an effective viscosity according to RMLT reasonably approximates the data. This minimum also corresponds to the threshold value above which an LSV appears \citep[][]{CelineLSV,FavierLSV}.

We apply these theoretically-predicted and empirically-fitted scaling laws to determine an effective viscosity in Fig.~\ref{fig:RMLTRascaling}. The blue line corresponds to the low frequency regime in Eq.~\ref{eq:effviscscalings}, the black line corresponds to the intermediate frequency regime, and the red line to the high frequency regime, with orange points indicating the simulations. Varying Ra in this figure also means varying the ratio of tidal to convective frequencies, which can change which regime might be predicted in Eq.~\ref{eq:effviscscalings}. The low and intermediate frequency predictions agree well with the simulations at high Ra. At low Ra the simulations agree with the high frequency prediction, though there is a departure for the smallest Ra for which the simulations are no longer sufficiently turbulent.

\begin{figure}
     \centering
     \begin{subfigure}[b]{0.5\textwidth}
         \centering
    \includegraphics[width=\linewidth,
    trim=1.7cm 0cm 2cm 0cm,clip=true]{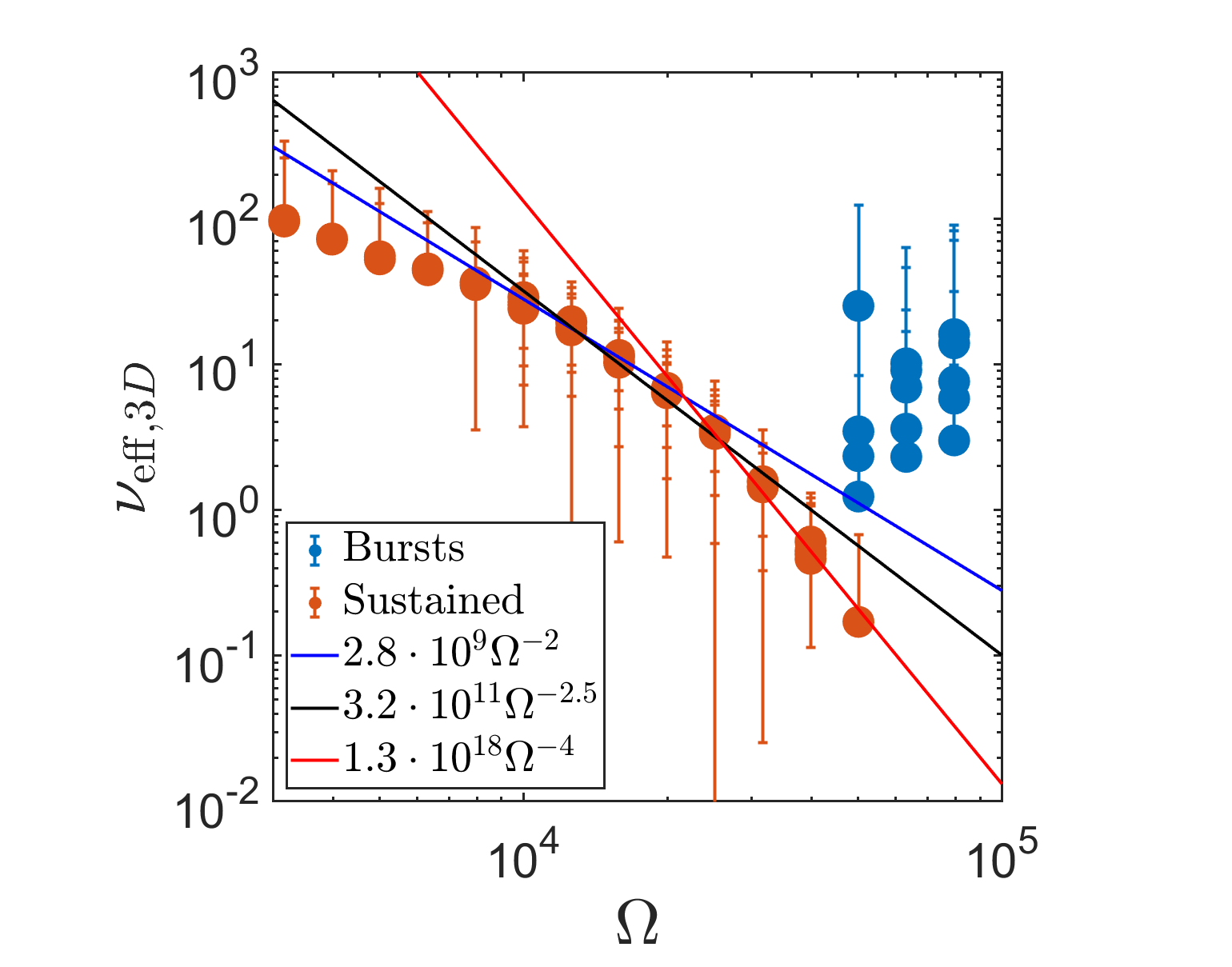}
    \caption{Fixed Ra}
     \end{subfigure}
     \hfill
     \begin{subfigure}[b]{0.5\textwidth}
        \centering
    \includegraphics[width=\linewidth,
    trim=1.7cm 0cm 2cm 0cm,clip=true]{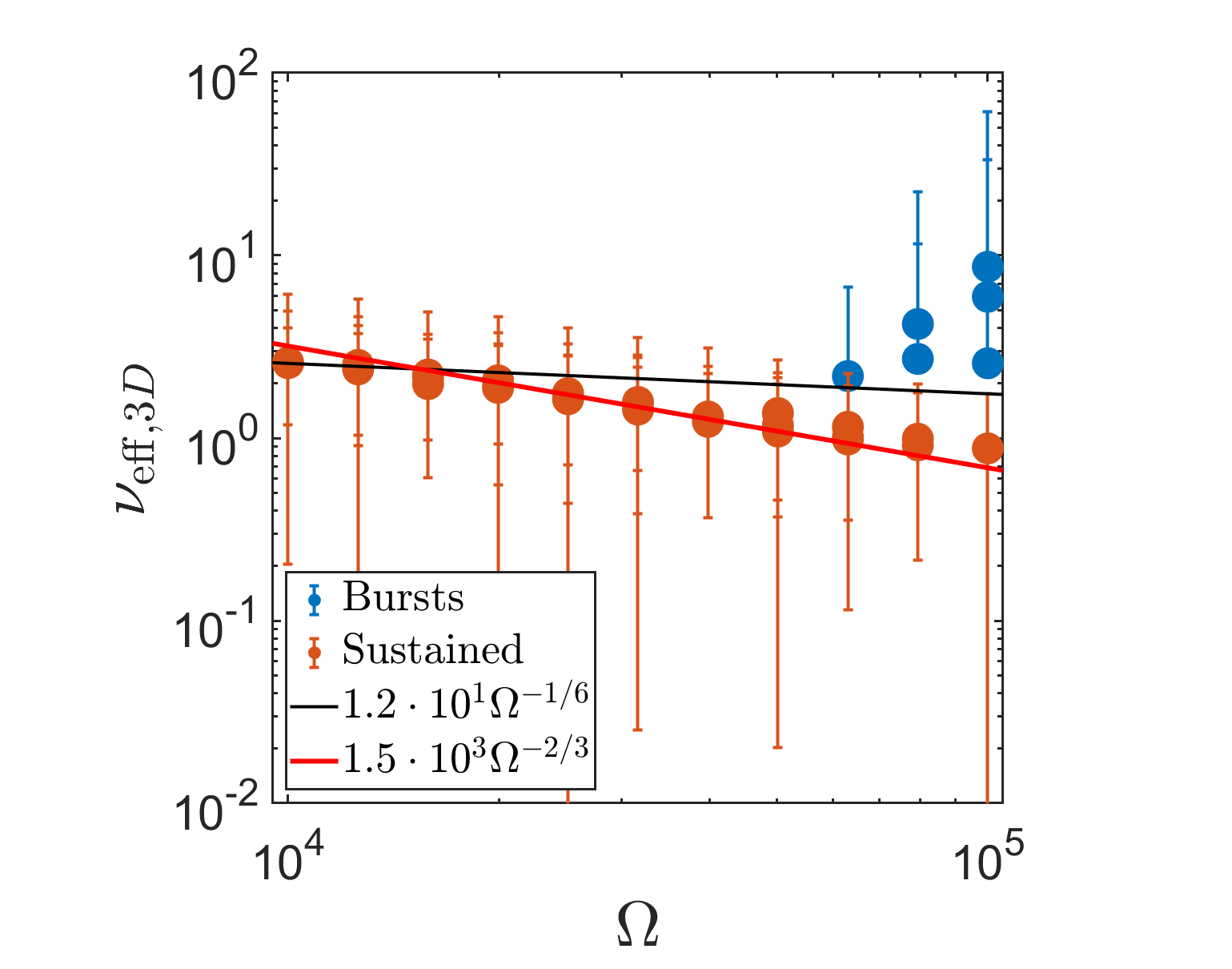}
    \caption{Fixed R (supercriticality)}
     \end{subfigure}
     \caption{Top: Effective viscosity at fixed Rayleigh number $\textrm{Ra}=1.3\cdot10^8$ and $\epsilon\in[0.02,0.05]$ as a function of rotation rate, together with all three predictions based on RMLT and the scalings obtained in \citet[][]{Craig2020effvisc}. Bottom: Same as above but at constant supercriticality $R=6$ and $\epsilon\in[0.02.0.05]$.}
    \label{fig:RMLTnueffomegascaling}
\end{figure}

The top panel of Fig.~\ref{fig:RMLTnueffomegascaling} shows instead the effective viscosity as a function of the rotation rate $\Omega$ at fixed $\text{Ra}=1.3\cdot10^8$, corresponding to $\text{Ra}=6\text{Ra}_c$ at $\textrm{Ek}=5\cdot10^{-5.5}$. At fixed Ra we expect the effective viscosity to rapidly decrease as the rotation rate increases. Since we set $\gamma=\Omega$ in these simulations the tidal frequency is $\omega=2\gamma=2\Omega$. The scalings of the effective viscosity obtained using RMLT according to Eq.~\ref{eq:effviscscalings} in terms of $\Omega$ are then respectively $\Omega^{-2}$, $\Omega^{-2.5}$ and $\Omega^{-4}$ in the low, intermediate and high tidal frequency regime. In the top panel of Fig.~\ref{fig:RMLTnueffomegascaling} we over-plot these low, intermediate and high frequency regime scalings, which are in good agreement with the simulation results. Based on our results for the convective length scale from the simulations there is some uncertainty around the solid-red fit of $\Omega^{-4}$. According to the simulation data this should possibly scale as $\Omega^{-3.6}$ instead, as the scaling obtained for the convective length sale goes as $\Omega^{-0.6}$ instead of $\Omega^{-1}$. The difference in the results is negligible however, and for consistency with the RMLT prediction for the effective viscosity we opted to keep instead the $\Omega^{-4}$ scaling in the plot. 

In the bottom panel of Fig.~\ref{fig:RMLTnueffomegascaling} we fixed $R=6$ at $\epsilon\in[0.02,0.05]$, which are the values of $\epsilon$ used for all subsequent results at fixed $R$. We examined the variation of the effective viscosity with $\Omega$. Again, we observe a decrease as the rotation rate is increased, though this is a weaker trend than we found when fixing Ra. We also observe two possible scaling regimes. When we compare with those expected by RMLT  we again find good agreement with our simulation results. We find that even when fixing the convective supercriticality, we obtain bursts of elliptical instability for sufficiently large $\Omega$. This is perhaps because the suppressive effect of convection on the elliptical instability is diminished for larger $\Omega$ because the effective viscosity is lowered, while the increased rotation rate enhances the growth rate of the elliptical instability (relative to the viscous damping rate). 

In this section we have generally found good agreement with both the predictions of RMLT for convective velocities and length scales, and with their application to the scaling laws for the effective viscosity acting on (tidal) oscillatory shear flows in \citet{Craig2020effvisc}. Based on our fits of RMLT scaling laws to the data in Figs.~\ref{fig:RMLTRascaling} and \ref{fig:RMLTnueffomegascaling}, we find the following effective viscosity regimes:
\begin{equation}
  \nu_{\mathrm{eff}} =
    \begin{cases}
      6.4\cdot10^{-3}\text{Ra}^{3/2}\text{Ek}^2\text{Pr}^{-1/2}\kappa& \textrm{low freq.},\\
     0.012\text{Ra}^{7/4}\text{Ek}^{2}\text{Pr}^{-1/4}\kappa^{3/2}d^{-1}\omega^{-1/2} & \textrm{interm. freq.},\\
      0.11\text{Ra}^{5/2}\text{Ek}^{2}\text{Pr}^{1/2}\kappa^{3}d^{-4}\omega^{-2}& \textrm{high freq.}.
    \end{cases} 
    \label{eq:effvisc_fits}
\end{equation}

\subsection{Regime transitions}
\begin{figure*}
    \centering\begin{subfigure}[b]{0.45\textwidth}
         \centering
         \includegraphics[width=\textwidth,
    trim=1.7cm 0cm 2cm 0.5cm,clip=true]{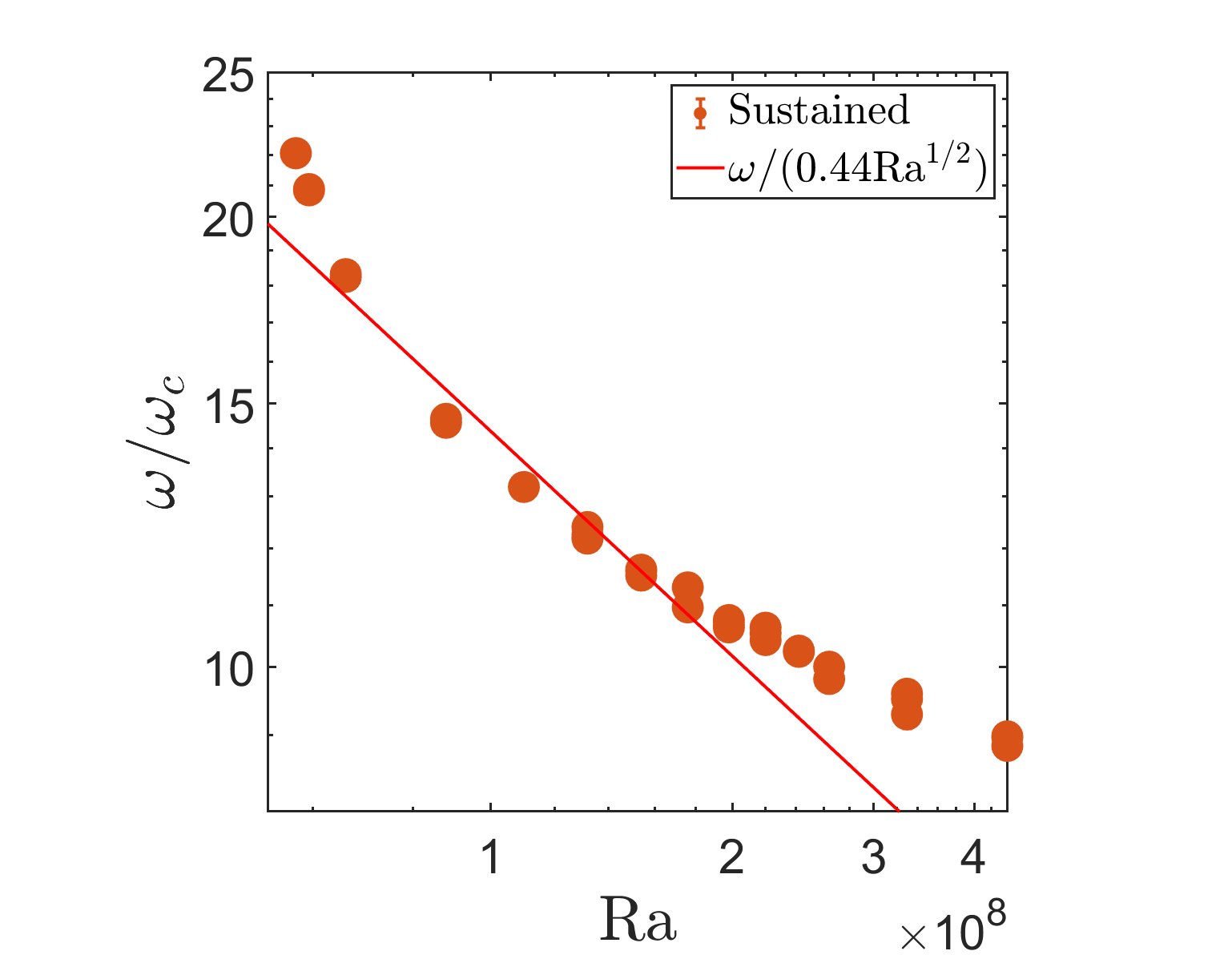}
         \label{fig:uconvscalings}
     \end{subfigure}
     \hfill
     \begin{subfigure}[b]{0.45\textwidth}
         \centering
         \includegraphics[width=\textwidth,
    trim=1.7cm 0cm 2cm 0.5cm,clip=true]{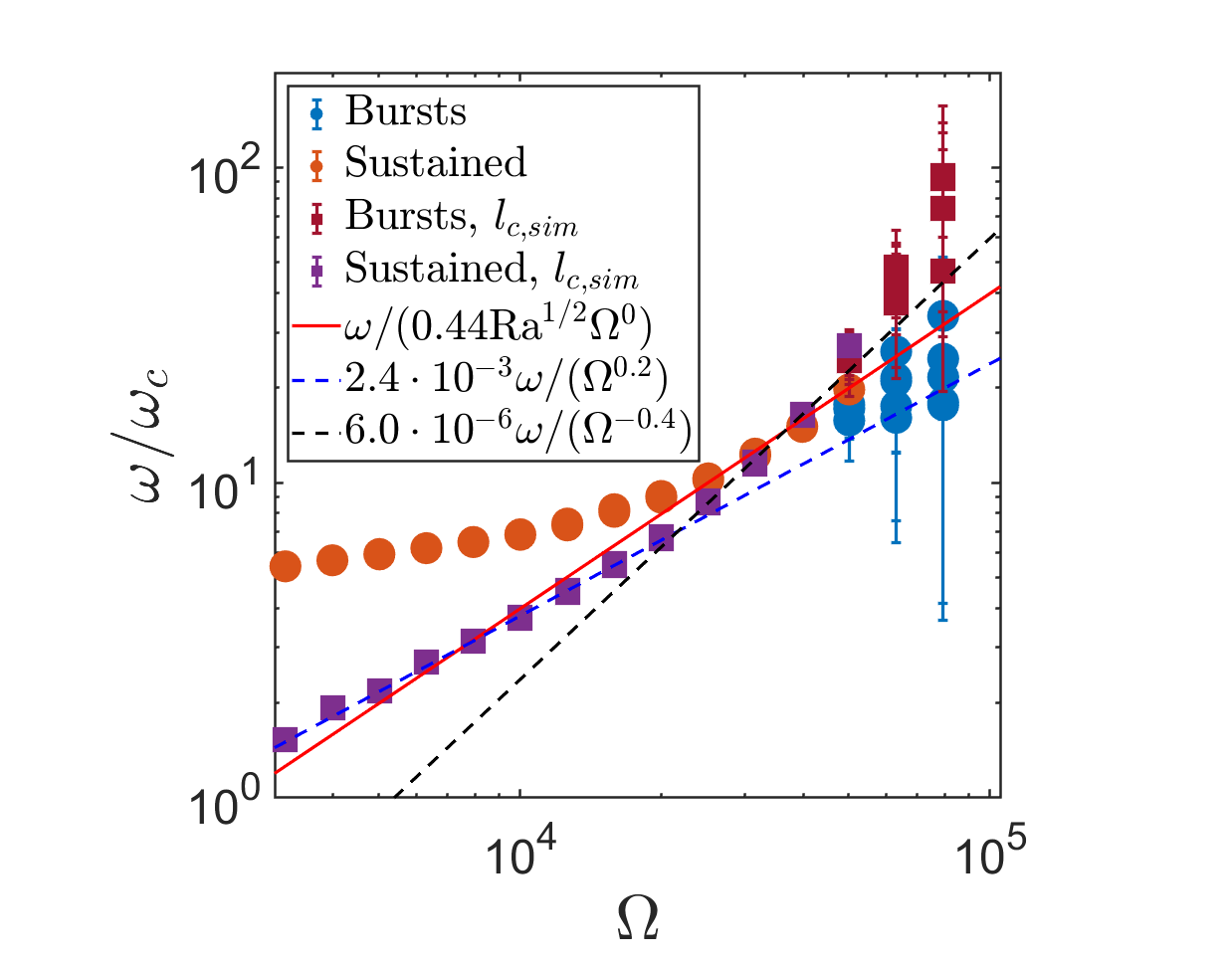}
         \label{fig:lconvscalings}
     \end{subfigure}
     \caption{Left: Ratio of the tidal to convective frequencies as a function of Ra compared with the RMLT prediction at fixed $\textrm{Ek}=5\cdot10^{-5.5}$. The data for $u_c$ is obtained from simulations, while $l_c$ is calculated using Eq.~\ref{eq:lc_scaling}. The predicted result based on Eq.~\ref{eq:omega_c_sims} is plotted in solid-red. The change from the RMLT to MLT scaling occurs around $\text{Ra}\approx2\cdot10^8$ in the left panel of Fig.~\ref{fig:uconv_scalings}, which matches the departure observed here and occurs at $\omega/\omega_c\approx10$. Right: Same at fixed $\textrm{Ra}=1.3\cdot10^{8}$ as a function of $\Omega$ in orange and blue circles. Here the change from MLT to RMLT occurs around $\Omega\approx10^{4.5}$ in the top panel of Fig.~\ref{fig:RMLTnueffomegascaling}, again matching the departure here, corresponding to $\omega/\omega_c\approx10$. The purple and burgundy squares represent $\omega_c$ calculated using both $u_c$ and $l_c$, which stays closer to the prediction of $\omega_c$ independent of the rotation rate, and therefore attains lower values than the RMLT prediction, crossing below the $\omega/\omega_c=5$ threshold.}
     \label{fig:ratio_omega_omegac}
\end{figure*}

The previous section tentatively suggests we can use MLT and RMLT and the tidal frequency regimes observed in simulations to interpret (and make predictions for) the effective viscosity. However, to understand the full picture, one would need to understand when transitions between different regimes occur. As described in \S~\ref{sec:RMLTscalings}, by virtue of setting $\Omega=\gamma$ in our simulations, the transitions are likely to occur for similar values of the Rossby number. Therefore, the occurrence of these combined transitions (MLT/RMLT and the different tidal frequency regimes) obfuscates the results in Fig.~\ref{fig:RMLTRascaling} and Fig.~\ref{fig:RMLTnueffomegascaling}. One way to separate these two transitions is to first consider the quantity $\omega/\omega_c$, which is important because it controls the regime transitions of the effective viscosity. However, it is also controlled by the transition from MLT to RMLT, because $\omega_c$ depends on $u_c$ and $l_c$. In Fig.~\ref{fig:ratio_omega_omegac} the ratio $\omega/\omega_c$ is plotted as a function of the Rayleigh number in the panel on the left, at constant $\textrm{Ek}=5\cdot10^{-5.5}$, and as a function of the rotation rate in the panel on the right, at constant $\textrm{Ra}=1.3\cdot10^8$. In the left panel we calculate $\omega_c$ using the convective velocities obtained from simulations, whilst basing the convective length scale on Eq.~\ref{eq:lc_scaling}. In addition, the prediction of $\omega/\omega_c$ according to RMLT simulation results, with $\omega_c$ given by Eq.~\ref{eq:omega_c_sims}, is plotted in solid-red. By forcing the convective length scale to follow the RMLT prediction, i.e. $l_c\sim\textrm{Ra}^{1/2}$, $\omega_c$ will no longer scale as $\textrm{Ra}^{1/2}$ when $u_c$ deviates from the RMLT prediction, and the scaling consequently changes from $u_c\sim\textrm{Ra}$ to $u_c\sim\textrm{Ra}^{1/2}$. This in turn forces the scaling of $\omega_c$ to go from $\omega_c\sim\textrm{Ra}^{1/2}$ to $\omega_c\sim\textrm{Ra}^0$. In the figure, this change is manifested by the data points deviating from the solid-red prediction as their slope decreases when $\textrm{Ra}\gtrsim2\cdot10^8$, in accordance with what is observed in Fig.~\ref{fig:RMLTRascaling}. Thus, by fixing the length scale but plotting the simulation data for the convective velocity we can easily identify at what values of $\omega/\omega_c$ this transition from RMLT to MLT occurs. From this panel, we find the transition at $\omega/\omega_c\approx10$, or a convective Rossby number $\textrm{Ro}_c\approx0.1$.

In the right panel of Fig.~\ref{fig:ratio_omega_omegac} we show the ratio $\omega/\omega_c$ as a function of $\Omega$ using orange and blue (with elliptical instability bursts) circles, which is computed in the same way as in the left panel. In addition, $\omega/\omega_c$ is calculated using the simulation data directly for both $u_c$ and $l_c$ in purple and burgundy squares. Purple squares indicate simulations without the elliptical instability, and burgundy squares indicate simulations with bursts of the elliptical instability. The prediction for $\omega/\omega_c$ in the RMLT regime is again plotted in solid-red. The deviation of the orange data points from this solid-red line occurs for $\Omega\lesssim 10^{4.4}$ like in the top panel of Fig.~\ref{fig:RMLTnueffomegascaling}. Furthermore, this deviation coincides with $\omega/\omega_c\approx10$, as  indicated in the left panel of Fig.~\ref{fig:ratio_omega_omegac}.

The convective frequency calculated directly using the simulation results for both $u_c$ and $l_c$ in the purple and burgundy squares illustrates how the transition from RMLT to MLT occurs in our simulations. First of all, the purple squares and some of the burgundy squares in the range $\Omega=[10^{4.5},\ 10^5]$ match the dashed-black fit of $\omega/\Omega^{-0.4}\sim\Omega^{1.4}$, illustrating that indeed according to simulations $\omega_c\sim u_c/l_c\sim\Omega^{-1}/\Omega^{-0.6}\sim\Omega^{-0.4}$. The purple squares in the interval $\Omega=[10^{3.5},\ 10^{4.4}]$ do not deviate as much from the solid-red prediction as the pure RMLT convective length scale results in orange on the same interval. This implies that when the convective velocity becomes independent of $\Omega$, so does the convective length scale. As a result $\omega_c$ is maintained to be almost independent of $\Omega$, which is indicated by scaling as $\omega_c\sim\Omega^{0.2}$ according to the dashed-blue fit. Note also that the value of $\omega/\omega_c$ using simulation results decreases to $\approx1$, suggesting that the effective viscosity in this range should transition from the high tidal frequency to the intermediate tidal frequency regime according to the transition found in the non-rotating simulations of \citet[][]{Craig2020effvisc}, if these hold here.

Fig.~\ref{fig:ratio_omega_omegac} indicates that care must be taken to first identify the regime of rotational influence on the convection (i.e. MLT vs RMLT) to predict the value of $\omega_c$ before calculating the ratio $\omega/\omega_c$, and thus determining which frequency regime is relevant for the effective viscosity. The deviation from the RMLT prediction for these quantities in both figures occurs roughly when $\textrm{Ro}_c^{-1}\approx10$, so we conclude that when $\textrm{Ro}_c<0.1$ RMLT is the correct prescription for the rotating convection, and that $\textrm{Ro}_c\approx0.1$ is where the transition from RMLT to MLT begins and the rotational influence diminishes.

\begin{figure*}
    \centering\begin{subfigure}[b]{0.45\textwidth}
         \centering
         \includegraphics[width=\textwidth,
    trim=1.7cm 0cm 2cm 0.5cm,clip=true]{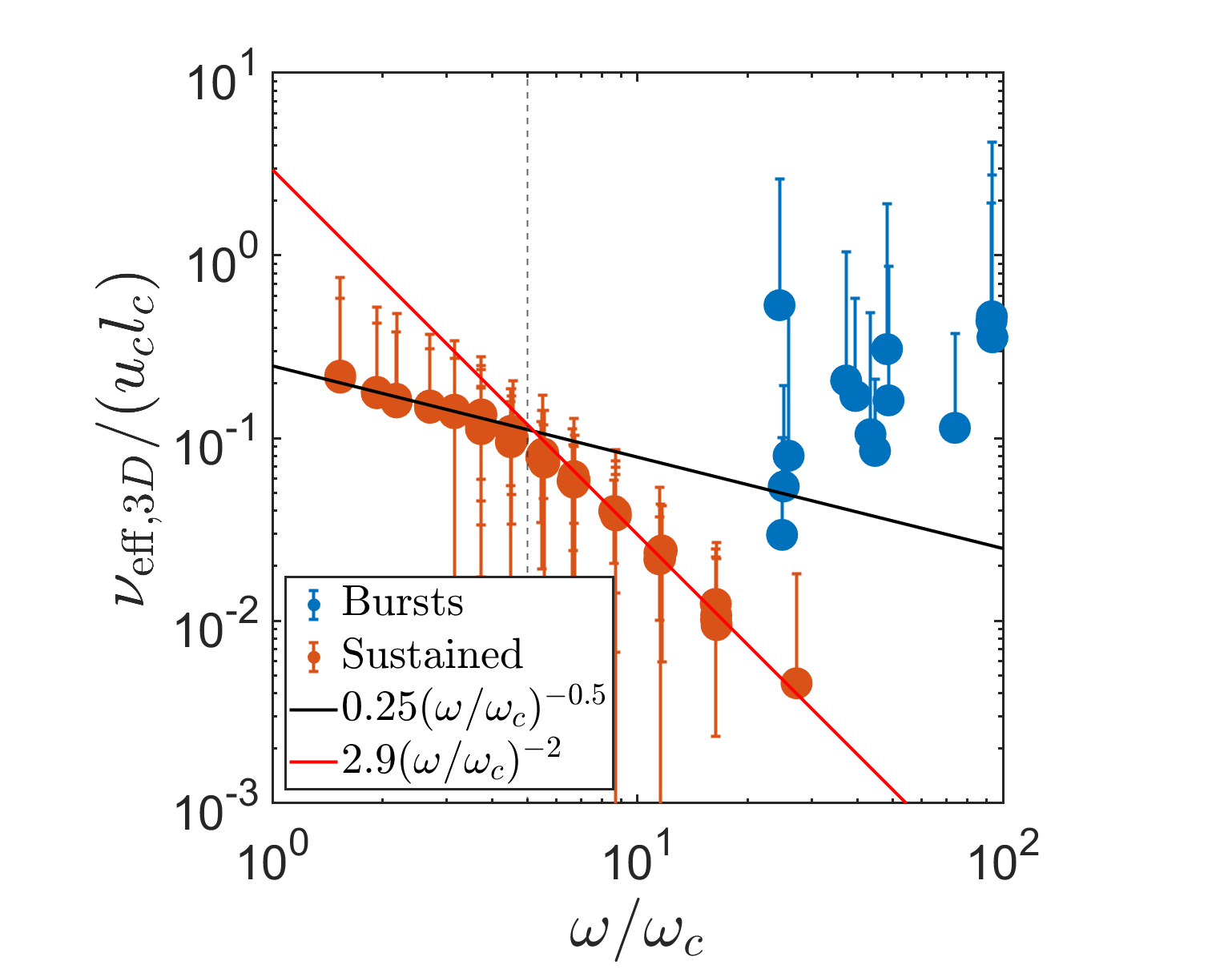}
    \caption{Fixed Ra}
         \label{fig:nueff_nrmlsd_Ra}
     \end{subfigure}
     \hfill
     \begin{subfigure}[b]{0.45\textwidth}
         \centering
         \includegraphics[width=\textwidth,
    trim=1.7cm 0cm 2cm 0.5cm,clip=true]{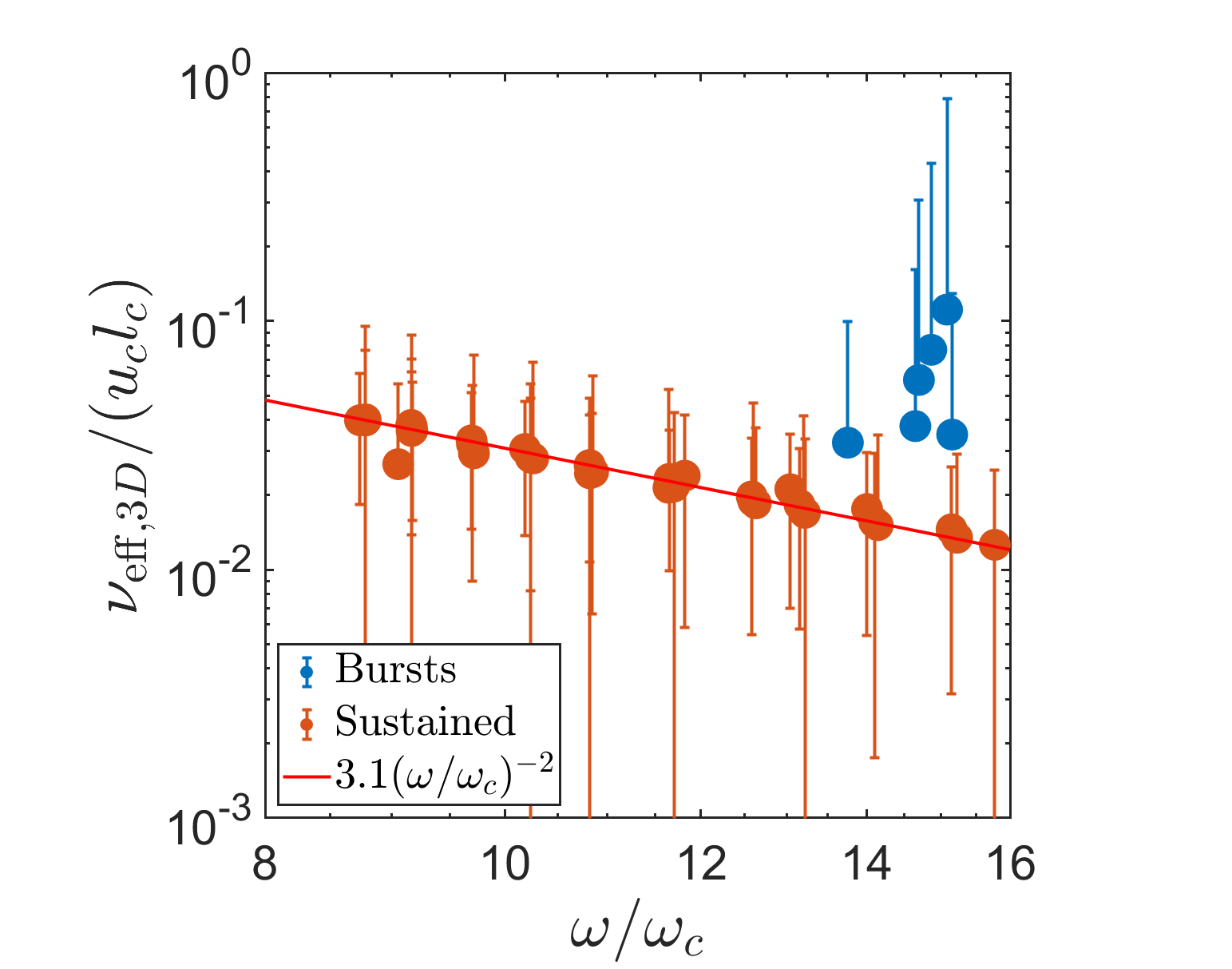}
    \caption{Fixed R (supercriticality)}
         \label{fig:nueff_nrmlsd_R}
     \end{subfigure}
     \caption{Left: Effective viscosity at fixed Rayleigh number $\textrm{Ra}=1.3\cdot10^8$ as a function of $\omega/\omega_c$, after dividing by $u_cl_c$. The high frequency prediction is plotted in solid-red, and the intermediate frequency prediction is plotted in black. The vertical dashed-black line indicates the transition between these regimes at $\omega/\omega_c=5$ found previously \citep[without rotation][]{Craig2020effvisc}, which matches the transition in our data well. Right: The same at fixed supercriticality $R=6$, where only the high frequency regime is present in this data.}
     \label{fig:nueff_nrmlsd}
\end{figure*}

To fully disentangle and interpret the effective viscosity and its dependence on $\Omega$ and $\omega$ separately, we should also calculate the effective viscosity as a function of the ratio $\omega/\omega_c$. To this end we use values of $\omega_c$ obtained from the simulations, i.e.\ corresponding to the square markers in the right panel of Fig.~\ref{fig:ratio_omega_omegac}. The results for $\nu_{\mathrm{eff},3D}$ are plotted in Fig.~\ref{fig:nueff_nrmlsd}. These figures are closely related to Fig.~\ref{fig:RMLTnueffomegascaling}, but are specifically designed to explore the $\omega/\omega_c$ dependence. In the left panel of Fig.~\ref{fig:nueff_nrmlsd}, we show results with fixed $\textrm{Ra}=1.3\cdot10^8$, while in the right panel simulations with fixed $R=6$ are plotted. The effective viscosity is divided by the factor of $u_cl_c$ which is present in all expressions for this quantity. By eliminating this factor the dependence of the effective viscosity on the ratio of $\omega/\omega_c$ is therefore directly measured. It is important to note that due to the transition from MLT to RMLT in the left panel and us fixing the supercriticality in the right panel, $\omega/\omega_c$ in general depends on the Ekman number. 
In the left panel both the intermediate and high frequency regimes are observed. The high frequency regime is plotted in solid-red line, while the intermediate frequency regime is plotted in solid-black. Both scalings agree well with simulation data. The transition from the high frequency to the intermediate frequency regime found previously at $\omega/\omega_c\approx 5$ \citep[without rotation in][]{Craig2020effvisc} is plotted using a vertical dashed line in the left panel. The location of this transition agrees remarkably well with our data. In the right panel, only the high frequency regime is observed. We thus conclude that we have not observed the low tidal frequency regime in our simulations. Moreover, we find that the intermediate regime in \citet[]{Craig2020effvisc} is reproduced and the transition to this seems to occur at the same value of $\omega/\omega_c$, even when the convective velocity and length scale are influenced by rotation. The prefactors are however different from those found in \citet[][]{Craig2020effvisc}, both lower by approximately a factor of two. Reproducing Eq.~\ref{eq:effviscCraig} with these altered prefactors:
\begin{equation}
  \nu_{\mathrm{eff}} =
    \begin{cases}
      5u_{c}l_{c} & \frac{| \omega|}{\omega_c}\lesssim 10^{-2},\\
      0.25u_{c}l_{c}\bigg(\frac{\omega_c}{\omega}\bigg)^{\frac{1}{2}} & \frac{| \omega|}{\omega_c}\in[10^{-2},5],\\
      3u_{c}l_{c}\bigg(\frac{\omega_c}{\omega}\bigg)^2 & \frac{| \omega|}{\omega_c}\gtrsim 5.
    \end{cases} 
    \label{eq:effvisc_altered}
\end{equation}

In summary, to correctly interpret and make predictions for the effective viscosity, one must first determine whether or not the convection is strongly influenced by rotation (i.e.\ whether RMLT or MLT is an appropriate description) using the convective Rossby number. Then the ratio of $\omega/\omega_c$, i.e.\ the ``tidal Rossby number", can be used to determine which of the low, intermediate or high tidal frequency regimes are appropriate. Upon plugging in the results for $u_c$ and $l_c$ from Eq.\ \ref{eq:ucresultRMLT} and Eq.\ \ref{eq:lc_sims}:
\begin{equation}
  \nu_{\mathrm{eff}} =
    \begin{cases}
      0.88\mathrm{Ra}^{3/2}\text{Ek}^2\text{Pr}^{-1/2}\kappa & \frac{| \omega|}{\omega_c}\lesssim 10^{-2},\\
     0.029\text{Ra}^{7/4}\text{Ek}^{2}\text{Pr}^{-1/4}\kappa^{3/2}d^{-1}\omega^{-1/2} & \frac{| \omega|}{\omega_c}\in[10^{-2},5],\\
      0.10\text{Ra}^{5/2}\text{Ek}^{2}\text{Pr}^{1/2}\kappa^{3}d^{-4}\omega^{-2}& \frac{| \omega|}{\omega_c}\gtrsim 5.
    \end{cases} 
\end{equation}

These scalings are likely to be more robust than the scalings in Eq. \ref{eq:effvisc_fits}, because the numerical coefficient of the scaling for the low frequency regime is based on a measured result in \citet[][]{Craig2020effvisc}{}{} and the scaling for the intermediate frequency regime is no longer obfuscated by the two transitions occurring at the same time.

\section{Astrophysical applications}
\label{sec:astroapplication}

In the previous section we obtained scaling laws to describe our simulation results for tidal energy transfer rates and effective viscosities, as well as convective velocities, length scales and frequencies. In this section we strive to apply these scaling laws to `real' parameters of astrophysical bodies to make predictions for these quantities in giant planets. This is possible because we have shown that the diffusion-free scaling laws of MLT and RMLT are applicable to most of our simulations, and if we assume they also apply in reality, we can therefore readily extrapolate our results. 

\subsection{Simple estimates}
\label{simple}

We start by reporting parameter estimates from the literature for Jupiter, obtained using models before 
\citep[][hereafter GSHS04]{Jupiterparams2004} and after \citep[][hereafter GW21]{JupiterparamsGastine2020} the Juno mission \citep[e.g.][]{Bolton2017}. We report these in Table \ref{table:Jupiter_params}. We calculate from this data the ratio of tidal to convective frequencies ($2\gamma/\omega_c$) to allow us to determine if we are in the high-frequency regime for the effective viscosity. This ratio is found to be, upon setting\footnote{This is appropriate for circularisation of weakly eccentric orbits in spin-synchronised planets, and can be thought of as a representative value for estimates of synchronisation tides with a circular orbit.} $\omega/2=\gamma={2\pi}/{P_{\mathrm{orb}}}$, 
\begin{equation}
    \omega/\omega_c=
    \begin{cases}
   9.4\cdot10^1 \bigg(\frac{P_{\mathrm{orb}}}{1\text{ d}}\bigg)^{-1}& \textrm{GSHS04},\\
      3.7\cdot10^2 \bigg(\frac{P_{\mathrm{orb}}}{1\text{ d}}\bigg)^{-1}& \textrm{GW21 at } R=0.196R_J,\\
       2.4\cdot10^2 \bigg(\frac{P_{\mathrm{orb}}}{1\text{ d}}\bigg)^{-1}& \textrm{GW21 at } R=0.98R_J.
    \end{cases}
\end{equation}
Thus we conclude that we are firmly in the high-frequency tidal regime ($\omega/\omega_c\gg 1$) for the orbital periods associated with Hot Jupiters, which is the regime explored in most of our simulations. This is also likely to be the case in Jupiter due to tidal forcing from its moons \citep[e.g.][]{GoldreichNicholson1977}.

\begin{table}
\caption{Table of dimensional and nondimensional parameters reproduced from \citet[][]{Jupiterparams2004} (GSHS04), \citet[][]{JupiterparamsGastine2020} (GW21).}
\label{table:Jupiter_params}
\begin{tabular}{|l|l|l|l|}
\hline
          & GSHS04 & GW21 & GW21 \\
          & & $R=0.196R_J$ & $R=0.98R_J$ \\
          \hline
$u_{c} \, (ms^{-1})$ & $0.1$                                     & $0.01-0.1$                              & $1$                               \\ \hline
$\Omega \, (s^{-1})$     & $1.75\cdot10^{-4}$                            & $1.75\cdot10^{-4}$                       & $1.75\cdot10^{-4}$                      \\ \hline
$d$    $(m)$      & $3\cdot10^6$                              & $5.5\cdot10^7$                         & $5.5\cdot10^7$                        \\ \hline
$\nu$ ($m^2s^{-1}$)     & $10^{-6}$                                & $2.66\cdot10^{-7}$                       & $3.92\cdot10^{-7}$                      \\ \hline
$\kappa$ ($m^2s^{-1}$)   & $10^{-5}$                              & $2.7\cdot10^{-5}$                        & $1.32\cdot10^{-6}$                      \\ \hline
Pr & 0.1 & 0.01 & 0.3 \\ \hline
Ek         & $10^{-15}$                              & $10^{-18}$                           & $10^{-18}$                          \\ \hline
Ra         & $10^{25}$                            & $10^{28}$                            & $10^{31}$                           \\ \hline
\end{tabular}
\end{table}

The effective viscosity can be calculated using the parameters from Table \ref{table:Jupiter_params}, again setting $\gamma={2\pi}/{P}$. To evaluate the different regimes, we assume the transitions from the low to intermediate frequency regimes obtained by \citet[][]{Craig2020effvisc} to obtain the following. Using data from the left column of the table for the purposes of illustration, we find:
\begin{equation}
  \nu_{\mathrm{eff}} =
    \begin{cases}
      880 \quad m^2/s, &\frac{| \omega|}{\omega_c}<10^{-2},\\
     2.54 \bigg(\frac{P_{\mathrm{orb}}}{1\text{ d}}\bigg)^{1/2} \quad m^2/s, &\frac{| \omega|}{\omega_c}\in[10^{-2},5],\\
      6.1\cdot10^{-3} \bigg(\frac{P_{\mathrm{orb}}}{1\text{ d}}\bigg)^{2} \quad m^2/s,& \frac{| \omega|}{\omega_c}>5,
    \end{cases} 
\end{equation}
We have included the low frequency regime for completeness even though this hasn't been clearly probed with our simulations.

\subsection{Detailed planetary models using MESA}
\label{sec:Detailed}

To provide a more detailed estimate of the effective viscosity and resulting tidal dissipation in a Jupiter-like planet we require models for its internal structure, i.e. profiles of pressure and density (and other quantities) as a function of radius. To do so, we use a modified version of the test suite case \textit{make\_planets} of the Modules for Experiments in Stellar Astrophysics (MESA) code \citep[][]{Paxton2011, Paxton2013, Paxton2015, Paxton2018, Paxton2019, Jermyn2022} with the MESASDK \citep[][]{richard_townsend_2022_7457723} to generate 1D interior profiles. This code has been previously used to generate a range of planetary models \citep[e.g.][]{Muller2020,Muller2023}. However, some caveats reside in the applicability of this code to planets: since it is designed to model stars it uses equations of state based on H and He without heavy elements -- unless the EOS is modified \citep[][]{Muller2020} -- necessary to generate for example a dilute core which is expected based on Juno's gravity field measurements of Jupiter \citep[][]{Stevenson2020_Jupint,HELLED2022}. Furthermore, it treats the core itself as rigid and omits the possibility of stable layers produced by helium rain. These may be important for tidal dissipation \citep[e.g.][]{Pontin2023} but are outside the scope of our study.

MESA by default treats the convection using MLT \citep[for which we use the Cox prescription,][]{Cox1968} instead of RMLT (if we assume this to be valid even in the presence of magnetic fields). We have maintained the mixing length parameter at the standard value of two, and intend to convert the obtained MLT values of these models to RMLT later on in this work. Following \citet{Muller2020} who find it to be negligible for planetary structure and evolution, we omit semiconvection in our models.

Our initial Jupiter model has a radius of $2R_J$ and a mass of $1M_J$, of which $10$ Earth-masses are located in a core with density $10 g cm^{-3}$. We have evolved the model for 4.5 Gyr to mimic the age of Jupiter and we use a constant surface irradiation of $5\cdot10^4$ erg $cm^{-2} s^{-1}$, similar to what Jupiter receives from the Sun, which is deposited at a column depth of 300 $gcm^{-2}$ (about 0.7 bar).

We also create a Hot Jupiter model with the same parameters except that we increase the surface heating to represent the irradiation of a one-day planet around a Sun-like star of $10^9$ erg $cm^{-2}  s^{-1}$. Furthermore, we incorporate additional interior heating with uniform rate $0.05$ erg $cm^{-3} s^{-1}$ throughout the fluid envelope, which can be thought to represent the impact of tidal heating or Ohmic dissipation (or other mechanisms) that could possibly inflate a number of Hot Jupiters. In this way, whilst keeping all other parameters equal, we can determine the effects of the increased radius (and stronger convection) of a puffy Hot Jupiter on the effective viscosity and tidal dissipation rates. A summary of changes to the default inlists used to generate these models is provided in Appendix~\ref{sec:MESA_inlists}.

The convective velocities and length scales (mixing lengths) obtained using the MESA code are calculated using non-rotating MLT. Although the rotation rate -- and thus the introduction of RMLT -- is expected to affect convective length scales and velocities, the effect on the heat flux is likely to be negligible \citep{StevensonRMLT, Ireland2018}. Therefore, we assume that the heat flux is independent of rotation, and is therefore the same in both MLT and RMLT. We then convert $u_c$ and $l_c$ to RMLT using the scalings we have derived, but to do so we must use flux-based scalings instead of the temperature-based scalings used in the previous sections and in the simulations of this paper. On the other hand, the temperature difference (which is imposed in simulations), and as a result the buoyancy frequency, are expected to change under the influence of rotation, in order to carry the same flux. In these flux-based scalings the conversion from MLT to RMLT is defined differently to the temperature-based scalings used previously in this work. In the temperature-based scalings the corrections introduced for both $u_c$ and $l_c$ involve $\textrm{Ro}_c$ linearly, while in the flux-based scalings the corrections are respectively:
\begin{equation}
    u_c=\Tilde{\textrm{Ro}}_c^{1/5}  \Tilde{u}_c, \quad \text{and} \quad
    l_c=\Tilde{\textrm{Ro}}_c^{3/5}  \Tilde{l}_c,
\end{equation}
where the quantities with a tilde are those calculated using non-rotating MLT. We have also denoted the Rossby number in the above equations with a tilde ($\Tilde{\mathrm{Ro}}_c=\Tilde{u}_c/(2 \Omega \Tilde{l}_c)$) because flux-based scalings imply Rossby numbers calculated using MLT and RMLT are different, unlike for the temperature-based scalings where they are the same. In the low frequency regime the effective viscosity must therefore be scaled by
\begin{equation}
    \nu_{\mathrm{eff}}\sim u_c l_c \sim \Tilde{u}_c \Tilde{l}_c \Tilde{\textrm{Ro}}_c^{4/5}.
\end{equation}
 This correction factor of $\Tilde{\textrm{Ro}}_c^{4/5}$ was also employed by \citet[][]{Mathis2016RMLT}. 

In the high tidal frequency regime the effective viscosity is instead scaled by
\begin{equation}
    \nu_{\mathrm{eff}}\sim u_c l_c \left(\frac{u_c}{l_c}\right)^2 \sim \Tilde{u}_c \Tilde{l}_c \Tilde{\textrm{Ro}}_c^{4/5} \left(\frac{\Tilde{u}_c}{\Tilde{l}_c}\right)^2 \Tilde{\textrm{Ro}}_c^{-4/5} \sim \Tilde{u}_c \Tilde{l}_c \left(\frac{\Tilde{u}_c}{\Tilde{l}_c}\right)^2.
\end{equation}
Combining these, we find in RMLT:
\begin{equation}
  \nu_{\mathrm{eff}} \propto
    \begin{cases}
      5\Tilde{u}_c\Tilde{l}_c\Tilde{\textrm{Ro}}_c^{4/5}  & \frac{| \omega|}{\omega_c}\lesssim 10^{-2},\\
      0.25\Tilde{u}_c\Tilde{l}_c\Tilde{\textrm{Ro}}_c^{3/5} \bigg(\frac{\Tilde{u}_c/\Tilde{l}_c }{\omega}\bigg)^{\frac{1}{2}} & \frac{| \omega|}{\omega_c}\in[10^{-2},5],\\
      3\Tilde{u}_c\Tilde{l}_c \bigg(\frac{\Tilde{u}_c/\Tilde{l}_c }{\omega}\bigg)^2 & \frac{| \omega|}{\omega_c}\gtrsim 5.
    \end{cases} 
    \label{eq:effviscCraig_RMLT}
\end{equation}

Hence, while the effective viscosity in the low tidal frequency regime is strongly affected by rotation, it is entirely unaffected by rotation in the high tidal frequency regime according to RMLT (assuming a fixed flux independent of rotation). This follows when considering the scaling laws in Eq.~\ref{eq:effviscscalings} in terms of flux-based RMLT:
\begin{equation}
\label{effvisrot}
  \nu_{\mathrm{eff}} \propto
    \begin{cases}
      F^{3/5}\Omega^{-4/5}d^{4/5}& \textrm{low frequency},\\
      F^{7/10}d^{3/5}\Omega^{-3/5}\omega^{-1/2} & \textrm{intermediate freq.},\\
      F\omega^{-2}& \textrm{high frequency}.
    \end{cases} 
\end{equation}
The equivalent relations written using flux-based MLT would be:
\begin{equation}
  \Tilde{\nu}_{\mathrm{eff}} \propto
    \begin{cases}
      F^{1/3}d^{4/3}& \textrm{low frequency},\\
      F^{1/2}d\omega^{-1/2} & \textrm{intermediate freq.},\\
      F\omega^{-2}& \textrm{high frequency}.
    \end{cases} 
\end{equation}
The scaling laws in the high tidal frequency regime with and without rapid rotation (i.e. according to MLT or RMLT) are therefore identical when written using flux-based scalings. However, the regime transitions may not be the same in both cases because the flux-based scalings for $\omega_c$ differ between MLT and RMLT. Convective frequencies are typically smaller in MLT, and as such the high tidal frequency regime is generally entered for lower tidal frequencies than in RMLT.

\begin{figure*}
     \centering
     \begin{subfigure}[b]{0.495\textwidth}
         \centering
    \includegraphics[width=\linewidth,trim=2cm 0cm 3cm 0.5cm,clip=true]{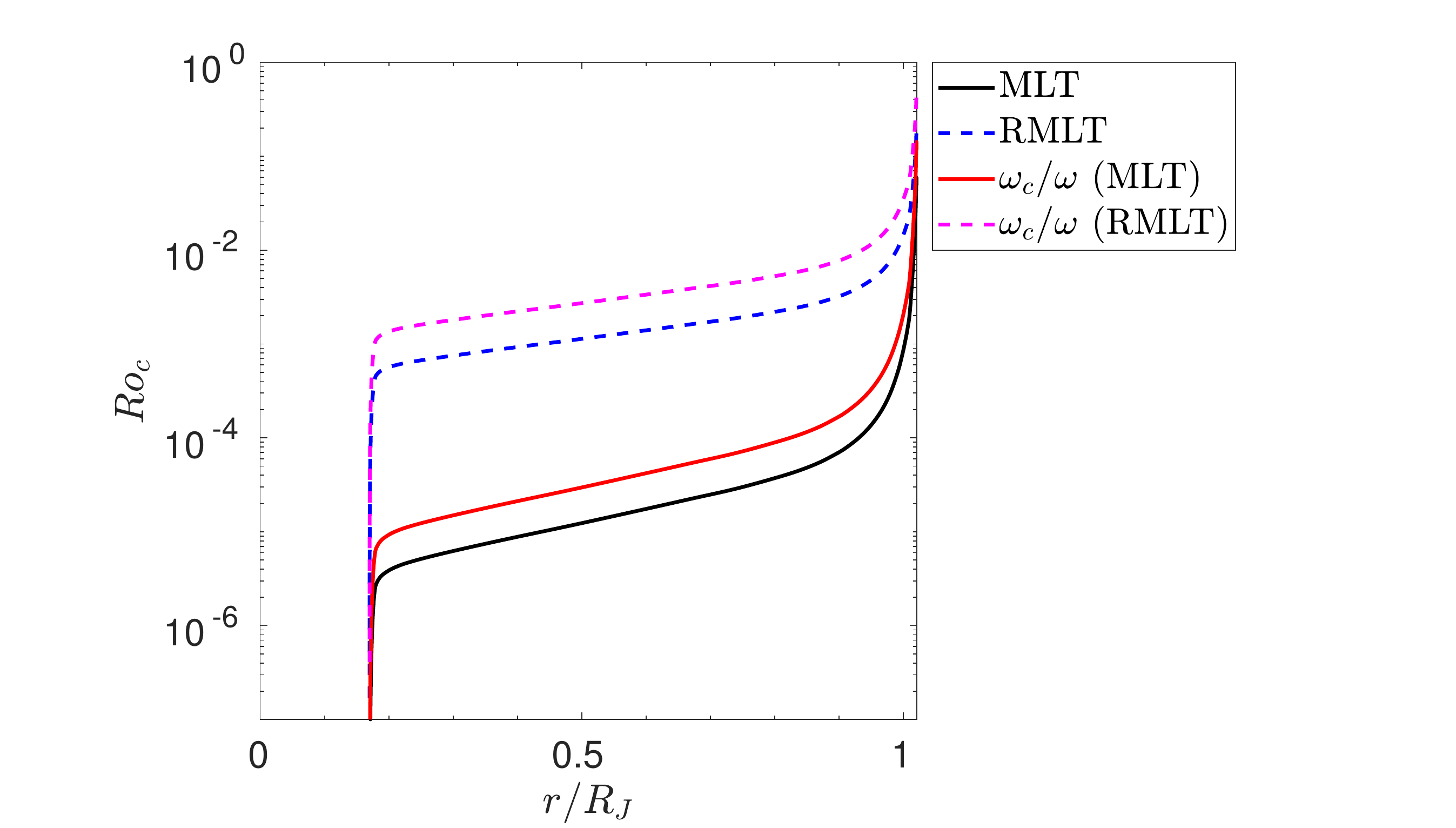}
     \end{subfigure}
     \hfill
     \begin{subfigure}[b]{0.495\textwidth}
        \centering
    \includegraphics[width=\linewidth,trim=2cm 0cm 3cm 0.5cm,clip=true]{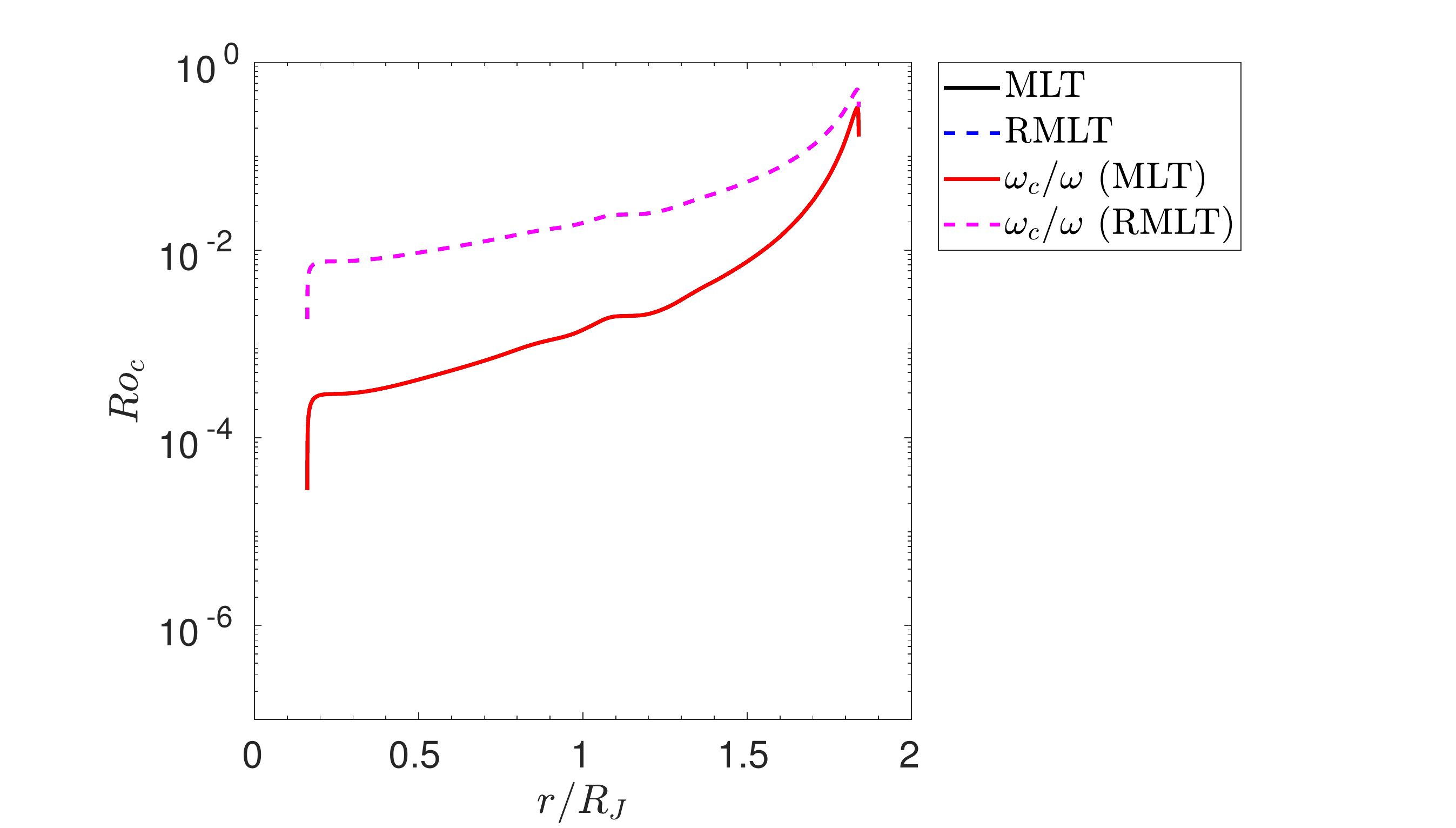}
     \end{subfigure}
     \caption{Left: Flux-based MLT (black) and RMLT (dashed-blue) Rossby numbers as a function of radius for the Jupiter-like planet with $P_{\mathrm{rot}}=10$ hrs, $P_{\mathrm{orb}}=P_{\mathrm{tide}}=1$ day after evolving the model for $4.5$ Gyr. This is much smaller than one in the whole of the interior according to both prescriptions, i.e. the interior is strongly rotationally constrained. The ratio of convective to tidal frequencies (``tidal Rossby number"), is also much smaller than one for these parameters, indicating that the planet is in the fast tides regime. Right: Same but for the inflated Hot Jupiter with $P_{\mathrm{rot}}=P_{\mathrm{orb}}=P_{\mathrm{tide}}=1$ day. Convection is stronger in this model but the same regimes (rapid rotation and fast tides) hold as in the left panel. The ratio of $\omega_c/\omega$ is equal to the convective Rossby number here, hence the lines overlap.}
    \label{fig:Roc_Jup}
\end{figure*}

\begin{figure*}
     \centering
     \begin{subfigure}[b]{0.495\textwidth}
         \centering
    \includegraphics[width=\linewidth, trim=1.7cm 0cm 2cm 1cm,clip=true]{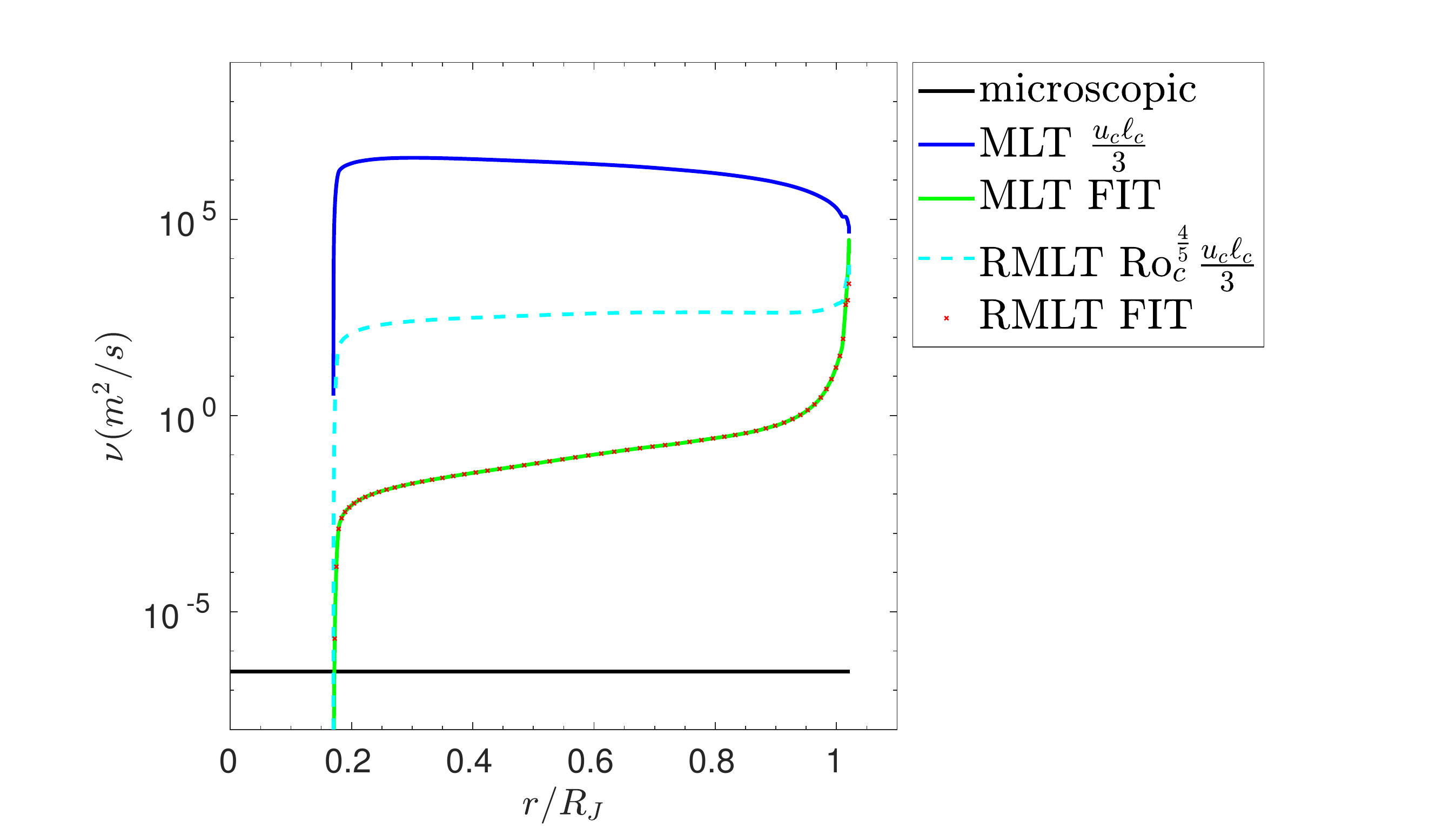}
     \end{subfigure}
     \hfill
     \begin{subfigure}[b]{0.495\textwidth}
        \centering
    \includegraphics[width=\linewidth,trim=1.7cm 0cm 2cm 1cm,clip=true]{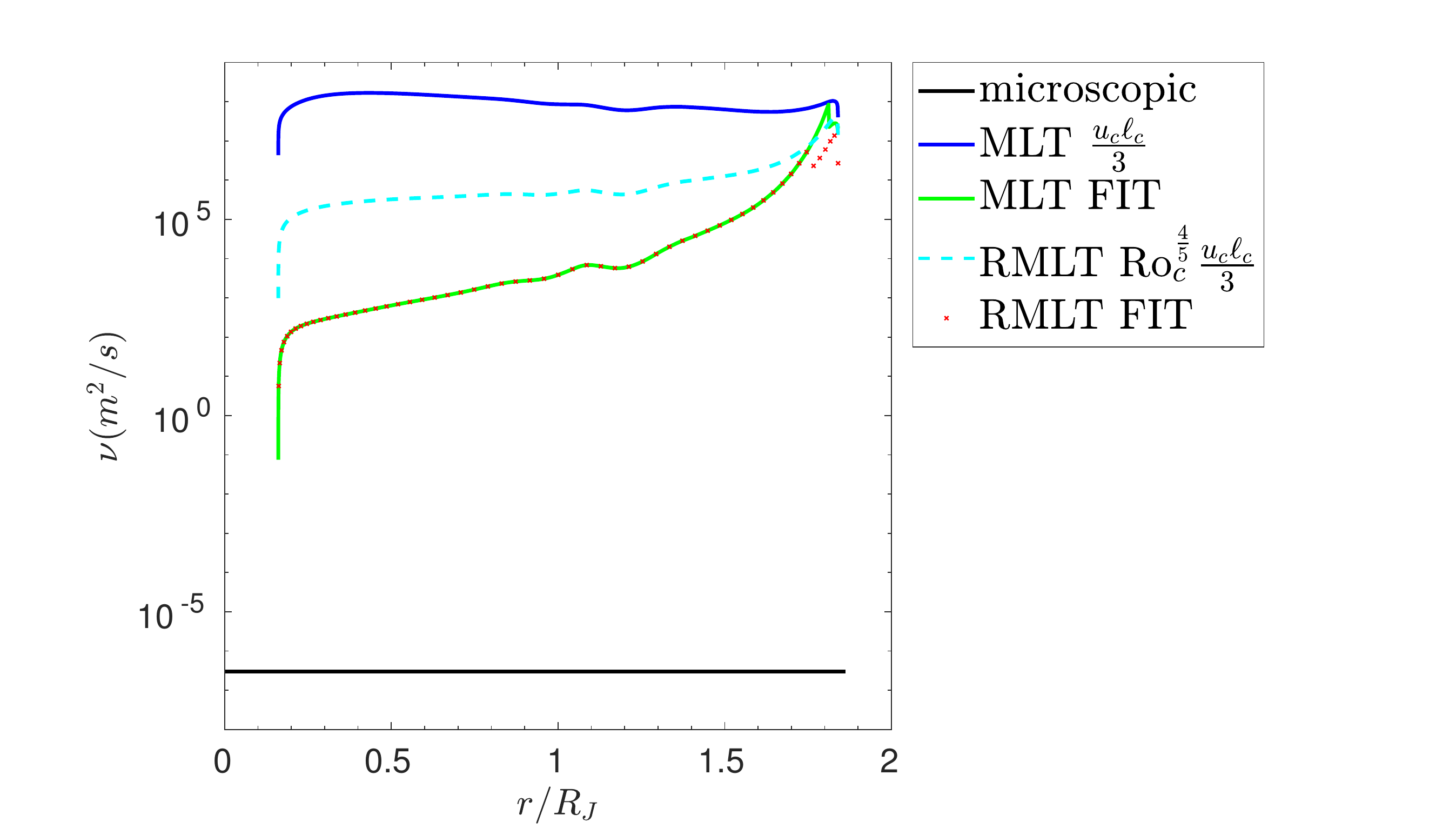}
     \end{subfigure}
     \caption{Left: Effective viscosity as a function of radius for the Jupiter-like planet with $P_{\mathrm{rot}}=10$ hrs, $P_{\mathrm{orb}}=P_{\mathrm{tide}}=1$ day after evolving the model for $4.5$ Gyr. We show the microscopic viscosity $3\cdot10^{-7}\ m^2/s$ reproduced from \citet[][]{French2012} (solid-black) for reference, the MLT prediction in the low frequency regime (solid-blue), the MLT prediction in the fast tides regime (solid-green), the RMLT prediction in the slow tides regime (dashed-cyan) and the RMLT prediction in the fast tides regime (dotted-red). The fast tides predictions overlap regardless of regime whereas applying RMLT in the slow tides regime drastically reduces the effective viscosity. Right: Same but for the inflated Hot Jupiter with $P_{\mathrm{rot}}=P_{\mathrm{orb}}=P_{\mathrm{tide}}=1$ day. The Hot Jupiter model has more efficient convection and larger effective viscosity in all regimes.}
    \label{fig:nu_Jup}
\end{figure*}

We next present our results for Rossby numbers and the corresponding effective viscosities -- in both the fast tide and slow tide regimes, using both MLT and RMLT -- as a function of radius in our two planetary models. For these illustrative calculations we set $P_{\mathrm{orb}}=1$ day and $P_{\mathrm{rot}}=10$ h for the Jupiter model, mimicking a planet similar to Jupiter but orbiting its star with a period of 1 day. For the Hot Jupiter model we instead set $P_{\mathrm{orb}}=P_{\mathrm{rot}}=1$ day, representing spin-orbit synchronisation. The tidal period is $P_{\mathrm{tide}}=1$ day for both figures. This can be thought to represent the eccentricity tide in a spin-orbit synchronised planet, as opposed to being based on $\gamma=\Omega-n$, but is only chosen for illustration in the first model. 

In Fig.~\ref{fig:Roc_Jup} the Rossby numbers are plotted in the Jupiter model on the left and the Hot Jupiter model on the right. The MLT Rossby number as calculated from the data is plotted in solid-black; the one calculated from RMLT is plotted in dashed-blue. The MLT Rossby numbers are clearly smaller, but even in RMLT they are much smaller than one, indicating that the convection is strongly rotationally-constrained. Note that the lower densities and stronger convection in the inflated Hot Jupiter model produce larger Rossby numbers, but they are still much smaller than one. This justifies the use of RMLT (over MLT) in giant planets. 

The ratio of convective to tidal frequencies $(\omega_c/\omega)$ is also plotted as a function of radius in Fig.~\ref{fig:Roc_Jup}. The MLT prediction for this ``tidal Rossby number" is plotted in solid-red and the RMLT prediction is plotted in dashed-magenta, and these only differ by a factor of $\Omega/|\gamma|$. In the Hot Jupiter model this factor equals one for our chosen parameters, and as such $\omega_c/\omega=\textrm{Ro}_c$. For both models $\text{Ro}_c\ll 1$, such that RMLT is the appropriate description of the convection, and hence for the convective frequency. This figure indicates that the fast tides regime is relevant inside both models (except for perhaps the final percent or so of the radius where we approach the surface stable layer).

The effective viscosity as a function of radius is shown in Fig.~\ref{fig:nu_Jup} in both planetary models. In the left panel, we show the effective viscosity in the Jupiter model for our chosen rotational and tidal periods, which demonstrates that this is much larger than the microscopic viscosity (solid-black) for all predictions. To compute the kinematic viscosity in Jupiter requires sophisticated calculations outside the scope of our models (and not calculated within MESA), so we use the typical value obtained by \citet[][]{French2012} for reference, of $\nu=3\cdot10^{-7}\ m^2/s$, in both panels.

There are large differences between the various predictions for $\nu_{\mathrm{eff}}$ in Fig.~\ref{fig:nu_Jup}. The MLT prediction in the slow tides regime in solid-blue predicts $\nu_{\mathrm{eff}}\approx 10^6\ m^2/s$, while the RMLT prediction in the same slow tides regime in dashed-cyan only attains values of $\approx10^2\ m^2/s$. The MLT prediction for this regime decreases slightly from the interior to the surface, which is because the convective length scale decreases faster than the convective velocity increases from the core to the surface. On the other hand, the RMLT prediction increases towards the surface, because the Rossby number rapidly increases there. The fast tides regime prediction according to both RMLT and MLT (strictly obtained using all three regimes in Eq.~\ref{eq:effviscCraig_RMLT} and the uncorrected version respectively, but the fast tides one is most relevant) are plotted in solid-green and dotted-red respectively. The two lines overlap because the effective viscosity is independent of rotation according to both theories, as we have demonstrated above. The effective viscosity in the fast tides regime is however several orders of magnitude smaller still than both predictions in the slow tides regime, with a value of only $\approx10^{-2}\ m^2/s$ except for close to the surface. This value is much larger than the microscopic viscosity, but is probably negligibly small for damping tidal flows. This would imply an effective Ekman number in the fast tides regime of $\textrm{Ek}\approx10^{-2}/(2\cdot10^{-4}\cdot(10^4)^2)=\mathcal{O}(10^{-7})$, where we've set $d$ to be a similar order of magnitude as the RMLT convective length scale, which is $\mathcal{O}(10^4)$ throughout most of the interior, except very close to the surface. This value is several orders of magnitude larger than the microscopic value, but is smaller than what is often used in numerical simulations.

The right panel of Fig.~\ref{fig:nu_Jup} shows the effective viscosity as a function of radius for our inflated Hot Jupiter model. We observe that all values for $\nu_{\mathrm{eff}}$ have shifted upwards compared to our Jupiter model. However, even in this model we expect to be in the fast tides regime throughout (almost) the entire planet, which would predict $\nu_{\mathrm{eff}}\approx 10^2\ m^2/s$. Thus the increased irradiation and internal heating introduced here results in significantly larger effective viscosities, and therefore smaller values of $Q'$.

\subsection{Tidal dissipation rates in Jupiter and Hot Jupiters}
\label{tidaldiss}

\begin{figure*}
     \centering
     \begin{subfigure}[b]{0.495\textwidth}
         \centering
    \includegraphics[width=\linewidth, 
    trim=0cm 0cm 2cm 0.5cm,clip=true]{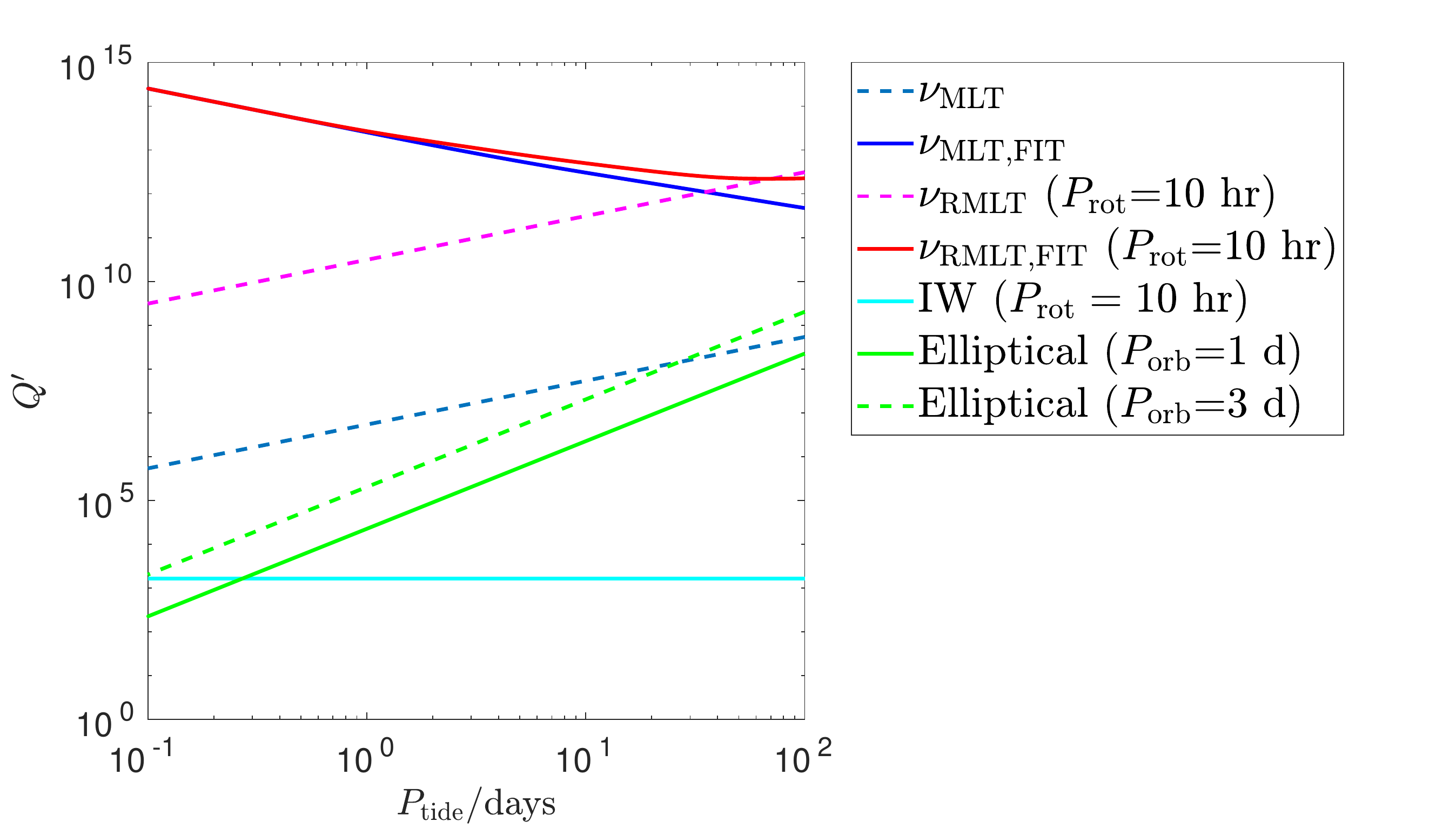}
     \end{subfigure}
     \hfill
     \begin{subfigure}[b]{0.495\textwidth}
        \centering
    \includegraphics[width=\linewidth, 
    trim=0cm 0cm 2cm 0.5cm,clip=true]{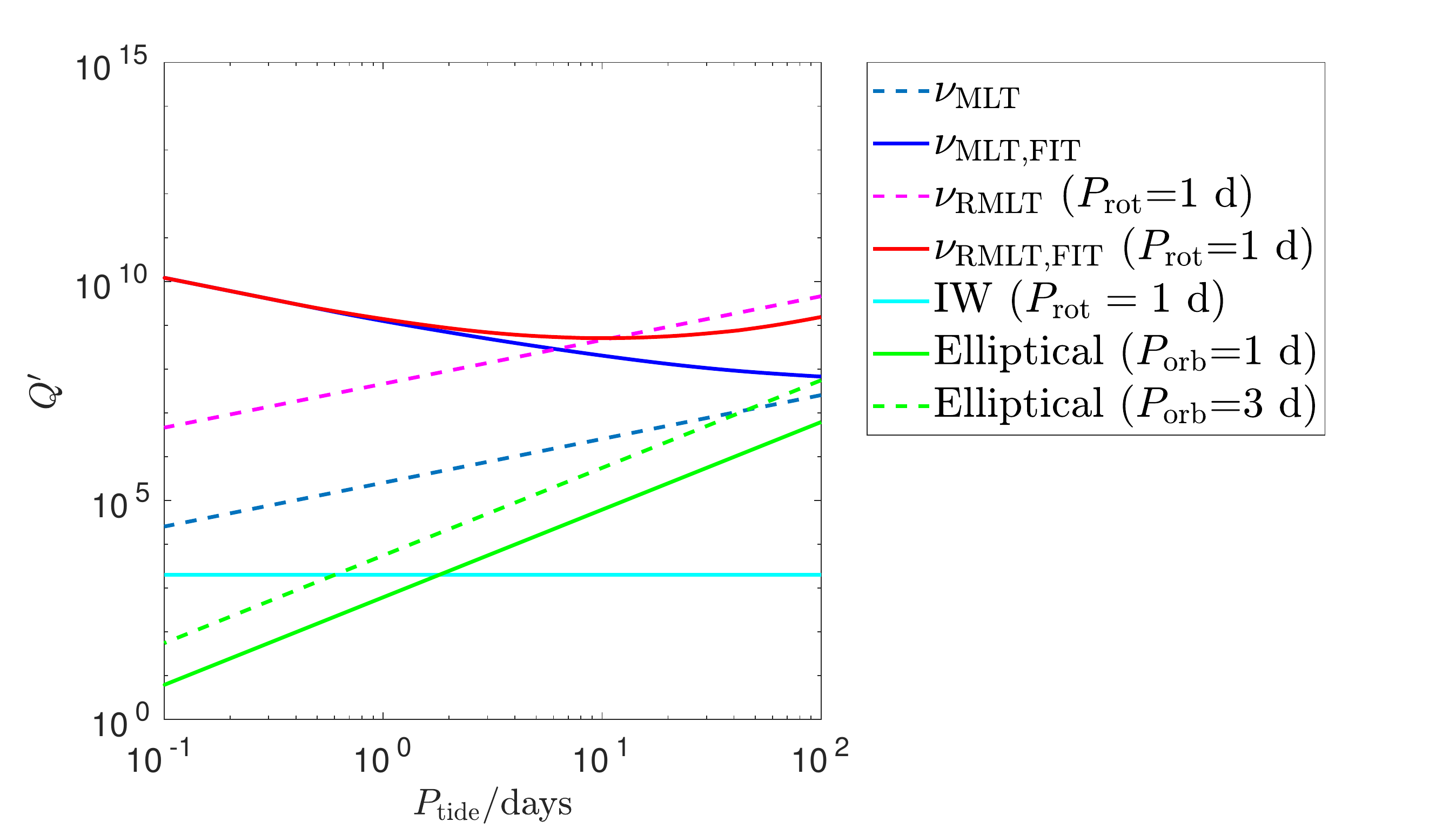}
     \end{subfigure}
     \caption{Tidal quality factor $Q'$ as a function of tidal period for a myriad of mechanisms. Left: Jupiter model. Right: Inflated Hot Jupiter model. In both panels, MLT and RMLT predictions for $Q'$ due to convective damping of equilibrium tides using an effective viscosity with no tidal frequency reduction (low frequency regime) are shown in dashed-blue and -magenta respectively. The frequency-reduced effective viscosities in solid-blue and -red for MLT and RMLT respectively indicate that the frequency reduction significantly reduces the effectiveness of the dissipation. The elliptical instability in solid-green and dashed-green lines for two different orbital periods, and the (linear) frequency-averaged inertial wave dissipation in solid-cyan are also plotted. Inertial waves are considerably more dissipative than equilibrium tide damping by turbulent viscosity, whether they are linearly or nonlinearly (i.e.\ via elliptical instability) excited. Elliptical instability is predicted to be dominant for the shortest tidal periods, and linear excitation of inertial waves is dominant for longer periods. The Hot Jupiter model has smaller $Q'$ (hence more efficient dissipation) for all dissipation mechanisms due to the larger radius and slower rotation.}
\label{fig:Q_Jup}
\end{figure*}

Now that we have obtained radial profiles of $\nu_{\mathrm{eff}}$ we can use these to compute the resulting damping of the equilibrium tide and the associated tidal quality factor $Q'$ in our planetary models. We follow the approach described in \citet[][]{Barker2020} to calculate the equilibrium tidal flow and its resulting dissipation and omit details here. To do so, we first calculate the irrotational equilibrium tide (more specifically the dominant quadrupolar $l=2$ component with azimuthal wave number $m=2$) defined in their section 2, since this is likely to be the correct one in giant planets\footnote{This should be used in preference to the equilibrium tide of e.g.~\citet[][]{Zahn1989Turbvisceqtide} in convective regions of planets since $|N^2|\ll \omega^2$ \citep{Terquem1998IGW}. We neglect the action of rotation on this component by considering Coriolis forces on the equilibrium tide to drive the wavelike tide. This equilibrium/dynamical or non-wavelike/wavelike splitting of the tidal response is formally valid in linear theory for low frequency (relative to the dynamical frequency) tidal forcing \citep{OgilvieIW}.}. The dissipation of this tidal flow is computed assuming an effective viscosity that acts like an isotropic microscopic kinematic viscosity but with a local value $\nu_{\mathrm{eff}}(r)$ to damp the equilibrium tide. This requires performing the integral over radius in Eq.~20 of \citet{Barker2020} to obtain the dissipation rate $D_\nu$. The only modification here is we account for the rotational dependence of $\nu_{\mathrm{eff}}$ and $\omega_c$ as described above, otherwise we employ their Eq.~27 to obtain $\nu_{\mathrm{eff}}(r)$ in the various different frequency regimes (the slightly different pre-factors we have obtained lead to negligible differences here). The resulting tidal quality factor is then obtained by:
\begin{equation}
    Q'=\frac{3(2l+1)R^{2l+1}}{16\pi G}\frac{|\omega||A|^2}{D_\nu},
\end{equation}
where $A\propto \epsilon$ is the amplitude of the tidal perturbation (so that the ratio $D_\nu/|A|^2$ and hence $Q'$ is independent of tidal amplitude in linear theory), $G$ is the gravitational constant, $R$ is the planetary radius, and $\omega=2\pi/P_{\mathrm{tide}}$.

To obtain $Q'$ for the elliptical instability we can use Eq.~\ref{eq:epsiloncubed} and find \citep[]{Barker2013}:
\begin{equation}
     Q'\approx\frac{10^5}{\chi/0.05}\left(\frac{m_1+m_2}{m_2}\right)\bigg(\frac{P_{\mathrm{orb}}}{1\textrm{ d}}\bigg)^4,
\end{equation}
where $\chi$ is fit from our simulations. This is particularly crude because it equates the size of our Cartesian box with the planetary radius, but in the absence of a better approach it provides us with an estimate that is broadly consistent with simulations. This represents a ``nonlinear" mechanism of tidal dissipation because $Q'$ depends on tidal amplitude and has a strong dependence on the orbital period $P_{\mathrm{orb}}$.

To put our results for these two mechanisms in context, we also compute $Q'$ resulting from the dissipation of linearly excited inertial waves in this planetary model by applying the frequency-averaged formalism of \citet{OgilvieIW}. We follow the approach outlined in Section 3.1 and Eq.~30 of \citet{Barker2020} to obtain $Q'$ in our planetary models, fully accounting for the planetary structure. This prediction for $Q'$ provides a tidal frequency-independent ``typical level of dissipation" due to inertial waves according to linear theory. This method necessarily ignores the potentially complicated frequency-dependence of the dissipation in linear theory and any possible modifications of this by nonlinear effects \citep[e.g.][]{OL2004,AstoulBarker2022}. However, it is thought to be representative of the dissipation of inertial waves excited by linear tidal forcing, i.e.~not via elliptical instability (which also excites inertial waves, but nonlinearly in this regard).

We show $Q'$ in Fig.~\ref{fig:Q_Jup} as a function of tidal period for each of these mechanisms. The spin frequency is fixed by setting $P_{\mathrm{rot}}=10$ hr for the Jupiter model in the left panel and $P_{\mathrm{rot}}=1$ day for the Hot Jupiter model in the right panel. For the elliptical instability, we provide two predictions, one with $P_{\mathrm{orb}}=1$ day and the other with $P_{\mathrm{orb}}=3$ day. Note that when $\nu_{\mathrm{eff}}$ is independent of tidal frequency (in the low frequency regime), $D_\nu\propto \omega^2\propto P_{\mathrm{tide}}^{-2}$ and $Q'\propto \omega^{-1}\propto P_{\mathrm{tide}}$, while in the high frequency regime where $\nu_{\mathrm{eff}}\propto \omega^{-2}\propto P_{\mathrm{tide}}^2$, $D_\nu$ is independent of $\omega$ and $Q'\propto\omega\propto P_{\mathrm{tide}}^{-1}$. In addition, we expect $Q'$ due to elliptical instability to scale as $\omega^3\propto P_{\mathrm{tide}}^{-3}$ and the frequency-averaged inertial wave prediction to be independent of $\omega$ by definition.

The left panel of Fig.~\ref{fig:Q_Jup} demonstrates that convective damping of equilibrium tides by an effective viscosity is indeed an inefficient tidal dissipation mechanism in giant planets and leads to large $Q'$. The low tidal frequency regime in dashed-blue and dashed-magenta for MLT and RMLT, respectively, indicate their strongest dissipation when the tidal frequency is large. Note that these predictions are calculated using the classical prefactor of $1/3$ for the effective viscosity for illustration. These lines indicate that if RMLT applies, as is expected, $Q'$ is still $\mathcal{O}(10^9)$ if we neglect the frequency-reduction of $\nu_{\mathrm{eff}}$, thus the dissipation (and resulting tidal evolution) is weak. The combination of low, intermediate and high tidal frequency regimes for $\nu_{\mathrm{eff}}$ with the fitted prefactors dubbed $\nu_{\mathrm{FIT}}$ in solid-blue and solid-red indicates that the high tidal frequency regime impacts the effective viscosity significantly, particularly when $P_{\mathrm{tide}}$ is small. These predictions approximately connect to the frequency-independent MLT and RMLT predictions for large $P_{\mathrm{tide}}$ where there is a transition to the intermediate and low frequency regimes. The prefactors obtained using fits to simulations are larger than the dashed-magenta prediction, thus resulting in a slightly lower $Q'$ when transitioning into the low tidal frequency regime. This is because the factor $1/3$ often utilised, as plotted here for the MLT and RMLT lines, is essentially arbitrary, unlike our numerical fits.

The elliptical instability on a 1 day orbit (solid-green) on the other hand is an efficient dissipation mechanism, particularly when the tidal frequency is high. It is significantly more effective than convective damping of equilibrium tides according to each prediction for the entire range of tidal periods considered. The elliptical instability prediction on a 3 day orbit (dashed-green) is weaker than the 1 day orbit prediction, but would still predict more effective dissipation even than the slow-tides MLT effective viscosity for almost all of the parameter range considered. The most efficient mechanism in this model, except for the very highest tidal frequencies, is the frequency-averaged dissipation due to inertial waves shown in solid-cyan, which produces a $Q'= \mathcal{O}(10^3)$ for our chosen rotation period. Since the rotation period is known, we would thus predict a typical value 
\begin{align}
    Q' \approx 2 \cdot 10^3 \left(\frac{P_{\mathrm{rot}}}{10 \mathrm{hr}}\right)^2,
\end{align}
for tidal dissipation due to inertial waves. Indeed, this is sufficiently dissipative to explain tidal dissipation rates in Jupiter and Saturn \citep{Lainey2009JupiterQ,Lainey2012SaturnQ,Lainey2017SaturnQ_Cassini}, without requiring any resonance-locking scenario \citep[e.g.][]{F2016}.

The Hot Jupiter model on the other hand has a larger radius, stronger convection, and is rotating somewhat more slowly, so it has much higher effective viscosities and is impacted to a lesser extent by rotation. As a result, all mechanisms except the dissipation of (linear) inertial waves are more efficient. The elliptical instability is predicted to be particularly efficient for short orbital periods, e.g.~1 day orbit prediction for $Q'=\mathcal{O}(10^2)$ when the tidal period is 1 day. The increase in dissipation here due to the elliptical instability stems from the large radius of the Hot Jupiter, resulting in $\epsilon\approx0.095$. Radius inflation and internal heating, as well as the marginally decreased rotation rate, allows the convective damping of equilibrium tides to operate more efficiently than in the Jupiter-like model in the left panel. However, once again the inertial wave mechanisms are predicted to be substantially more dissipative than effective viscosity acting on equilibrium tides. Linear dissipation of inertial waves occurs with a similar order of magnitude to the Jupiter-like model, and is predicted to be dominant for $P_{\mathrm{tide}}\gtrsim 2$ days.

\section{Discussion and conclusion}
\label{sec:discussion}
\subsection{Comparison with previous work}

We find tidal dissipation rates due to the elliptical instability that are roughly equivalent to those observed in prior work \citep[][]{Barker2013,Barker2014,Barker2016} when it operates. Indeed, our efficiency factor $\chi$ is consistent with a similar value ($\chi\in[0.01,0.1]$) and is independent of Rayleigh number when the elliptical instability operates. We also potentially observe the $\epsilon^6$ scaling found in \citet{Barker2013}, and find that if this scaling holds true only the very closest Hot Jupiters experience significant tidal dissipation due to the elliptical instability (because this would effectively imply a much smaller value of $\chi$ for realistic $\epsilon$ values). 

The scaling laws we have confirmed using temperature-based RMLT match those obtained from Coriolis-Inertial-Archimedean (CIA) triple balance arguments \citep[e.g.][and many others]{IP1982,Aubert2001,Jones2015,G2016,Celine2019,Aurnou2020_scalings,Bouillaut2021} and the applicability of these temperature-based scalings reinforce the applicability of the diffusion-free flux-based scalings confirmed previously using simulations \citep[][]{Barker2014RMLT,CurrieconvRMLT}. We observe the transition from RMLT to MLT to begin around $\textrm{Ro}_c\approx0.1$ as in \citet[][]{Barker2014RMLT}{}{}, and find RMLT is the appropriate description of (sufficiently turbulent) convection for $\textrm{Ro}_c\lesssim 0.1$. Similar to \citet[][]{Celine2019}{}{} we find sufficiently strongly supercritical (turbulent) convection is required for the convective length scale to agree with the diffusion-free predictions of RMLT. Our results for the length scale depend strongly on how it is calculated, but they are (when properly interpreted) not generally consistent with the predictions from the linear onset of convection.

One caveat to the above is that different methods to calculate the convective length-scale give vastly different results, and simulations must be turbulent enough and sufficiently rotationally-constrained to obtain reasonable agreement with RMLT. We favour definitions for the length-scale based on either the peak or the integrated ``centroid" wavenumber for the heat flux spectrum as a function of horizontal wavenumber, which give better agreement with RMLT than e.g.~the temperature fluctuation spectrum. The most challenging case for testing RMLT is when measuring the convective length scale as a function of rotation rate (Ekman number) at constant Rayleigh number, where only a narrow range of simulations are in the appropriate regime (sufficiently turbulent but rotationally-constrained).
Furthermore, we find that when fixing supercriticality, i.e. similar to measuring the convective length scale as a function of $\textrm{RaEk}^{4/3}$, the length-scale scales proportional to $\textrm{Ek}^{1/3}\propto \Omega^{-1/3}$. While this superficially agrees with the linear onset prediction (in which the scale is viscously-controlled), we demonstrate that this coincides with the diffusion-free prediction of RMLT when supercriticality is fixed, and we find a different pre-factor than predicted by the former. Furthermore, we find that the appropriate regime in terms of the convective Rossby number for RMLT to be a valid description of convection is $\textrm{Ro}_c<0.1$, with a transition in scaling laws from RMLT to MLT starting at $\textrm{Ro}_c\approx0.1$.

Regarding the tidal frequency dependence of the effective viscosity of turbulent convection in damping the equilibrium tide, our results are consistent with the same three regimes of tidal frequency as the non-rotating simulations of \cite{Craig2020effvisc}, even though they used an oscillating shear flow and we use a more realistic equilibrium tidal (elliptical) flow. However, we have studied rotating convection and thus obtained different prescriptions in terms of the dimensionless parameters that are described well by our heuristic application of RMLT. Despite these differences, our results are consistent with the intermediate tidal frequency scaling of $\left(\omega_c/\omega\right)^{-1/2}$ as \citet[][]{Craig2020effvisc,Vidal_Barker_2020}{}{}. The prefactors in the intermediate and high tidal frequency regimes are lower by approximately a factor of two. However, we observe a transition from the high tidal frequency to the intermediate tidal frequency at the same value $\omega/\omega_c\approx 5$.

\subsection{Future work}
One avenue for future work would be to perform simulations varying $\gamma$ and $\Omega$, to fully disentangle the different dependencies on $\textrm{Ro}_c$ and $\textrm{Ro}_\omega$. Changing $\gamma$ and $\Omega$ independently would allow the realistic scenario of a planet orbiting with a nonzero orbital frequency in the inertial frame to be studied. This would be likely to impact the strength of the elliptical instability as it changes its linear growth rate. This would be expected to cause suppression of the elliptical instability for different strengths of convective driving (or for a different $\epsilon$ for fixed Ra). However, we do not expect any of our conclusions will be substantially modified in this case.

Furthermore, in our current setup the Cartesian box is situated at the poles of the planet, with the gravity and rotation axis both pointing in the $z$-direction. The latitudinal location of the box, and thus the relative directions of gravity and the rotation axis could affect the resulting tidal dissipation. If the box is moved to a lower latitude, the directions of gravity and the rotation axis will be misaligned, causing convective motions subjected to rapid rotation to change angle \citep[][]{Novi2019,CurrieconvRMLT}. At lower latitudes the vortices introduced by rotating convection turn into zonal flows, which could modify dissipation due to the elliptical instability as well as the effective viscosity of convection. In addition, \citet{CurrieconvRMLT} demonstrated that the predictions of RMLT hold from the poles to mid-latitudes, but at low-latitudes deviations were observed due to the presence of both zonal flows and because boundary conditions constrain the flow in the latitudinal direction. Hence, future work should focus on obtaining a theoretical understanding of convection and of the effective viscosity at mid and low latitudes, in the presence of strong zonal flows.

There are strong magnetic fields present in Jupiter, and it is expected that Hot Jupiters would also have strong fields. This expectation is supported by observations tentatively inferring that a number of Hot Jupiters possess strong magnetic fields \citep[][]{Cauley_HJ_strongB}. Therefore it is important to study the inclusion of magnetic fields, as they could have significant effects on tidal dissipation. Magnetic fields may prevent LSV formation by the elliptical instability, and therefore allow a continuous operation of the resulting energy transfers \citep[][]{Barker2014}{}{}. It is likely that they also prevent the formation of the convective LSV \citep[e.g.][]{Maffei2019}, and if so could allow continuous operation of the elliptical instability while convection is present in the system, potentially allowing for enhanced tidal dissipation. In addition, sufficiently strong magnetic fields will modify the properties of the convection and therefore the effective viscosity, and it remains to be seen how valid the predictions of \cite{StevensonRMLT} would be in this case.

Also on the topic of magnetic fields, in a similar fashion to convection acting as an effective viscosity, an effective turbulent magnetic resistivity might arise \citep[][]{Tobias_2013_eta_eff,Cattaneo2013_eta_eff}. The turbulent magnetic resistivity has been explored previously in accretion discs \citep[][]{Lesur2009_eta_eff}{}{}, but not in the context of tidal dissipation. It is entirely unknown whether an effective resistivity acting on a tidal flow features the same frequency-reduction as the effective viscosity \citep[as assumed by][]{Wei2022}, and whether it might be an effective dissipation mechanism of the equilibrium tide for high tidal frequencies. 

Another question lies in the applicability of effective turbulent diffusivities like the effective viscosity and effective resistivity. The effective viscosity as calculated here is purely representative of the interaction of rotating convection with the tidal background flow. It is unclear if, for instance, the interaction between inertial waves generated by the elliptical instability (or more directly by tidal forcing) and convection can be modelled in the same way. So studying the interaction of convection with inertial waves, and calculating whether (and if so how) this can be modelled using an effective viscosity is an important topic for future work. In addition, the possible role of alternative energy transfer routes for fast tides, such terms involving correlations between tidal flow components and gradients of the convective flow (which identically vanish in our model) should be explored in global models to determine if they are ever important \citep[e.g.][]{Terquem2021,BA2021}.

A final avenue of future work is related to the analysis of tidal dissipation rates using planetary models. It would be worthwhile to modify the equation of state in a manner akin to \citet[][]{Muller2020}{}{}, which would allow us to obtain an extended dilute core and to measure the impact of such a core on tidal dissipation rates. Furthermore, a stably stratified dilute core might provide an important additional contribution to tidal dissipation by permitting the excitation of internal (inertia-)gravity waves \citep[e.g.][]{F2016,Andre2019,Christina_2020,Pontin2023,Lin2023,Dewberry2023}. Finally, studying how $Q'$ evolves with planetary evolution for each of these mechanisms would be worthwhile. For self-consistency, one might then consider also evolving orbital parameters and irradiation fluxes in tandem with the structural evolution.

\subsection{Conclusion}

We have studied interactions between the elliptical instability and rotating turbulent convection in a local model representing a small patch of a giant planet (or star), building upon the simulations and analysis in \citetalias[][]{deVries2023}{}{}. We have found the elliptical instability to provide time-averaged tidal dissipation rates consistent with an $\epsilon^3$ scaling when it operates (where $\epsilon$ is proportional to the dimensionless tidal amplitude), which would lead to tidal quality factors $Q'\propto P_{\mathrm{orb}}^4$ \citep[consistently with][]{Barker2013,Barker2014,Barker2016}. We find a dissipation rate sufficient to suggest this tidal mechanism could be the dominant one for the very shortest-period Hot Jupiters, with orbital periods shorter than two days. In this work we find that the observed efficiency factor ($0.05\approx \chi \lesssim 0.18$ as an upper bound, defined such that in our units the dissipation rate $D\equiv \chi \epsilon^3\gamma^3$) seems to be independent of the convective driving (Rayleigh number) as long as the elliptical instability operates. Some of our results are also consistent with a steeper $\epsilon^6$ scaling, which, if robust, would significantly weaken tidal dissipation for realistic values of $\epsilon$, restricting the effectiveness of this mechanism except for the very shortest orbital periods.

Our simulations have also obtained a sustained energy injection rate scaling as $\epsilon^2$ for smaller values of $\epsilon$ than those for which the elliptical instability is observed. This can be interpreted as an effective viscosity arising from the interaction between rotating convection and the equilibrium tidal flow that is independent of $\epsilon$ (as would be predicted by a linear tidal mechanism). On the other hand, this effective viscosity is observed to depend on the convective velocity, length scale and tidal frequency. In this work we have obtained scaling laws for convective velocities and length scales, which are used to find predictions for the convective frequency and the effective viscosity, using both (temperature-based) MLT and RMLT prescriptions. We find very good agreement between the predictions of RMLT and our simulation data. Our simulations confirm the applicability of the diffusion-free scalings of RMLT \citep[e.g.][]{StevensonRMLT,Barker2014RMLT,CurrieconvRMLT,Aurnou2020_scalings} to describe sufficiently turbulent rapidly rotating convection.

We find that the scaling laws for the effective viscosity as a function of convective velocity, length scale and frequency -- when the rotational modification of these quantities is accounted for -- previously found in non-rotating simulations \citep[][]{Craig2020effvisc}{}{} largely hold true in our rotating simulations. Our results support the frequency-reduction of the effective viscosity for fast tides $(\omega_c/\omega)^{2}$ when $\omega\gg \omega_c$. We also confirm the presence of the intermediate frequency regime they identified in our simulations, and that the transition to this regime occurs at a similar ratio of $\omega/\omega_c\approx5$. Furthermore, when considering the more realistic flux-based scalings instead of temperature-based scalings we find that the MLT and RMLT predictions for the high frequency (fast tides) regime for the effective viscosity are identical and are independent of rotation rate (as long as the heat flux is independent of rotation rate, which is a reasonable first approximation).

Finally, we employed the MESA code to construct illustrative interior models of a Jupiter-like and an inflated Hot-Jupiter-like planet, subject to Jupiter-like irradiation and Hot Jupiter-like irradiation plus artificial interior heating, respectively. We compute the rotational modifications of convective velocities and length scales in these models, as well as the modifications of the effective viscosity to allow us to compute tidal dissipation resulting from convective damping of equilibrium tides according to the scaling laws we have derived and verified with simulations. In both models (even in inflated short-period Hot Jupiters), we find the convective Rossby numbers to be much smaller than one, indicating that the convection is strongly affected by rotation, therefore motivating our study of this regime in this paper. We find that for almost all applications to giant planets, the fast tides regime, in which the tidal frequency is much larger than the convective frequency, is highly likely to be the relevant one. In this regime the effective viscosity scales as $\nu_{\mathrm{eff}}\propto (\omega_c/\omega)^2$. The resulting tidal quality factors $Q'$ for equilibrium tide damping \citep[computed following][]{Barker2020} are estimated to be in excess of $10^9$ for tidal periods of interest, and this mechanism is therefore predicted to be an ineffective one in giant planets.

On the other hand, we predict the elliptical instability to be efficient for very short orbital and tidal periods (with $Q'\sim 10^2$ in Hot Jupiters for periods of order one day), but that it falls off rapidly with increasing (tidal and orbital) periods. 

We also compute for the first time $Q'$ arising from the frequency-averaged dissipation due to inertial waves in ``realistic models" of giant planets \citep[following][]{OgilvieIW,Barker2020}. This mechanism assumes these waves to be excited linearly by tidal forcing, as opposed to nonlinearly (with respect to tidal amplitude) by the elliptical instability. Inertial waves are by far the most efficient mechanism studied here, either those excited by the elliptical instability for short orbital and tidal periods, or by the linear frequency-averaged dissipation. The latter leads to $Q'\approx 10^3 (P_{\mathrm{rot}}/10 \mathrm{hr})^2$ for Jupiter-like rotation periods, which is consistent with the efficient tidal dissipation rates required to explain the observed orbital migration of the moons of Jupiter and Saturn \citep[e.g.][where tidal amplitudes are likely to be too small for the elliptical instability to operate effectively]{Lainey2009JupiterQ,Lainey2012SaturnQ}. 

All mechanisms except the frequency-averaged inertial wave mechanism are more efficient in the Hot Jupiter model due to its larger radius, weaker rotation and stronger convective driving. This allows the elliptical instability to be on par or even more efficient than linearly-excited inertial waves in the shortest-period Hot Jupiters. We conclude that inertial wave mechanisms are probably the most efficient ones for dissipating tidal energy in giant planets, at least those without extended stable layers.

\section*{Acknowledgements}
 NBV was supported by EPSRC studentship 2528559. AJB and RH were supported by STFC grants ST/S000275/1 and ST/W000873/1. RH would like to thank the Isaac Newton Institute for Mathematical Sciences, Cambridge, for support and hospitality during the programme ``Frontiers in dynamo theory: from the Earth to the stars'' where work on this paper was undertaken. This work was supported by EPSRC Grant No.~EP/R014604/1. RH's visit to the Newton Institute was supported by a grant from the Heilbronn Institute. Simulations were undertaken on ARC4, part of the High Performance Computing facilities at the University of Leeds, and the DiRAC Data Intensive service at Leicester, operated by the University of Leicester IT Services, which forms part of the STFC DiRAC HPC Facility (\href{www.dirac.ac.uk}{www.dirac.ac.uk}). The equipment was funded by BEIS capital funding via STFC capital grants ST/K000373/1 and ST/R002363/1 and STFC DiRAC Operations grant ST/R001014/1. DiRAC is part of the National e-Infrastructure.

\section*{Data Availability}
The simulation data used in this article will be shared on reasonable request to the corresponding author.


\bibliographystyle{mnras}
\bibliography{Manuscript_paper2}



\appendix

\section{Resolution}

\begin{table}
\caption{Table of resolutions used in the simulations with different Rayleigh numbers, Ekman numbers, and horizontal box size $L_x=L_y=4$ (unless otherwise specified). The same resolution was used for all ellipticities. The square brackets indicate all entries within are multiplied by the factor 5 in front.}
\centering
\begin{tabular}{|l|l|l|}
\hline
$\textrm{Ek}=5\cdot10^{-5.5}$                                   & $n_x,n_y$   & $n_z$  \\ \hline
$R=-6,-4,-3,-1,-0.8,0.3,0.8,2,3,4,5,6$ & 256x256 & 96  \\ \hline
$R=7,8$                        & 256x256 & 128 \\ \hline
$R=-10,9,10,11,12,15$                    & 256x256 & 160 \\ \hline
$R=20$                       & 256x256 & 224 \\ \hline

$\textrm{Ra}=1.3\cdot10^{8}$                                   & $n_x,n_y$   & $n_z$  \\ \hline
$\text{Ek}=5\cdot\left[10^{-5.6},10^{-5.7},10^{-5.8},10^{-5.9}\right]$ & 256x256 & 96  \\ \hline
$\text{Ek}=5\cdot10^{-5.4}$                      & 256x256 & 128 \\ \hline
$\text{Ek}=5\cdot\left[10^{-5.2},10^{-5.3}\right]$                       & 256x256 & 160 \\ \hline
$\text{Ek}=\textbf{5}\cdot\left[10^{-5},10^{-5.1}\right]$                       & 256x256 & 196 \\ \hline
$\text{Ek}=5\cdot\left[10^{-4.5},10^{-4.6},10^{-4.7},10^{-4.8},10^{-4.9}\right]$                     & 256x256 & 128 \\ \hline

$R=6$                                   & $n_x,n_y$   & $n_z$  \\ \hline
$\text{Ek}=5\cdot\left[10^{-5.6},10^{-5.7},10^{-5.8},10^{-5.9},10^{-6}\right]$ & 256x256 & 96  \\ \hline
$\text{Ek}=5\cdot\left[10^{-5.0},10^{-5.1},10^{-5.4}\right]$                      & 256x256 & 128 \\ \hline
$\text{Ek}=5\cdot\left[10^{-5.2},10^{-5.3}\right]$                       & 256x256 & 160 \\ \hline

$R=6$, $\epsilon=0$, $L_x=2$ (to determine $l_c$)                                   & $n_x,n_y$   & $n_z$  \\ \hline
$\text{Ek}=5\cdot\left[10^{-5},10^{-5.1},...,10^{-5.8},10^{-5.9},10^{-6}\right]$ & 512x512 & 128  \\ \hline

$\mathrm{Ra}=1.3\cdot 10^8$, $\epsilon=0$, $L_x=2$ (to determine $l_c$)                                   & $n_x,n_y$   & $n_z$  \\ \hline
$\text{Ek}=5\cdot\left[10^{-4.5},10^{-4.6},....,10^{-5.8},10^{-5.9},10^{-6}\right]$ & 512x512 & 128  \\ \hline
\end{tabular}
\label{tab:resolutiontable}
\end{table}

A table of resolutions used for our simulations is given in Table~\ref{tab:resolutiontable}. High Rayleigh number simulations were carried out with higher vertical resolutions. Simulations with higher ellipticities were not found to require higher resolutions. In addition, to fully resolve the heat flux spectrum for calculations of the convective length scale we opted to use a 2x2x1 ($L=2$) box, as indicated in the bottom two columns of Table~\ref{tab:resolutiontable}. 

\begin{figure}
    \centering
    \includegraphics[width=\linewidth]{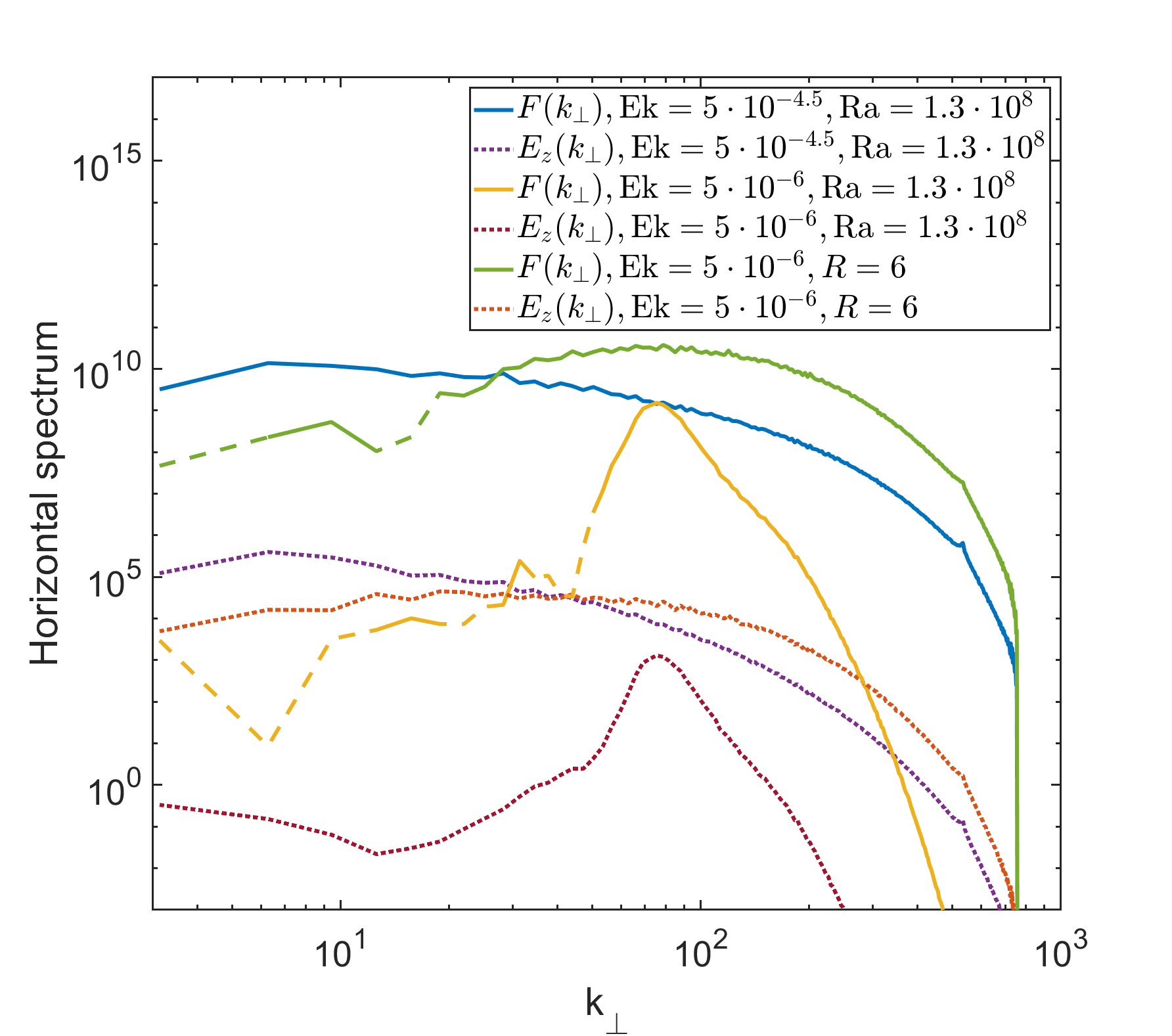}
    \caption{The heat flux $F(k_\perp)$ (solid and dashed lines) and the vertical kinetic energy $E_z(k_\perp)$ (dotted lines) spectra as a function of horizontal wavenumber $k_\perp$. The spectra are plotted at the most extreme values of the surveyed parameter space in Ekman number. The largest Ekman number, $\mathrm{Ek}=5\cdot10^{-4.5}$, at fixed $\mathrm{Ra}=1.3\cdot10^8$ is plotted in solid-blue and dotted-purple, the smallest Ekman number, $\mathrm{Ek}=5\cdot10^{-6}$, at fixed $\mathrm{Ra}=1.3\cdot10^8$ in solid-yellow and dotted-burgundy, and the smallest Ekman number at fixed $R=6$ in solid-green and dotted-orange. Dashed parts of the heat flux spectra indicate negative heat flux for these scales.}
    \label{fig:lc_spectracheck}
\end{figure}

The horizontal spectra used in the calculation of the length scale are shown in Fig.~\ref{fig:lc_spectracheck}. Both the heat flux $F(k_\perp)$ and the vertical kinetic energy $E_z(k_\perp)$ spectra are plotted. The heat fluxes are shown with solid (and dashed) lines, and the vertical kinetic energy spectra with dotted lines. The spectra from cases with the most extreme parameters we considered are displayed; the largest Ekman number, $\mathrm{Ek}=5\cdot10^{-4.5}$, at fixed $\mathrm{Ra}=1.3\cdot10^8$ is plotted in solid-blue and dotted-purple, the smallest Ekman number, $\mathrm{Ek}=5\cdot10^{-6}$, at fixed $\mathrm{Ra}=1.3\cdot10^8$ in solid-yellow and dotted-burgundy, and the smallest Ekman number at fixed $R=6$ in solid-green and dotted-orange. The heat flux spectra at small Ekman number contain dashed parts, indicating negative heat flux for these $k_\perp$. The heat flux tends to be concentrated towards larger wave numbers with broad distributions. The strong peak in solid-yellow and dotted-burgundy emerges because this simulation is close to onset, and thus strongly follows the linear onset wave number. The conventional rule that the power in the peak of the spectrum must be a factor of at least $10^3$ larger than at the anti-aliasing scale is maintained in the kinetic energy spectra as well as the heat flux spectra. Our simulations are therefore likely to be spatially converged in the horizontal plane.

\begin{figure}
    \centering
    \includegraphics[width=\linewidth,
    trim=0cm 0cm 0cm 0cm,clip=true]{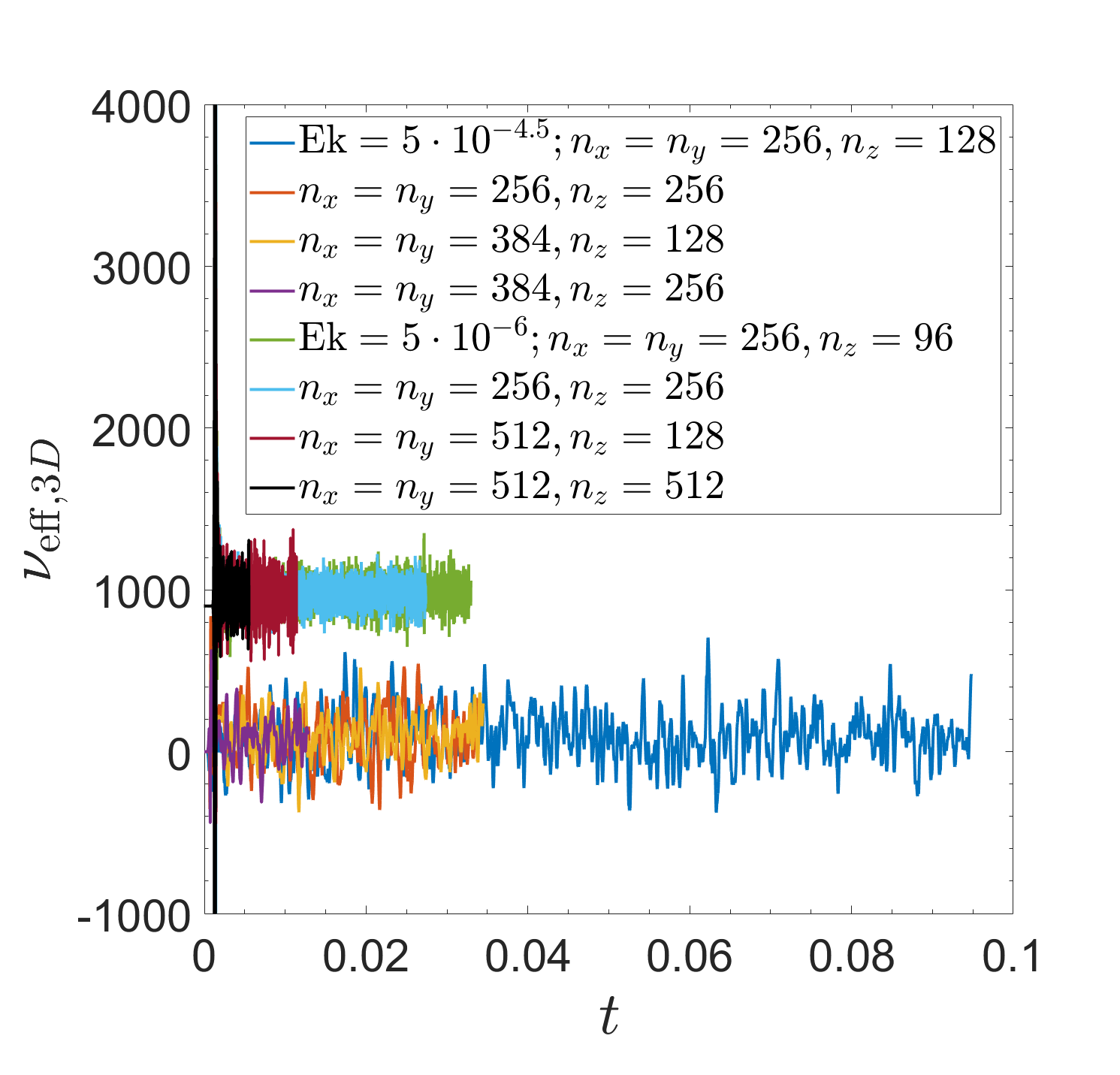}
    \caption{The effective viscosity at largest $\mathrm{Ek}=5\cdot10^{-4.5}$, $\mathrm{Ra}=1.3\cdot10^8$ in blue, orange, yellow and purple and at smallest $\mathrm{Ek}=5\cdot10^{-6}$, $R=6$ in green, cyan, burgundy and black. The effective viscosities at $\mathrm{Ek}=5\cdot10^{-6}$ are increased by a factor of 100 to offset them for clarity. Apart from fluctuations the effective viscosity is unchanged with increased resolution.}
    \label{fig:nueff_rescheck}
\end{figure}

The main parameter in this work is the energy injection rate and the resulting effective viscosity. Here we plot the effective viscosity at the two most extreme parameters used in our simulations in Fig.~\ref{fig:nueff_rescheck}. This shows the smallest Ekman number, fixed supercriticality $R=6$ and $\epsilon=0.03$, and largest Ekman number, fixed $\mathrm{Ra}=1.3\cdot10^8$ and $\epsilon=0.02$. Low ellipticities are specifically chosen to avoid bursts of the elliptical instability. The effective viscosities in these simulations take different values due to the different Rayleigh and Ekman numbers, but are further offset for clarity. There is no change as we increase the resolution for this quantity at either extreme of parameter space, so we conclude that the effective viscosities we measured are numerically converged with our adopted resolutions.

\section{MESA Inlists}
\label{sec:MESA_inlists}

The illustrative models in \S~\ref{sec:Detailed} are based on the MESA test suite case \textit{make\_planets}. We highlight here changes in the inlist files we used to obtain these models. Any parameters not mentioned here are unchanged from the test suite default values. The \textbf{\textit{inlist\_create}} and \textbf{\textit{inlist\_core}} are the same for both the Jupiter and Hot Jupiter model: \\
\noindent
\underline{\textbf{\textit{inlist\_create}}}\\
\textbf{\textit{max\_model\_number = 1020}}\\
\textbf{\textit{initial\_Y=0.27}}\\
\noindent
\underline{\textbf{\textit{inlist\_core}}}\\ 
\textbf{\textit{dlg\_core\_mass\_per\_step=0.002d0}}\\
The differences between these models lies in \textbf{\textit{inlist\_evolve}}; for the Jupiter model:\\
\noindent\textbf{\textit{max\_model\_number = 2500}}\\
\textbf{\textit{irradiation\_flux = 50000.d0}}\\
\textbf{\textit{inject\_uniform\_extra\_heat = 0.0d0}}\\
\textbf{\textit{max\_age=4.5d9}}\\
\noindent
and for the Hot Jupiter model:\\
\noindent\textbf{\textit{max\_model\_number = 2500}}\\
\textbf{\textit{irradiation\_flux = 1.d9}}\\
\textbf{\textit{inject\_uniform\_extra\_heat = 0.05d0}}\\
\textbf{\textit{max\_age=4.5d9}}.


\bsp	
\label{lastpage}
\end{document}